\documentclass[reprint,superscriptaddress,amsmath,amssymb,aps,prb,floatfix]{revtex4-2}

\usepackage{graphicx}
\usepackage{dcolumn}
\usepackage{bm}
\usepackage{amsmath} 
\usepackage{physics}
\usepackage{cancel}
\usepackage[colorlinks,
    linkcolor=blue,
    anchorcolor=red,
    citecolor=green
]{hyperref}

\usepackage{simpler-wick}
\usepackage{array}
\usepackage{times}
\usepackage[caption = false]{subfig}

\usepackage{soul}

\allowdisplaybreaks

\newcommand{\red}[1]{\textcolor{red}{#1}}

\newcommand{\mi}{\mathrm{i}}

\begin{document}

\title{Multichannel topological Kondo models and their low-temperature conductances}

\author{Guangjie Li}
\affiliation{%
Department of Physics and Astronomy, Purdue University, West Lafayette, Indiana 47907, USA.
}
\affiliation{%
Department of Physics and Astronomy, University of Utah, Salt Lake City, Utah 84112, USA.
}

\author{Elio J. K{\"o}nig}
\affiliation{%
Department of Physics, University of Wisconsin-Madison, Madison, Wisconsin 53706, USA
}%
\affiliation{%
Max-Planck Institute for Solid State Research, 70569 Stuttgart, Germany
}%

\author{Jukka I. V{\"a}yrynen}
\affiliation{%
Department of Physics and Astronomy, Purdue University, West Lafayette, Indiana 47907, USA.
}

\date{\today}

\begin{abstract}
In the multichannel Kondo effect, overscreening of a magnetic impurity by conduction electrons leads to a  frustrated exotic ground state. 
It has been proposed that multichannel topological Kondo (MCTK) model involving topological Cooper pair boxes with $M$ Majorana modes [SO($M$) ``spin”] and $N$ spinless electron channels exhibits an exotic intermediate coupling fixed point.
This intermediate fixed point has been analyzed through large-$N$ perturbative calculations, which gives a zero-temperature conductance decaying as $1/N^2$ in the large-$N$ limit. 
However, the conductance at this intermediate fixed point has not been calculated for generic $N$.
Using representation theory, we verify the existence of this intermediate-coupling fixed point and  find the strong-coupling effective Hamiltonian for the case $M=4$. 
Using conformal field theory techniques for SO($M$), we generalize the notion of overscreening and conclude that the MCTK model is an overscreened Kondo model. 
We find the fixed-point finite-size energy spectrum and the leading irrelevant operator (LIO). We express the fixed-point conductance  in terms of the modular S-matrix of SO($M$) for general $N$, confirming the previous large-$N$ result. 
We describe the finite-temperature corrections to the conductance by the LIO and find that they are qualitatively  different for the cases $N=1$ and $N\geq2$  due to the different fusion outcomes with the current operator. 
We also compare the multichannel topological Kondo model to the topological symplectic Kondo model.
\end{abstract}

\maketitle

\tableofcontents

\section{Introduction}
The Kondo model explains the local minimum resistivity at low temperatures in certain noble metals using a spin-spin exchange interaction between the magnetic impurity and conduction electrons~\cite{kondo1964resistance}. The multichannel generalization of it \cite{Nozieres1980} with fruitful experimental developments in mesoscopic devices~\cite{GoldhaberGordon1998, Pustilnik_2004, 2007Natur.446..167P, KellerGoldhaberGordon2015,2015Natur.526..233I,2018Sci...360.1315I,PouseGoldhaberGordon2022} leads Kondo physics to a new era due to not only the experimental measurement of conductance, but also the potential applications to topological quantum computations using emergent anyonic excitations \cite{affleck1995conformal, LopesSela2020, Komijani2020, PhysRevB.105.035151, PhysRevLett.129.227703,RanTsvelik2024}. 
The key requirement is overscreening: the impurity spin ($S$) is smaller than the largest possible local total spin ($N/2$) for $N$-channel conduction electrons. Overscreening indicates an intermediate coupling fixed point \cite{Nozieres1980}, whose existence was verified in the large-$N$ limit where the intermediate coupling fixed point moves toward weak coupling and becomes perturbatively accessible in the $1/N$ expansion \cite{PhysRevLett.70.686}.

The intermediate coupling fixed point of the $N\geq2$-channel Kondo model is an exotic state not described by free electrons. As first demonstrated using the Bethe Ansatz technique~\cite{AndreiDestri1984,TsvelickWiegmann1985} the impurity entropy is $S_{\text{imp}}=\ln[g(N)]$ where the generally irrational number $g(N)=2\cos[\pi/(N+2)]$ can be interpreted as the effective ground state degeneracy of the impurity problem. Crucially, it coincides with the quantum dimension of the anyon ``$1/2$'', labeled by its representation in the $\text{SU}(2)_N$ fusion category \cite{affleck1995conformal}. As shown
by Emery and Kivelson \cite{PhysRevB.46.10812} (see also Refs. \cite{1994PhRvB..4910020S,Coleman_1995,Rozhkov_1998}) for the two-channel Kondo (2CK) model, an emergent Majorana at the impurity spin will be decoupled from the conduction electrons.

Since the 1990s (multichannel) Kondo physics has also been realized in artificial quantum emulators. Of particular relevance are strongly interacting quantum dots coupled to electronic leads~\cite{Pustilnik_2004}. In this context, while there has been progress towards measuring the entropy of a mesoscopic device \cite{hartman2018directentropymeasurement, child2022qdotentropymeasurement, han2022fractionalentropy}, a somewhat more natural experimental observable than the impurity entropy is the (trans-)conductance across the dot. Crucially, at zero temperature it takes an $N$-dependent universal value, while the conductance correction near $T = 0$ is given by the leading perturbation, namely the least irrelevant operator (LIO). The first-order correction $T^{\Delta_{\text{LIO}}-1}$ with $\Delta_{\text{LIO}}=1+2/(N+2)$ dominates unless it vanishes in which case one considers the second-order correction $T^{2(\Delta_{\text{LIO}}-1)}$~\cite{2007Natur.446..167P,2015Natur.526..233I}. In any case, the temperature dependence demonstrates non-Fermi liquid behavior, i.e., non-$T^2$ behavior. 
Here, the LIO is formed by the first descendants of the adjoint (spin-1) primary operators~\cite{PhysRevLett.67.161,affleck1990current,affleck1991kondo,affleck1991critical,PhysRevB.48.7297,ludwig1994exact,affleck1995conformal,PhysRevLett.67.3160,PhysRevB.58.3794}.

Theoretically, further platforms of
Kondo effects in mesoscopic devices with non-Fermi liquid behavior have been proposed for setups with symmetry groups \cite{doi:10.7566/JPSJ.90.024708} other than SU(2). These include $\text{SO}(M)$, e.g.~the (multichannel) topological Kondo model whose impurity consists of $M$ Majorana zero modes on a floating topological superconducting island coupled to normal spin-polarized metallic leads \cite{PhysRevLett.109.156803,PhysRevLett.110.216803,PhysRevLett.130.066302,BollmannKoenig2024}, as well as $\text{Sp}(2k)$, e.g. the topological symplectic Kondo model whose impurity consists of $k$ one-dimensional topological end states that are coupled to spinful leads \cite{PhysRevB.107.L201401,KONIG2023169231,LotemAnisotropyTSK}. The existence of the intermediate coupling fixed point of the topological symplectic Kondo model has been demonstrated through the strong coupling analysis \cite{PhysRevB.107.L201401}. However the strong coupling of the $\text{SO}(M)$ topological Kondo model has not been well studied and thus the existence of the intermediate coupling fixed point has not been fully demonstrated generally for any $M$, although the mapping from the $M=3,4$ topological Kondo models to the MCK models and the large-$N$ results of the $N$-channel topological Kondo model both indicate its existence \cite{PhysRevLett.109.156803,PhysRevLett.130.066302}. 
In this paper we show that the strong-coupling fixed point of the $\text{SO}(M)$ topological Kondo model is indeed unstable, suggesting flow towards an intermediate fixed point. We will also clarify the meaning of overscreening in the topological Kondo model. 

Another open question that we will address in this work concerns the low-temperature conductance. The low-temperature conductance correction calculated in the large-$N$ limit is $T^{(M-2)/(2N)}$~\cite{PhysRevLett.130.066302}, which suggests $T^{\Delta}$ where $\Delta=(M-2)/(2N + M-2)$ is the scaling dimension for the adjoint primary operator. The first descendant of this primary operator gives the LIO with scaling dimension $1+\Delta$. However, it is also known that the $N=1$ topological Kondo model has a conductance correction $T^{2\Delta}$~\cite{PhysRevLett.109.156803,PhysRevLett.110.216803,PhysRevResearch.2.043228}, which seems contradictory to the above. A similar change in the temperature exponent of the conductance correction occurs in the multichannel charge-Kondo effect~\cite{2018Sci...360.1315I,bao2017quantumhallchargekondo}. In this paper, by developing the boundary conformal field theory for the topological Kondo model, we address the reason for this change in the temperature exponent. 

This manuscript is structured as follows. 
In Sec.~\ref{sec:existence}, we first analyze the $\text{SO}(M)$ topological Kondo exchange interaction using representation theory and show that it is an overscreened Kondo interaction which supports the existence of the intermediate coupling fixed point. In Sec.~\ref{sec:CFT}, we introduce the conformal field theory techniques and the Kac-Moody algebra for the topological Kondo intermediate coupling fixed point. In Sec.~\ref{sec:zeroTconduc}, by identifying the important operators from the intermediate fixed point, we give the analytical results of the zero temperature conductance and the finite-temperature conductance correction for the $N$-channel topological Kondo model with $M$ Majoranas. In Sec.~\ref{sec:Spconductance}, we further discuss the analogous conductance correction of the topological $\text{Sp}(2k)$ Kondo model.

\section{Instability of the strong coupling fixed point {for SO(M) Kondo models}}
\label{sec:existence}
\begin{figure}[tb]
\centering
\includegraphics[width=0.95\columnwidth]{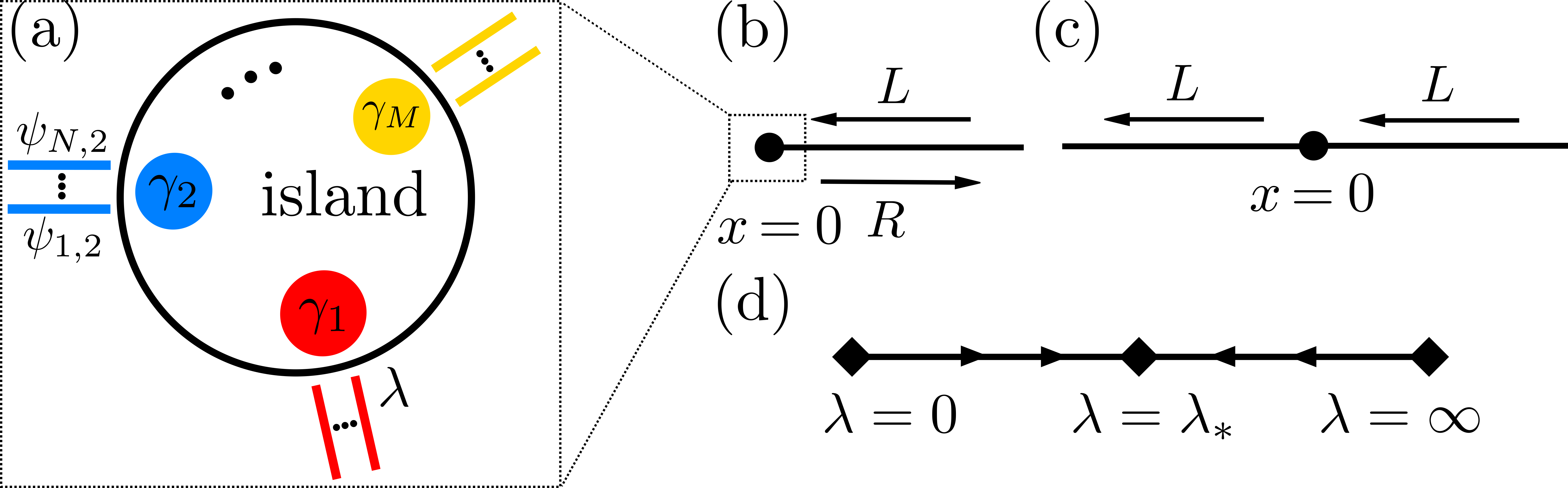}
\caption{(a) Setup of the multichannel topological Kondo model implemented as a  Coulomb blockaded Majorana island with $M$ Majoranas (flavors) coupled to $N$ metallic leads (channels) for each Majorana. 
(b) Each channel is assumed to host linearly dispersing left ($L$) and right ($R$) moving modes on a half axis, coupled to the impurity at $x  = 0$. 
(c) Unfolding to chiral model:  The right movers can be considered as a continuation of the left movers defined on the full axis, see Eq.~(\ref{eq:H0}). 
(d) The RG flow of the Kondo coupling $\lambda$.} 
\label{fig:conducMCTKsetup}
\end{figure}

The total Hamiltonian $H_0 + H_{\text{MCTK}}$ of the $N$ channel topological Kondo model consists of the linearized kinetic energy of the itinerant fermions ($H_0$), and the $N$-channel $\text{SO}(M)$ topological Kondo interaction $H_{\text{MCTK}}$ (see Fig.~\ref{fig:conducMCTKsetup}a):
\begin{align}
    & H_0 = \frac{v_{\text{F}}}{2\pi} \int_{-l}^l \mathrm{d}x\,\sum_{n=1}^N \sum_{\alpha =1}^M 
    \mathrm{i}\,\psi^\dagger_{n,\alpha}(x) \partial_x \psi_{n,\alpha}(x), \label{eq:H0}\\
    & H_{\text{MCTK}} = \lambda \sum_{n=1}^N \sum_A S^A J^A_n(x=0) , \label{eq:MCTK}
\end{align}
where $v_{\text{F}}$ is the Fermi velocity and $2l$ is the length of the system, which can be taken to infinity at the end~\cite{ludwig1994field}. Here, we considered the right mover as a continuation of the left mover to the negative axis \cite{affleck1995conformal,ludwig1994field} (see Fig.~\ref{fig:conducMCTKsetup}b-c).
The $\text{SO}(M)$ impurity operators are $S^{A=(\alpha,\beta)} = -\mi \gamma_{\alpha}\gamma_{\beta}/2$ with $\gamma_{1,...,M}$ denoting the $M$ Majorana zero modes. The conduction electron operators in the $n$th channel are $J^{A=(\alpha,\beta)}_n=-\mi (\psi^\dagger_{n,\alpha}\psi_{n,\beta}-\psi^\dagger_{n,\beta}\psi_{n,\alpha})$ and there are $M(M-1)/2$ $(\alpha, \beta)$ pairs that satisfy $\alpha < \beta$ when $\alpha,\beta = 1,\dots,M$ label flavors. 
The operators $S^A$ and $J^A_n$ are different representations of the $\text{SO}(M)$ generators.

A weak Kondo coupling $\lambda>0$ flows to the stronger coupling under renormalization group (RG) flow~\cite{PhysRevLett.109.156803} (see Fig.~\ref{fig:conducMCTKsetup}d). 
For simplicity we consider an idealized case of only one Kondo coupling, i.e.~fully isotropic exchange; while weak flavor anisotropies are irrelevant under the RG flow~\cite{PhysRevLett.109.156803,PhysRevLett.130.066302,matanrelevanceofanisotropy2024}, channel anisotropy is not~\cite{PhysRevLett.130.066302}.
In the large-$N$ limit, in Ref.~\cite{PhysRevLett.130.066302} some of us perturbatively calculated the RG equations and found that there is a stable intermediate coupling fixed point $\lambda_* = 1/(2N\rho)$ where $\rho = 1/( \pi v_F)$ is the constant density of states per length. 
This intermediate-coupling fixed point indicates an unstable strong-coupling fixed point (see Fig.~\ref{fig:conducMCTKsetup}d). 
However, the perturbative results can no longer be trusted in the case $N \sim \mathcal O(1)$, where the intermediate fixed point (if any) no longer resides in the weak coupling regime.

Here we first evince the presence of the intermediate non-trivial fixed point. Trivially, as for any Kondo problem, the fixed point $\lambda =0$ is unstable (because $\lambda$ is marginally relevant). Next we consider $\lambda = \infty$ and study if this fixed point is stable or not. 
To this end, we assume dominant $H_{\text{MCTK}}$,  Eq.~(\ref{eq:MCTK}), in the strong coupling regime and momentarily neglect kinetic energy, Eq.~(\ref{eq:H0})~\cite{Nozieres1980}. In the lattice version of Fig.~\ref{fig:conducMCTKsetup}b, $x$ takes discrete values $0,1,2,\dots$ and $0$ denotes the closest site to the impurity. At $x=0$, we notice that
\begin{equation}
J^A= -\mi(\gamma_{\alpha,R}\gamma_{\beta,R}+\gamma_{\alpha,I}\gamma_{\beta,I})/2 \equiv S^A_{R} +S^A_{I}\,, \label{eq:JAtwoSA}
\end{equation}
when introducing Majorana representation  $\psi_{\alpha}=(\gamma_{\alpha,R}+\mi \gamma_{\alpha,I})/2$ for each conduction electron channel $n$ (the channel index $n$ is omitted). 
Thus, the impurity represented by operator $S^A$ is screened by conduction electrons represented by $2N$ $S^A$s (see also \cite{tsvelik2014topological,MitchellAffleck2021,LibermanSela2021} for the case of single conduction $S^A$). Since two $S^A$s can combine into a singlet, see Eq.~\eqref{eq:RTA}, this  means that Eq.~(\ref{eq:MCTK}) is an overscreened Kondo interaction. This overscreening in terms of the number of $S^A$ indicates the instability of the strong-coupling fixed point and therefore leads to a stable intermediate-coupling fixed point. By analogy with the 2CK model, i.e., when the Majorana island is screened by two conduction Majorana channels, we expect that the ground states of Eq.~(\ref{eq:MCTK}) with $N=1$ form a new impurity that is represented by $S^A$ and will be further screened by the next nearest ($x=1$) conduction electrons.

The above analogy can be shown by solving the ground states of Eq.~(\ref{eq:MCTK}) using the following standard Casimir form:
\begin{equation}\label{eq:casimirops}
 \sum_A S^A J^A=\frac{1}{2}\qty[\sum_A (S^A+J^A)^2-(S^A)^2-(J^A)^2].
\end{equation}
These three operators on the right are the Casimir operators, and the Casimir invariant $c(R)$ for each Casimir operator is determined by its representation $R$. In terms of the Dynkin's labels $a_i$ (non-negative integers), i.e., $R=\sum_{i=1}^r a_i \mu_i$ where the integer $r$ is the rank of the Lie algebra [$r = m$ for SO$(M=2m)$ and SO$(M=2m+1)$] and $\mu_i$ are the fundamental representations, the Casimir invariant is
\begin{equation}\label{eq:casimirinv}
c(R)=\sum_{i,j=1}^{m}(\alpha_i\cdot \alpha_i)(a_i/2+1)(C^{-1})_{ij}a_j.   
\end{equation}
The vector $\alpha_i$ are the simple roots and the matrix $C_{ij}\equiv 2({\alpha_i \cdot \alpha_j})/({\alpha_i\cdot \alpha_i})$ is the Cartan matrix \cite{georgi2000lie,ma2007group,zee2016group,LieARTFeger_2020}.  Thus, the ground states and their energy of Eq.~(\ref{eq:casimirops}) are both determined by the representations of $S^A$ and $J^A$.

Before going to the representation theory, one can calculate the Casimir invariant of $S^A$ using only the commutation relations $\{\gamma_i, \gamma_j\}=2\delta_{ij}$:
\begin{equation}\label{eq:casimirSA}
\sum_A (S^A)^2=- \sum_{\alpha<\beta} \gamma_{\alpha} \gamma_{\beta}\gamma_{\alpha} \gamma_{\beta}/4=M(M-1)/8.
\end{equation}
This result can be verified by plugging the Dynkin's labels of $R(S^A)$ (see Appendix \ref{ap:so2m}), 
\begin{equation}
R(S^A)=
\begin{cases}
\mu_{m-1}\oplus\mu_{m},  & \text{if}\ M=2m \\
\mu_{m}, &  \text{if}\ M=2m+1
\end{cases}\label{eq:RSA}
\end{equation}
into Eq.~(\ref{eq:casimirinv}), which also results in the right side of Eq.~(\ref{eq:casimirSA}). For the following part in this section, we focus on the even case $M=2m$.
The representations $\mu_{m-1}$ and $\mu_{m}$ are the two spinor fundamental representations which have the same dimension $2^{m-1}$ but different total parities $\Pi_{\alpha=1}^{2m} \gamma_{\alpha} = \pm 1$.

In order to obtain the Casimir invariant for $J^A$, we note that $J^A (x=0)$ commutes with the number operator and thus is block diagonal in the single-site $x=0$ Fock space. Therefore $R(J^A)$ is a reducible representation. 
Consequently, we introduce the $K$-particle sector at the site $x=0$ and define $J^{A,K}$ the corresponding block in $J^{A}$.  
The dimension of the $K$-particle representation $R(J^{A,K})$ is $\binom{2m}{K}$. The total dimension of $R(J^A)$ is $\sum_{K=0}^{2m} \binom{2m}{K}=2^{2m}$ where $0\le K\le M$; this total dimension agrees with what we expect from Eq.~\eqref{eq:JAtwoSA}.  

The $K$-particle sector and $M-K$-particle sector are identical due to particle-hole symmetry. This can be seen from the Casimir operator for $J^A$ in the $K$-particle sector
\begin{equation}\label{eq:casimirJA}
\sum_A J^{A,K} J^{A,K} =\sum_{\alpha\neq\beta}\psi^{\dagger}_{\alpha}\psi_{\alpha}(1-\psi^{\dagger}_{\beta}\psi_{\beta}) = K(M-K).
\end{equation} 
The second equality above  is obtained by noting that for any $K$-particle state, in order to have a nonzero average for $\psi^{\dagger}_{\alpha}\psi_{\alpha}(1-\psi^{\dagger}_{\beta}\psi_{\beta})$, $\alpha$ should be one of the $K$ occupied flavors, and $\beta$ should be one of the $M-K$ unoccupied flavors. In total, there are $K(M-K)$ $(\alpha, \beta)$ pairs. Eq.~(\ref{eq:JAtwoSA}) means that the representation $R(J^A)$ can be given by $R(S^A_{R})\otimes R(S^A_{I})$ and thus we get the representations of $J^A$ for $\text{SO}(M=2m)$ (see Appendix \ref{ap:so2m}): 
\begin{align}
R(J^A) = R(S^A)\otimes R(S^A) \equiv\bigoplus_{K=0}^{2m} R(J^{A,K})\label{eq:RTA}
\end{align}
using the representation $R(S^A)$, i.e., Eq.~(\ref{eq:RSA}).
We check that Eq.~(\ref{eq:casimirJA}) matches the result by plugging the Dynkin's labels of $R(J^{A,K})$ [Eq.~(\ref{eq:RTA})] into Eq.~(\ref{eq:casimirinv}).

Now, we can decompose $R(S^A) \otimes R(J^{A,K})$ for each $K$-particle sector. 
While the corresponding Casimir invariant can be evaluated, its full expression is too complex to show here, but the reader can refer to Appendix~\ref{ap:repSJ} for more information. Nevertheless, we find that $\mu_{m-1}$ and $\mu_m$ have the smallest Casimir invariant $c(R)$ [Eq.~(\ref{eq:casimirSA})] except the trivial representation $R=0$ and always exist in the decompositions of $R(S^A) \otimes R(J^{A,K})$ for any $0 \leq K \leq 2 m$. Thus, the ground states of Eq.~(\ref{eq:MCTK}) lie in the $K=m$-particle sector (half-filling), where Eq.~(\ref{eq:casimirJA}) is maximal and Eq.~(\ref{eq:casimirops}) has its  minimum. These ground states labeled by $\mu_{m-1}\oplus\mu_{m}$ form a new impurity [same representation as the original impurity, see Eq.~(\ref{eq:RSA})]. We thus expect by symmetry that it couples to conduction electrons as in Eq.~\eqref{eq:MCTK} and thus will be screened by the electrons at site $x=1$, which confirms that the leads can be considered as two-channel Majorana-lead according to Eq.~(\ref{eq:JAtwoSA}), which will overscreen the Majorana-island just as in the conventional 2CK effect. Moreover, the effective Hamiltonian of strong coupling is still a topological Kondo interaction but with a new Kondo coupling $\lambda' \sim t^2/\lambda$ where $t$ is the kinetic energy of the itinerant electrons. The strong coupling limit of $\lambda$ is the weak coupling limit of $\lambda'$, which we know to be unstable. This weak-strong duality predicts a self-dual point, which is the intermediate fixed point (see Fig.~\ref{fig:conducMCTKsetup}d).
We show an explicit example in the Appendix~\ref{ap:so4Heff} for $M=4$ and $N=1$. The ground states in the strong coupling fixed point are in the $K=m=2$-particle sector, which can be represented by two decoupled spin-1s (a six-dimensional space formed by placing $K=2$ particles into $M=4$ available orbitals). The impurity can be represented by two decoupled spin-1/2s. Thus, the 1-channel topological $\text{SO}(4)$ Kondo model is equivalent to 2CK and is overscreened. This equivalence comes from the fact $\text{SO}(4) \sim \text{SU}(2)\times \text{SU}(2)$ (see Appendix \ref{ap:so4gs} and the supplementary material of Ref.~\cite{PhysRevLett.130.066302}). Similar arguments about overscreening from the strong-coupling perspective are also used in Refs.~\cite{Nozieres1980,PhysRevB.50.17732,PhysRevB.54.14918}. 

The multichannel case ($N> 1$) with more screening is also an overscreened Kondo model. In conclusion, we have shown the instability of the strong-coupling fixed point and thus the existence of the intermediate-coupling fixed point (see the $\lambda=\lambda_*$ point in Fig.~\ref{fig:conducMCTKsetup}d).

\section{Conformal Field Theory at the intermediate-coupling fixed point}\label{sec:CFT}
The powerful boundary conformal field theory method was successfully applied to the intermediate-coupling fixed point of the multichannel $\text{SU}(2)$ Kondo model \cite{affleck1990current,affleck1991critical,affleck1991kondo,affleck1995conformal,ludwig1994field,ludwig1994exact}. As we verified the strong-coupling instability of the multichannel $\text{SO}(M)$ topological Kondo model in the last section, suggesting the existence of an intermediate-coupling fixed point, we introduce in this section the $\text{SO}(M)$ generalization of the conformal field theory description.

\subsection{The definition of overscreening}
Below Eq.~\eqref{eq:JAtwoSA} we argued that the topological Kondo model is overscreened. In this section we make this notion more precise. 
Recall that overscreening in the $N$-channel (spin-1/2 conduction electrons) $\text{SU}(2)$ Kondo model with impurity spin-$s$ means  $2s<N$~\cite{Nozieres1980}. This constraint for the impurity spin is also the cut-off for the allowed spins (representations) in the $\text{SU}(2)_N$ affine Lie algebra, i.e., the Kac-Moody algebra, which is part of the conformal field theory describing the intermediate-coupling fixed point \cite{affleck1995conformal}. 
Thus, we can think of overscreening as the requirement that the impurity representation is allowed in the affine Lie algebra. 
We note however the following two exceptions: (i) the case $s=N/2$  corresponding to exact or perfect screening, and (ii) the (trivial) case of a scalar impurity [spin-$0$], which are both allowed in the Kac-Moody algebra but flow to the same free fermion theory as the free fixed point. Next, we generalize the definition of overscreening using the representation cut-off for the $\text{SO}(M)$ case. 

The symmetry current operator for the $N$-channel $\text{SO}(M)$ Kondo model, Eq.~\eqref{eq:MCTK}, is defined as
\begin{equation}
\widetilde{J}^{A}(x)=\sum_{n=1}^{N}J_{n}^{A}=\sum_{n=1}^{N}\sum_{\alpha,\beta=1}^{M}\psi_{n,\alpha}^{\dagger}(x)(T^{A})_{\alpha\beta}\psi_{n,\beta}(x) ,\label{eq:currentopreal}
\end{equation}
where $T^{A}$s are the traceless generators of the $\text{SO}(M)$ group,  defined as $(T^{A})_{\alpha\beta}\equiv(T^{rs})_{\alpha\beta}=\mathrm{i}(\delta_{\alpha}^{r}\delta_{\beta}^{s}-\delta_{\beta}^{r}\delta_{\alpha}^{s}) $.
Here the generators are labeled by the pair $r,s$ of indices $r<s = 1, ..., M$. We may alternatively label the generators by a single integer $A=1,\dots,M(M-1)/2$. The generators 
satisfy $\mathrm{Tr}(T^{A}T^{B})=2\delta^{AB}$ and $[T^{A},T^{B}]=\mathrm{i}\sum_{C}f^{ABC}T^{C}$
where $f^{ABC}=(-\mathrm{i}/2)\mathrm{Tr}([T^{A},T^{B}]T^{C})$ is
the structure constant. The Kac-Moody algebra for the Fourier components of Eq.~(\ref{eq:currentopreal}) is
\begin{equation}
[\widetilde{J}_{p}^{A},\widetilde{J}_{p'}^{B}]=\mathrm{i}\sum_{C}f^{ABC}\widetilde{J}_{p+p'}^{C}+2N\,p\delta^{AB}\delta_{p,-p'} , \label{eq:SOM2NKacMoody}
\end{equation}
with the prefactor in the second term defining the level $2N$ (see Appendix \ref{ap:SOM2N}). Thus, the affine Lie algebra for the $N$-channel topological Kondo model is $\text{SO}(M)_{2N}$ (see Ref.~\cite{PhysRevLett.113.076401} for the $N=1$ case)  because each screening fermion channel is two screening Majorana channels as shown by Eq.~(\ref{eq:JAtwoSA}). For $N$ Majorana channels one finds  $\text{SO}(M)_{N}$ \cite{tsvelik2014topological}.

By requiring primary states to have non-negative norm, we can find that the allowed representations $R=\sum_{i=1}^m a_i \mu_i=\vec{\ell}$ for the $\text{SO}(M)_{2N}$ Kac-Moody algebra must satisfy (see Appendix~\ref{ap:repcutSO2mk} and \ref{ap:repcutSO2mp1k} for derivations)
\begin{equation}
\begin{cases}
\ell_{i}\pm\ell_{j}\leq 2N,  & \text{if } M=2m \\
\ell_{i}\pm\ell_{j}\leq 2N \ \text{and} \ \ell_{j}\leq2N, & \text{if } M=2m+1
\end{cases}
\label{eq:repcutoff}
\end{equation}
where $1\leq i<j=1,...,m$, $a_{i}=\ell_{i}-\ell_{i+1},\,i=1,...,m-1$ and
\begin{equation}
a_m=
\begin{cases}
\ell_{m-1}+\ell_{m}, & \text{if } M=2m \\
 2\ell_{m}, & \text{if } M=2m+1 
\end{cases}.
\end{equation}
The necessary requirements are thus that the Dynkin labels 
\begin{equation}
a_{i=1,...,m-1}\leq 2N,\ \text{and}, a_m \leq
\begin{cases}
2N, & \text{if } M=2m \\
4N, & \text{if } M=2m+1 
\end{cases}.
\end{equation}
When the impurity representation satisfies all inequalities in Eq.~(\ref{eq:repcutoff}) but upon fusion with the conduction electrons (see Sec.~\ref{subsec:fusion rules}) still gives the free-fermion spectrum~\cite{ludwig1994field}, we define this case as ``trivial screening''. The overscreening is when the representation satisfies Eq.~(\ref{eq:repcutoff}) but does not belong to trivial screening. For the $\text{SU}(2)_N$ case, the trivial screening occurs when the impurity spin $s=0$ or $N/2$. All other cases with spin $s=1/2, 1,\dots, (N-1)/2$ are overscreening. 
We can easily check that the impurity representation $R(S^A)$, i.e., Eq.~(\ref{eq:RSA}) satisfy Eq.~(\ref{eq:repcutoff}) for any $N\geq 1$ and does not belong to the trivial screening.  
Thus, we conclude that the multichannel topological Kondo model is an overscreened Kondo model which matches our conclusion in the last section that the strong-coupling fixed point is unstable.

\subsection{Fusion rules, quantum dimensions,  and scaling dimensions \label{subsec:fusion rules}}
The fusion rule for two allowed representations $R_1$ and $R_2$ of the Kac-Moody algebra is $R_1 \times R_2 = \sum_{R_3} N^{R_3}_{R_1 R_2} R_3$ where $N^{R_3}_{R_1 R_2} = \sum_{R_4} S_{R_1, R_4} S_{R_2, R_4} S^*_{R_3, R_4}/S_{0, R_4}$ by the Verlinde formula~\cite{VERLINDE1988360}, $S_{R_1, R_4}$ is the modular S-matrix for $\text{SO}(M)_{2N}$~\cite{hung2018linking} and $*$ denotes complex conjugation. (These $\text{SO}(M)_{2N}$ fusion rules for all pairs of representations define a fusion category~\cite{kong2022invitationtopologicalorderscategory}.) 
Based on the fusion rule (see Appendix~\ref{ap:PrimstatesFusion} for example), one can calculate the effective dimension for each representation, that is, the quantum dimension which is generally noninteger because of the cut-off Eq.~(\ref{eq:repcutoff}). The quantum dimension is given by the modular S-matrix $g[R]=S_{0,R}/S_{0,0}$, and it further gives the impurity entropy~\cite{affleck1995conformal,Kimura2021} 
\begin{equation}
S_{\text{imp}}=\ln g[R(S^A)]=\ln g[\mu_m]   \,,
\end{equation}
for the $\text{SO}(M=2m)_{2N}$ and $\text{SO}(M=2m+1)_{2N}$ topological Kondo models, which matches with the $N=1$ result in Ref.~\cite{Altland_2014}. For example, the quantum dimensions for all the allowed representations of $\text{SO}(4)_2$ by Eq.~(\ref{eq:repcutoff}) are listed in Tab.~\ref{table:so42qdimlist}. Because the representation of the impurity is $\mu_1$ or $\mu_2$ as shown by Eq.~(\ref{eq:RSA}), the impurity entropy is $\ln \sqrt{2} $ for the $\text{SO}(4)_2$ topological Kondo model. 
\begin{table}
\begin{centering}
\begin{tabular}{|c|c|c|c|c|c|c|c|c|}
\hline 
 $0$ & $\mu_{1}$ & $\mu_{2}$ & $\mu_{1}+\mu_{2}$ & $2\mu_{1}$ & $2\mu_{2}$ & $2\mu_{1}+\mu_{2}$ & $\mu_{1}+2\mu_{2}$ & $2\mu_{1}+2\mu_{2}$\tabularnewline
\hline 
\hline 
 $1$ & $\sqrt{2}$ & $\sqrt{2}$ & $2$ & $1$ & $1$ & $\sqrt{2}$ & $\sqrt{2}$ & $1$\tabularnewline
\hline 
\end{tabular}
\par\end{centering}
\caption{The quantum dimensions for all the allowed representations in $\text{SO}(4)_{2}$, see Sec.~\ref{subsec:fusion rules}. The anyons denoted by $\mu_{1,2}$ appear at the topological Kondo intermediate fixed point.
The presence of quantum dimensions larger than one indicate overscreening. 
The trivial screening cases are given by the representations with quantum dimension $1$.}
\label{table:so42qdimlist}
\end{table}

In order to obtain the scaling dimensions of leading operators in the Kondo intermediate fixed point, we now proceed with the construction of the conformal field theory description.
The kinetic energy $H_0$ [Eq.~(\ref{eq:H0})] in its Sugawara form after non-Abelian bosonization [Eq.~(\ref{eq:currentopreal})]  becomes~\cite{gogolin2004bosonization,Kimura2021,ludwig1994field}:
\begin{align}
H_{0}=  & \frac{\pi v_{\text{F}}}{l}\sum_{p=-\infty}^{\infty}\left[\sum_A \frac{:\widetilde{J}_{-p}^{A}\widetilde{J}_{p}^{A}:}{M+2N-2}+\sum_a \frac{:\widetilde{J}_{-p}^{a}\widetilde{J}_{p}^{a}:}{2N+M-2}\right],\label{eq:H0SO42}
\end{align}
where $\widetilde{J}_{p}^{A}$ [$\widetilde{J}_{p}^{a}$] are the $p$th component from the Fourier transformation of the $\text{SO}(M)_{2N}$ [$\text{SO}(2N)_M$] generators; the normal ordering $:\dots:$ moves operators with $p>0$ to the right.  The Kondo interaction $H_{\text{MCTK}}$ [Eq.~(\ref{eq:MCTK})] becomes
\begin{equation}
H_{\text{MCTK}}=\tilde{\lambda}\,\frac{\pi v_\text{F}}{l}\sum_{p=-\infty}^{\infty}\sum_A\widetilde{J}^A_{p} S^A \,, \label{eq:HMCTKFourier}
\end{equation}
where $\tilde{\lambda} = \rho \lambda = \lambda/ (\pi v_{\text{F}}) $. At a critical value (the intermediate-coupling fixed point) $\tilde{\lambda}=\tilde{\lambda}_{*}=2/(M+2N-2)$, the Kondo interaction $H_{\text{MCTK}}$ can be absorbed into Eq.~(\ref{eq:H0SO42}) by defining the currents $\mathcal{J}^A_p=\widetilde{J}^A_{p}+S^A$ 
which still satisfy the $\text{SO}(M)_{2N}$ Kac-Moody algebra Eq.~(\ref{eq:SOM2NKacMoody}) \cite{affleck1995conformal,ludwig1994field}. 
However, this will only work if the representation obtained by fusing the impurity and conduction electron representations is in the Kac-Moody algebra, i.e., satisfies Eq.~(\ref{eq:repcutoff}). 

The most fundamental states in the spectrum of Eq.~(\ref{eq:H0SO42}) are called primary states which are annihilated by $\widetilde{J}_{p>0}^{A}$ and $\widetilde{J}_{p>0}^{a}$. Excited states can be obtained from them by acting with  $\widetilde{J}_{p<0}^{A}$ and $\widetilde{J}_{p<0}^{a}$ \cite{LudwigCFTinCMP,ludwig1994field}. 
Thus, the primary states can be labeled by the representations of the $\text{SO}(M)_{2N}$ and $\text{SO}(2N)_M$ Lie algebra. 
Another way to see this is from Eq.~(\ref{eq:H0SO42}) which is in the standard Casimir form, so that its eigenvalues can be labeled by the representations of the $\text{SO}(M)_{2N}$ and $\text{SO}(2N)_M$ Lie algebra. For example, when $M=4$ and $N=1$, the primary states (see Tab.~\ref{tab:primary}) are labeled by the representations (Dynkin's label) of the $\text{SO}(4)$ ($a_1, a_2$) and $\text{SO}(2)$ ($Q$) Lie algebra. In the free fixed point, only vector representations (even $a_1 + a_2$) are allowed because the current operators Eq.~(\ref{eq:currentopreal}) are defined using fermions and not Majoranas. The primary-state energies are [for $(M=4, N=1)$]
\begin{equation}
E_0^{(a_1, a_2, Q)}=\frac{\pi v_{\text{F}}}{l} \qty[\frac{1}{8}\qty(a_{1}+\frac{a_{1}^{2}}{2}+a_{2}+\frac{a_{2}^{2}}{2})+\frac{1}{8} Q^{2} ]. \label{eq:E0SO42}
\end{equation}
If we take $a_{1}=2j$ and $a_{2}=2j_{f}$, Eq.~(\ref{eq:E0SO42}) becomes identical to the $\text{SU}(2)_{2}$ conformal data \cite{LudwigCFTinCMP, ludwig1994field}. The conformal scaling dimension of the primary operator that creates the primary state $(a_1, a_2, Q)$ when applied to the vacuum is $E_0^{(a_1, a_2, Q)}l/(\pi v_{\text{F}})$ (see Tab.~\ref{tab:primary})~\cite{francesco2012conformal}.
\begin{table}
\begin{centering}
\begin{minipage}[t]{0.45\textwidth}
\begin{tabular}{|c|c|c|c|c|}
\hline 
$a_{1}$ & $a_{2}$ & $Q\,\text{mod}4$ & $E_0^{(a_1,a_2,Q)}l/\pi v_{\text{F}}$ & multiplicity\tabularnewline
\hline 
\hline 
$0$ & $0$ & $0$ & $0$ & $1$\tabularnewline
\hline 
$1$ & $1$ & $1$ & $1/2$ & $1$\tabularnewline
\hline 
$2$ & $0$ & $2$ & $1$ & $1$\tabularnewline
\hline 
$0$ & $2$ & $2$ & $1$ & $1$\tabularnewline
\hline 
$2$ & $2$ & $0$ & $1$ & $1$\tabularnewline
\hline 
$1$ & $1$ & $3$ & $3/2$ & $1$\tabularnewline
\hline 
\end{tabular}
\end{minipage}
\begin{minipage}[t]{0.45\textwidth}
\begin{tabular}{|c|c|c|c|c|}
\hline 
$a_{1}$ & $a_{2}$ & $Q\,\text{mod}4$ & $E_0^{(a_1,a_2,Q)}l/\pi v_{\text{F}}$ & multiplicity\tabularnewline
\hline 
\hline 
$1$ & $0$ & $0$ & $3/16$ & $1$\tabularnewline
\hline 
$0$ & $1$ & $1$ & $5/16$ & $1$\tabularnewline
\hline 
$2$ & $1$ & $1$ & $13/16$ & $1$\tabularnewline
\hline 
$1$ & $0$ & $2$ & $11/16$ & $1$\tabularnewline
\hline 
$1$ & $2$ & $2$ & $19/16$ & $1$\tabularnewline
\hline 
$1$ & $2$ & $0$ & $11/16$ & $1$\tabularnewline
\hline 
$0$ & $1$ & $3$ & $21/16$ & $1$\tabularnewline
\hline 
$2$ & $1$ & $3$ & $29/16$ & $1$\tabularnewline
\hline 
\end{tabular}
\end{minipage}
\par\end{centering}
\caption{
Primary states/operators $(a_1, a_2 , Q)$ and their scaling dimensions $E_0^{(a_1,a_2,Q)}l/\pi v_{\text{F}}$ of the  $\text{SO}(4)_2$ Kac-Moody algebra. 
Upper table: free fixed point. Lower table: Kondo fixed point obtained from free fixed point after  single fusion with the impurity representation  $\mu_1$. The single fusion results with impurity $\mu_2$ can be found by interchanging $a_1 \leftrightarrow a_2$.}\label{tab:primary}
\end{table}

By the Affleck-Ludwig \textit{fusion hypothesis}~\cite{affleck1991critical,PhysRevB.48.7297}, the Kondo fixed point (the intermediate-coupling fixed point, see Fig.~\ref{fig:conducMCTKsetup}d) primary states are determined by fusion of the free fixed point primary states \cite{ludwig1994field} with the representation of the impurity. For the case $\text{SO}(4)_2$, where the impurity representation is $R(S^A)=\mu_1 \oplus \mu_2$, see Tab.~\ref{tab:primary}. 
The allowed boundary operators are found from the free fixed point primary operators  after  double fusion with the impurity~\cite{ludwig1994field}. 
We are particularly  interested in the leading irrelevant boundary operator (LIO) that characterizes for instance finite-temperature corrections to observables near the Kondo intermediate fixed point. 
Although $R(S^A)$ is reducible, we only consider the fusions relevant to fixed fermion parity sectors, $\mu_1 \otimes \mu_1 \otimes (a_1 \mu_1 + a_2 \mu_2)$ and $\mu_2 \otimes \mu_2 \otimes (a_1 \mu_1 + a_2 \mu_2)$ (see the Appendix~\ref{ap:doublefusion} for the cross terms). The leading irrelevant operators are obtained from $a_1 = a_2 =0$ and $a_1=a_2=2$ with $Q=0$. When $a_1=a_2=0$ (the $a_1=a_2=2$ case gives identical results), we get
\begin{equation}
\mu_{1,2} \otimes \mu_{1,2} \otimes 0 = 2\mu_{1,2} \oplus 0  , \label{eq:dfusion0}
\end{equation}
where $2\mu_{1}$ $(a_1=2, a_2=0, Q=0)$ and $2\mu_{2}$ $(a_1=0, a_2=2, Q=0)$ both have scaling dimension $1/2$,  calculated from Eq.~(\ref{eq:E0SO42}). 
These resulting primaries $2\mu_{1,2}$ transform under the adjoint representation of the $\text{SO}(4)$ Lie algebra.

In the next section, we discuss the conductance correction given by the LIO,  which is the first descendant of adjoint primary fields $\vec{\phi}$, i.e., $\vec{J}_{-1} \cdot \vec{\phi}$ \cite{affleck1991critical,doi:10.7566/JPSJ.90.024708}. In the $\text{SO}(4)_2$ case, these operators are $\vec{J}^{2\mu_1}_{-1}\cdot\vec{\phi}^{2\mu_1}$ and $\vec{J}^{2\mu_2}_{-1}\cdot\vec{\phi}^{2\mu_2}$ where $\vec{\phi}^{2\mu_{1,2}}$ are the dimension $1/2$ primary operators that correspond to the representations $2\mu_{1,2}$ in Eq.~(\ref{eq:dfusion0}). The current operators $\vec{J}^{2\mu_{1,2}}_{-1}$ are the generators that transform under the representation $2\mu_{1,2}$ and their scaling dimensions are $1$. Thus, the scaling dimension of the LIO $\vec{J}_{-1} \cdot \vec{\phi}$ for the case $\text{SO}(4)_2$ is $1+1/2=3/2$. Generally for any $M,N$, the scaling dimension for the LIO of the $\text{SO}(M)_{2N}$ case is \cite{Kimura2021}
\begin{equation}
\Delta_1=1+\Delta_0 = 1+\frac{M-2}{2N+M-2} \label{eq:LIOscald}
\end{equation}
where $\Delta_n=n+\Delta_0$ is the scaling dimension of the corresponding $n$th descendant of the primary operator with scaling dimension $\Delta_0$. Special cases of Eq.~(\ref{eq:LIOscald}) can be found in the $\text{SO}(3)_2$ Kondo model~\cite{PhysRevLett.109.156803} which was mapped to the $\text{SU}(2)_{4}$ Kondo model, in the $\text{SO}(M)_2$ model~\cite{Altland_2014, PhysRevLett.113.076401,PhysRevB.96.205403}, in the $N$-channel $\text{SO}(4)_{2N}$ model which was mapped to the $\text{SU}(2)_{2N}$ Kondo model~\cite{PhysRevLett.130.066302}, and in the perturbative large-$N$ limit of the multichannel topological Kondo model~\cite{PhysRevLett.130.066302}. 

\section{Low-temperature conductance}\label{sec:zeroTconduc}
The low-temperature conductance near the intermediate-coupling fixed point can be calculated using conformal field theory techniques~\cite{affleck1995conformal, PhysRevLett.105.226803, PhysRevLett.109.156803, PhysRevB.85.045120,LibermanSela2021, Pustilnik_2004}. The charge transport is determined by the conduction electron current in the leads;  for  flavor $\alpha$ and channel $j$, the current operator is $I_{j,\alpha} = I_{R;j,\alpha} - I_{L;j,\alpha}$ 
where the chiral currents are  
$I_{L,R;j,\alpha}(x)=ev_{\text{F}}\psi^\dagger_{L,R;j,\alpha}(x)\psi_{L,R;j,\alpha}(x)$ (see Fig.~\ref{fig:conducMCTKsetup}). One can calculate the linear response conductance by using the Kubo formula that involves the equilibrium current-current correlation functions, see Ref.~\cite{PhysRevB.107.L201401} for details on a fully analogous calculation. 
For simplicity and practical relevance~\cite{PhysRevLett.130.066302} we focus on charge transport in one of the channels, say $j=1$. 
Because of charge conservation, it is sufficient to evaluate the off-diagonal elements of the conductance matrix $G_{\alpha \beta} = \lim_{V_\beta \to 0} \langle I_{1,\alpha} \rangle / V_{\beta}$ which characterizes the current in lead $\alpha$ generated by a weak voltage $V_{\beta}$ in lead $\beta$.    
The presence of a Kondo impurity modifies only the correlation functions $\langle I_{L;1,\alpha}I_{R;1,\beta}  \rangle$ between opposite chirality currents~\cite{PhysRevB.85.045120,PhysRevLett.105.226803}. 
We suppress the position and time arguments in the correlation functions for brevity~\footnote{See the supplementary material of Ref.~\cite{PhysRevB.107.L201401} for similar calculations in full detail.}; this omission is not important for our derivations below.

\subsection{Zero-temperature conductance} \label{sec:T0conduc}
The channel and flavor symmetries exhibited by the Hamiltonian Eqs.~(\ref{eq:H0})-(\ref{eq:MCTK}) in both the free and Kondo fixed points restrict the current-current correlation functions to only two independent ones: $\langle I_{L;j,\alpha}I_{R;k,\alpha} \rangle= \delta_{jk}\langle I_{L;1,1}I_{R;1,1}  \rangle $ and $\langle I_{L;j,\alpha}I_{R;k,\beta} \rangle = \delta_{jk}\langle I_{L;1,1}I_{R;1,2}\rangle $ for $\alpha \neq \beta$. 
At the Kondo fixed point, we can relate these two correlation functions to a single correlator evaluated at the free fixed point. In order to do this, we first define chiral ($\chi = L,R$) densities that transform properly under $\text{SO}(M)$ rotations: $J^{(c)}_{\chi}(x)=\sum_{j}\vec{\psi}^\dagger_{\chi;j}(x)\cdot \vec{\psi}_{\chi;j}(x)=\sum_{j,\alpha} I_{L,R;j,\alpha}(x)$ which transforms as the representation $0$ (singlet) and $J^{(d)}_{\chi}(x)=\sum_{j}\vec{\psi}^\dagger_{\chi;j}(x)D_1\vec{\psi}_{\chi;j}(x)= \sum_{j} [I_{\chi;j,1}(x) - I_{\chi;j,2}(x)]$ which transforms as the representation $4\mu_1$ (or spin-2, see the supplementary materials of Ref.~\cite{PhysRevLett.109.156803}) for $\text{SO}(3)$, $2\mu_1 + 2\mu_2$ for $\text{SO}(4)$, and $2\mu_1$ for $\text{SO}(M\geq 5)$. 
Here, $\vec{\psi}_{\chi;j}=(\psi_{\chi;j,1},\dots,\psi_{\chi;j,M})^T$ and $D_1=\text{diag}(1,-1,0,\dots,0)$ (a symmetric traceless $M\times M$ matrix). 
Note that the density $J^{(d)}$ is one of the $M-1+M(M-1)/2$ densities given by the symmetric traceless $M\times M$ matrices (see Appendix~\ref{ap:rep2mu1}).

\begin{figure}[tb]
\raggedright
\includegraphics[width=0.95\columnwidth]{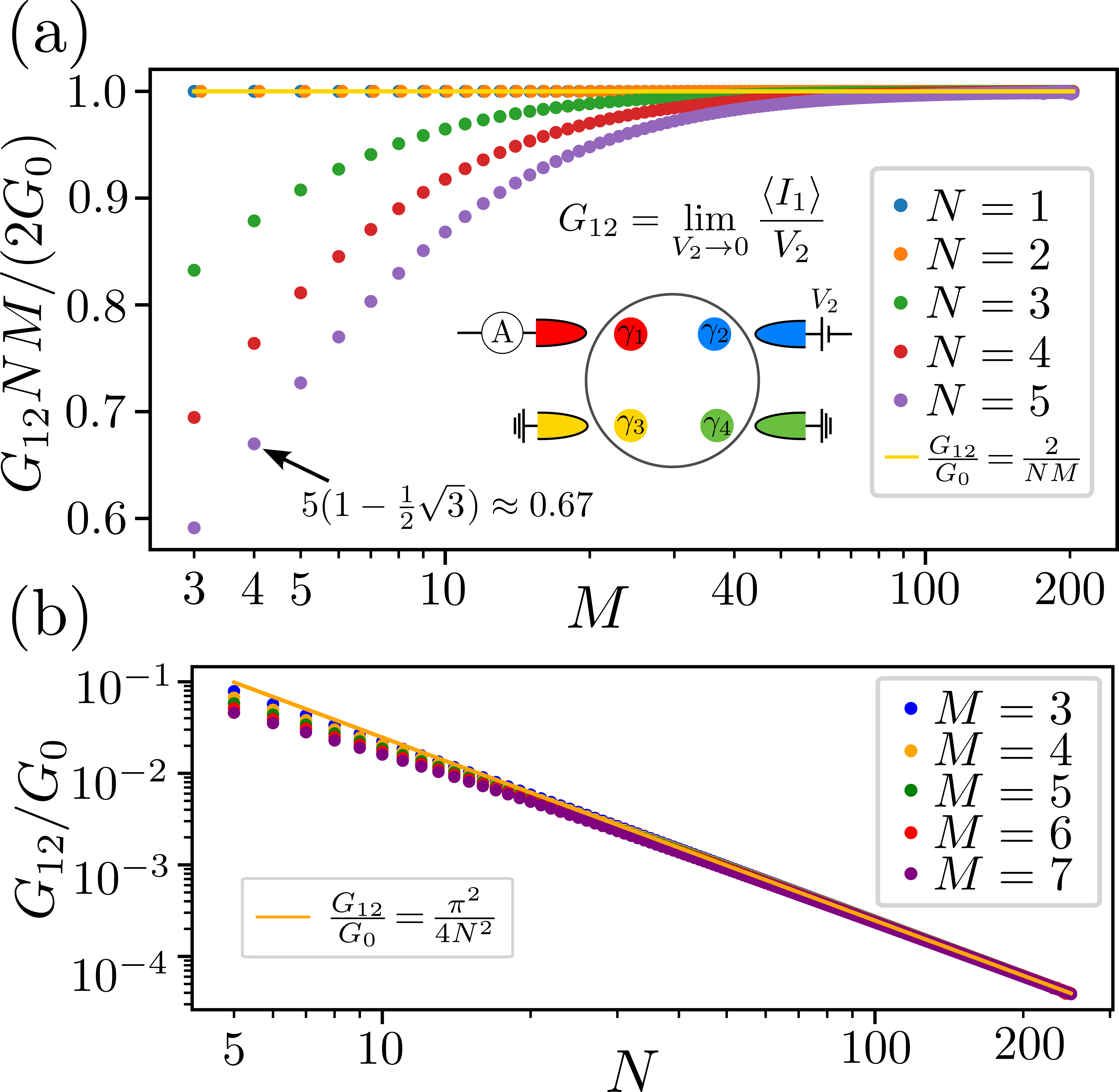}
\caption{(a) The zero-temperature conductance for $N=1,\dots,5$ and $M=3,\dots,203$. All the points above are given by analytical results Eq.~(\ref{eq:zeroTG12}) and Eq.~(\ref{eq:SN}) [for illustration the $(N,M)=(5,4)$ point is shown explicitly]. The $N=1,2$ points are  degenerate but we horizontally shifted the $N=2$ points in order to show both of them.
The inset shows the definition of an off-diagonal conductance matrix element for the case $N=1,\,M=4$. In the large-$M$ limit, the conductance $G_{12}/G_0$ is asymptotically $2/(MN)$. (b) In the large-$N$ limit, the conductance is asymptotically $\pi^2/(4N^2)$.
} 
\label{fig:conducCFT}
\end{figure}

The current-current correlation functions can be written in terms of the above density-density correlations functions:
\begin{align}
& \langle I_{L;1,1}I_{R;1,1}\rangle = \frac{1}{M^{2}N}\langle J_{L}^{(c)}J_{R}^{(c)}\rangle+\frac{M-1}{2MN}\langle J_{L}^{(d)}J_{R}^{(d)}\rangle, \label{eq:II=JJ} \\
& \langle I_{L;1,1}I_{R;1,2}\rangle= \frac{1}{M^{2}N}\langle J_{L}^{(c)}J_{R}^{(c)}\rangle-\frac{1}{2MN}\langle J_{L}^{(d)}J_{R}^{(d)}\rangle. \label{eq:II=JJ2}
\end{align} 
These equations are true with or without Kondo interaction. At the free fixed point (denoted from hereon by subscript f), we have no  flavor mixing so $\langle I_{L;1,1}I_{R;1,2}\rangle_\text{f}=0$ (leading to $G_{12} = 0$) and 
\begin{align}
    & \langle J^{(c)}_{L}J^{(c)}_{R} \rangle_\text{f} = N M \langle I_{L;1,1}I_{R;1,1}\rangle_\text{f}, \label{eq:JJff=IIff}  \\
    & \langle J^{(d)}_{L}J^{(d)}_{R} \rangle_\text{f} =  2 N \langle  I_{L,R;1,1} I_{L,R;1,1}\rangle_\text{f}. \label{eq:JJff=IIff2} 
\end{align}
At the Kondo intermediate-coupling fixed point (denoted by subscript K), the density-density correlation function $\langle J^{(c,d)}_{L}J^{(c,d)}_{R} \rangle_\text{K}$ is given by its free fixed point value multiplied by a constant factor that is fully determined by representation theory~\cite{affleck1995conformal}. The density $J^{(c)}$ is a singlet and thus its factor will be $1$; its correlation function is not modified by Kondo interaction.  
The density $J^{(d)}$ transforms non-trivially with a representation $R(J^{(d)})$ that depends on $M$ [see above Eq.~(\ref{eq:II=JJ})]. The corresponding factor is given by the following modular S-matrix values~\cite{affleck1995conformal,francesco2012conformal}:
\begin{equation}
\mathcal{S}_{N}(M)=\frac{ S_{R(J^{(d)}), R(S^A)} / S_{R(J^{(d)}),0}}{S_{R(S^A),0}/S_{0,0} }. \label{eq:SN} 
\end{equation}
In other words, $\langle J_{L}^{(c)}J_{R}^{(c)}\rangle_\text{K}= \langle J_{L}^{(c)}J_{R}^{(c)}\rangle_\text{f}$ and $\langle J_{L}^{(d)}J_{R}^{(d)}\rangle_\text{K}= \mathcal{S}_N \langle J_{L}^{(d)}J_{R}^{(d)}\rangle_\text{f}$. By using these relation  in  Eqs.~(\ref{eq:II=JJ})-(\ref{eq:JJff=IIff2}), we get 
\begin{align}
&\langle I_{L;1,1}I_{R;1,1}\rangle_\text{K}= \frac{1+(M-1)\mathcal{S}_N}{M}\langle I_{L;1,1}I_{R;1,1}\rangle_\text{f},  \\
&\langle I_{L;1,1}I_{R;1,2}\rangle_\text{K}= \frac{1-\mathcal{S}_N}{M}\langle I_{L;1,1}I_{R;1,1}\rangle_\text{f}. 
\end{align}
which relate the Kondo fixed point correlation functions to a single free fixed point correlation function. A similar relation holds for the zero-temperature conductance matrix elements~\cite{PhysRevB.107.L201401}. Thus, the transconductance $G_{12}$ of the $N$-channel $\text{SO}(M)$ topological Kondo model is 
\begin{equation}
   \frac{G_{12}(N,M)}{G_0} = \frac{1-\mathcal{S}_N(M)}{M} , \quad G_0=e^2/h \,. \label{eq:zeroTG12}
\end{equation}
Unfortunately there is no simple closed formula for $\mathcal{S}_N(M)$ with general $N$ and $M$~\cite{hung2018linking}. 
We can nevertheless evaluate $\mathcal{S}_N(M)$ and $G_{12}$ for any given $N,M$, see Fig. \ref{fig:conducCFT} for some small values. For $M=3,4,5,6,7$ and generic $N$, we have 
\begin{align}
&\mathcal{S}_{N}(3)=
1-2 \cos \left(\frac{\pi }{2 N+1}\right)+2 \cos \left(\frac{2 \pi }{2 N+1}\right),\label{eq:SNso34} \\
&\mathcal{S}_{N}(4)=
2 \cos \left(\frac{\pi }{N+1}\right) -1,
\label{eq:SNso342}\\
&\mathcal{S}_{N}(5)=2 \cos \left(\frac{2 \pi }{2 N+3}\right)+\Big[1-2 \cos \left(\frac{\pi }{2N+3}\right)\Big]^{-1}, \label{eq:SNso5} \\
& \mathcal{S}_{N}(6)=\cos \left(\frac{2 \pi }{N+2}\right) \Big/ \cos \left(\frac{\pi }{N+2}\right), \label{eq:SNso6} \\
& \mathcal{S}_{N}(7) = 2 \cos \left(\frac{2 \pi }{2 N+5}\right)\nonumber \\
&+\Big[2 \cos \left(\frac{\pi }{2
   N+5}\right)-2 \cos \left(\frac{2 \pi }{2 N+5}\right)-1 \Big]^{-1}.\label{eq:SNso7}
\end{align}

We will mention some special cases of Eq.~(\ref{eq:zeroTG12}) that have been previously discussed in the literature. 
In the conventional $N=1$ topological Kondo model, one has $\mathcal{S}_{N=1}=-1$ and $G_{12}=(2/M)(e^2/h)$ for all $M$, which matches with previous  results~\cite{PhysRevLett.109.156803,PhysRevLett.110.216803, PhysRevLett.113.076401,PhysRevB.89.045143,PhysRevB.94.235102,PhysRevB.96.205403,PhysRevResearch.2.043228}.
The result for $M=3,4$ and generic $N$ can also be verified by mapping to previous results on the $\text{SU}(2)$ multichannel Kondo effect. 
The $N$-channel $\text{SO}(3)$ topological Kondo model can be mapped to the $4N$CK, i.e., $\text{SU}(2)_{4N}$ Kac-Moody algebra and the $N$-channel $\text{SO}(4)$ topological Kondo model can be mapped to the $2N$CK, i.e., $\text{SU}(2)_{2N}$ Kac-Moody algebra at the intermediate coupling fixed point. After the mapping, we use the modular S-matrix for $\text{SU}(2)_{k}$ \cite{doi:10.7566/JPSJ.86.084703,affleck1995conformal,hung2018linking}
\begin{equation}
S_{j,s}=\sqrt{\frac{2}{k+2}}\sin\left[\frac{\pi(2j+1)(2s+1)}{k+2}\right]. \label{eq:mSmatrixSU2k}
\end{equation}
The representation of $J^{d}$ is spin-$2$ $(j=2)$ for the $\text{SU}(2)_{4N}$ case and spin-$1$ $(j=1)$ for the $\text{SU}(2)_{2N}$ case. The representation of impurity is spin-$1/2$ ($s=1/2$) for both cases. The $\mathcal{S}_{N}$s calculated from Eq.~(\ref{eq:mSmatrixSU2k}) match with Eqs.~(\ref{eq:SNso34})-(\ref{eq:SNso342}) that are calculated using the modular S-matrix of $\text{SO}(M=3,4)_{2N}$.

Finally, we can comment on the limits $N,M\to\infty$ of Eq.~(\ref{eq:zeroTG12}). 
The perturbative large-$N$ limit~\cite{PhysRevLett.130.066302} can be seen from Eqs.~(\ref{eq:SNso34})-(\ref{eq:SNso7})   which yield $\lim_{N\to\infty}G_{12}(N,M=3,\dots,7)/G_0=\pi^2/(4N^2)$  (see Fig.~\ref{fig:conducCFT}b),  matching  with the large-$N$ conductance of Ref.~\cite{PhysRevLett.130.066302}. 
For a fixed $N$,  we numerically show that the conductance vanishes as $2/(MN)$ in the large-$M$ limit (see Fig. \ref{fig:conducCFT}a). 
However, in perturbation theory in the Kondo interaction, the conductance $G_{12}$ is proportional to $\tilde{\lambda}_*^2$~\cite{PhysRevLett.130.066302}, which is $\tilde{\lambda}_*^2 \sim 1/M^2$ because $\tilde{\lambda}_*=2/(M+2N-2)\sim 1/M$ in the large-$M$ limit [see below Eq.~(\ref{eq:HMCTKFourier})]. This mismatch of the large-$M$ conductance between the conformal field theory method and the perturbation theory indicates that, unlike the large-$N$ limit, the large-$M$ limit is not perturbatively accessible.

\subsection{Leading finite-temperature correction to the conductance} 
Next, we discuss the finite-temperature correction  to the conductance, Eq.~(\ref{eq:zeroTG12}). 
This correction arises from the deviation $\tilde{\lambda} - \tilde{\lambda}_*$ of the Kondo coupling from its fixed point value, also known as correction to scaling~\cite{ludwig1994field}. 
We will show that the conductance correction depends on the channel number $N$ as,  
\begin{equation}
G_{12}(T)-G_{12}(0) \sim
  \begin{cases}
    T^{2(\Delta_{1} - 1)}, & N=1 \\
    T^{(\Delta_{1} - 1)}, & N \geq 2 
  \end{cases}, \label{eq:GTcorrection}
\end{equation}
where $\Delta_1$ is given by Eq.~(\ref{eq:LIOscald}). 
The result Eq.~(\ref{eq:GTcorrection}) agrees with the cases $N=1$ and $N\to \infty$ that were previously studied in the literature~\cite{PhysRevLett.109.156803,PhysRevLett.110.216803, PhysRevLett.113.076401,PhysRevB.94.235102,PhysRevB.89.045143,PhysRevB.96.205403, PhysRevResearch.2.043228,PhysRevLett.130.066302}.

The piecewise result Eq.~(\ref{eq:GTcorrection}) is due to different fusion results of the operator $J^{d}$ for $N=1$ and $N \geq 2$ cases, i.e., 
\begin{equation}
R(J^{(d)}) \times R(J^{(d)}) =
  \begin{cases}
    0, & N=1 \\
    0 + \dots, & N \geq 2
  \end{cases}.\label{eq:fusionJd}
\end{equation}
The operator $J^{d}$ is used to calculate the fixed point conductance by using Eqs.~(\ref{eq:II=JJ})-(\ref{eq:II=JJ2}) and the conductance correction is given by the correction to $\langle J^d J^d\rangle$ due to the LIO $\vec{J}_{-1} \cdot \vec{\phi}$ with scaling dimension shown in Eq.~(\ref{eq:LIOscald}). The first-order conductance correction is
\begin{align}\label{eq:GT1st}
&\lim_{\omega\to 0}\frac{-1}{\omega L}\int_{-\beta/2}^{\beta/2} d\tau\int_0^L dx'\int_{-\beta/2}^{\beta/2} d\tau_{1}\, \mathrm{e}^{\mathrm{i}\omega\tau}\nonumber \\
&\!\langle\mathcal{T}_{\tau}J^{(d)}(x,\tau)J^{(d)}(x',0)
\vec{J}_{-1} \cdot \vec{\phi}(0,\tau_{1})\rangle \,,
\end{align}
where $\beta=1/T$ and $\tau=\mathrm{i}t$~\cite{Oshikawa_2006,bao2017quantumhallchargekondo,PhysRevB.107.L201401}. By power counting, each integral contributes $T^{-1}$ and the correlation function contributes $\sim T^{2+\Delta_1}$. Thus, the first-order correction Eq.~(\ref{eq:GT1st}) gives conductance correction $\sim T^{(\Delta_1-1)}$. 
This is the case in the multichannel $N\geq 2$ topological Kondo
model~\cite{PhysRevLett.130.066302} which allows fusion results other than $0$. However, the first-order correction vanishes at $N=1$ because the fusion rule, Eq.~(\ref{eq:fusionJd}), of two $J^{d}$s gives only $0$, i.e., $J^{d}J^{d}\sim \mathbb{I}$ and thus $\langle J^{d} J^{d} \vec{J}_{-1} \cdot \vec{\phi} \rangle\sim \langle \vec{J}_{-1} \cdot \vec{\phi} \rangle=0$  (see also the supplementary materials of Ref.~\cite{PhysRevLett.109.156803}). 
It means that the temperature dependence of the conductance correction for the 1-channel topological Kondo model arises from the second order in the LIO,
\begin{align}\label{eq:GT2nd}
&\lim_{\omega\to 0}\frac{-1}{\omega L}\int_{-\beta/2}^{\beta/2} d\tau\int_0^L dx'\int_{-\beta/2}^{\beta/2} d\tau_{1}\int_{-\beta/2}^{\beta/2} d\tau_{2}\, \mathrm{e}^{\mathrm{i}\omega\tau}\nonumber \\
&\! \langle\mathcal{T}_{\tau}J^{(d)}(x,\tau)J^{(d)}(x',0)
\vec{J}_{-1} \cdot \vec{\phi}(0,\tau_{1})\vec{J}_{-1} \cdot \vec{\phi}(0,\tau_{2})\rangle,
\end{align}
and is of order $T^{2(\Delta_{1} - 1)}=T^{2(M-2)/M}$~\cite{PhysRevLett.113.076401,PhysRevB.96.205403,PhysRevResearch.2.043228}.

\section{Comparison to the symplectic case}\label{sec:Spconductance}
In the above sections, we mainly discussed the application of conformal field theory to calculate the conductance of the multichannel SO$(M)$ topological Kondo model.
One may wonder how these techniques are applicable to other Lie groups  such as the recently studied symplectic $\text{Sp}(2k)$ Kondo model~\cite{PhysRevB.107.L201401,KONIG2023169231}. 
In the symplectic Kondo case, the low-temperature conductance correction was found to be the Fermi liquid $T^2$~\cite{PhysRevB.107.L201401,KONIG2023169231}. However, as we show below,  this is only true for the $\text{Sp}(4)_1$ case.

Due to a similar representation cutoff as Eq.~(\ref{eq:repcutoff}), the allowed representations of the $\text{Sp}(2k)_1$ Kac-Moody algebra are the fundamental ones $\mu_{i=1,\dots,k}$ where $\mu_1$ represents the impurity. The double fusion rule of the impurity $\mu_1$ gives
\begin{equation}
    \mu_1 \otimes \mu_1 \otimes 0 = \cancel{2\mu_1} \oplus \mu_2 \oplus 0 \label{eq:mu1fusion}
\end{equation}
where \cancel{$2\mu_1$} means this adjoint representation $2\mu_1$ is not allowed by the Kac-Moody algebra (see also the supplemental material of Ref.~\cite{PhysRevB.107.L201401}). Thus, the LIO $\vec{J}_{-1} \cdot \vec{\phi}^{2\mu_1}$ is not allowed in the boundary operators.  
However, one can consider high-order descendants and contract them with the representations allowed by the $\text{Sp}(2k)_1$ Kac-Moody algebra. It turns out that the LIO is $(\overrightarrow{J_{-1} J_{-1}})^{\mu_2} \cdot \vec{\phi}^{\mu_2}$ with scaling dimension $\Delta_2^{\mu_2} = 2 + \Delta_{0}^{\mu_2}$ where $\Delta_{0}^{\mu_2}=k/(k+2)$. The operators $(\overrightarrow{J_{-1} J_{-1}})^{\mu_2}$ are the $\mu_2$ operators constructed from the product of two $2\mu_1$ (the adjoint representation) operators $J_{-1}$. This construction always exits because of the following representation decomposition \cite{LieARTFeger_2020}:
\begin{equation}
    2\mu_1 \otimes 2\mu_1 = 4\mu_1 \oplus (2\mu_1 + \mu_2) \oplus 2\mu_1 \oplus 2\mu_2 \oplus \mu_2 \oplus 0.\label{eq:mu2construction}
\end{equation}
The $\mu_2$ at the right site of Eq.~(\ref{eq:mu2construction}) denotes the representation of $\overrightarrow{J_{-1} J_{-1}}$.

Similar to the $\text{SO}(M)$ Kondo model, the charge conductance in the symplectic $\text{Sp}(2k)$ case is determined by the operator $J^d$ [see Eq.~(\ref{eq:II=JJ}) for the $\text{SO}(M)$ case], which transforms as the representation $\mu_2$ of $\text{Sp}(2k)$~\cite{PhysRevB.107.L201401}.  
The first-order low-temperature conductance correction resulting from the LIO $(\overrightarrow{J_{-1} J_{-1}})^{\mu_2} \cdot \vec{\phi}^{\mu_2}$ is proportional to $T^{(\Delta_{2}^{\mu_2}-1)}$. As mentioned in Sec.~\ref{sec:zeroTconduc}, if the fusion rule of the operator $J^d$ gives only a trivial representation $0$, the first-order correction vanishes and one should consider the second-order perturbation. The fusion rule of the $\text{Sp}(2k)_1$ affine Lie algebra is identical to that of the $\text{SU(2)}_k$ affine Lie algebra: the impurity representation $\mu_1$ of $\text{Sp}(2k)_1$ is equivalent to the ``1/2'' of $\text{SU(2)}_k$ and the $\mu_2$ is equivalent to the ``1'' of $\text{SU(2)}_k$ [see also Eq.~(\ref{eq:mu1fusion}) as a verification]. 
At $k=2$, we have $\mu_2\otimes \mu_2 = 0$, which is equivalent to $1\times 1 =0$ in $\text{SU}(2)_2$. Thus, one considers the second-order perturbation which would lead to $T^{2(\Delta_{2}^{\mu_2}-1)}=T^3$. However, this  is subleading compared to the Fermi liquid correction $T^2$ given by the operator $\vec{J}^{2\mu_1}\cdot\vec{J}^{2\mu_1}$ \cite{doi:10.7566/JPSJ.90.024708}. When $k\geq3$, one considers the first-order correction. Thus, the conductance correction of the $\text{Sp}(2k)_1$ Kondo model is 
\begin{equation}
\delta G(T) \sim
  \begin{cases}
    T^{2}, & k=2 \\    T^{(\Delta_{2}^{\mu_2}-1)}=T^{2\frac{k+1}{k+2}}, & k \geq 3 
  \end{cases}. \label{eq:SpGTcorrection}
\end{equation}
Note that in the $\text{Sp(4)}_1$ setup of Ref.~\cite{MitchellAffleck2021,LibermanSela2021}, the first-order conductance correction does not vanish because their conductance is not calculated using $J^d$ and thus they obtain a non-Fermi liquid behavior $T^{(\Delta_{2}^{\mu_2}-1)}=T^{3/2}$ at $k=2$. This difference in conductance correction between their $\text{SO}(5)_1$ or $\text{Sp}(4)_1$ Kondo model \cite{MitchellAffleck2021,LibermanSela2021} and the recently studied symplectic $\text{Sp}(4)_1$ Kondo model \cite{PhysRevB.107.L201401,KONIG2023169231} is similar to the difference between the spin-2CK model and charge-2CK model \cite{bao2017quantumhallchargekondo}.

Furthermore, in the multichannel $\text{Sp}(2k)_{N\geq 2}$ case, the adjoint primary operator $\vec{\phi}^{2\mu_1}$ will be allowed. Its descendant $\vec{J}_{-1} \cdot \vec{\phi}^{2\mu_1}$ will compete with the operator $(\overrightarrow{J_{-1} J_{-1}})^{\mu_2} \cdot \vec{\phi}^{\mu_2}$. The former becomes the leading irrelevant operator since its scaling dimension satisfies 
$\Delta^{2\mu_1}_{1}< \Delta^{\mu_2}_{2}$ where
\begin{align}
&\Delta^{2\mu_1}_{1}=1+\Delta^{2\mu_1}_{0}=1+\frac{k+1}{N+k+1}, \\
&\Delta^{\mu_2}_{2}=2 + \Delta^{\mu_2}_{0} =2+ \frac{k}{N+k+1}.
\end{align}
Depending on the fusion rule, the conductance correction will exhibit a non-Fermi liquid behavior, $T^{\Delta_{0}^{2\mu_1}}$ or $T^{2\Delta_{0}^{2\mu_1}}$, similar to Eq.~(\ref{eq:GTcorrection}).

\section{Conclusions}
In this paper, we demonstrated the instability of the strong coupling fixed point in the $\text{SO}(M)$ topological Kondo model, demonstrating the existence of  the intermediate coupling fixed point. 
By developing the generalized Affleck-Ludwig conformal field theory technique, 
we extended the concept of overscreening to higher rank Lie groups, finding a set of inequalities for the impurity representation, see Eq.~(\ref{eq:repcutoff}). We also clarified the conductance correction of the multichannel topological Kondo model, showing that a first order correction given by the LIO exists in the multichannel model but not for single channel. Similarly, in the $N$-channel charge-Kondo model, the first-order correction exits for $N\geq3$ but not for $N=2$~\cite{2018Sci...360.1315I} because the fusion rule of two $J^d$s, spin-$1$s, gives only spin-$0$ at $N=2$. The method of determining the LIO can be applied to other compact Lie groups as we have shown with $\text{Sp}(2k)$. The analysis for getting the $\text{Sp}(2k)$ conductance correction, Eq.~(\ref{eq:SpGTcorrection}), is also useful if the spin-$1/2$ 2-flavor Kondo model in Ref.~\cite{MitchellAffleck2021,LibermanSela2021} is further generalized to a spin-$1/2$ $k$-flavor Kondo model, which is a possible regime for a similar spin-$k$CK quantum dot device.

\section{Acknowledgements}

We thank Taro Kimura, Yiwen Pan, Eran Sela, and Fan Zhang for useful correspondence. 
We used LieART \cite{LieARTFeger_2020} for decomposing tensor products. 
This work was supported by the U.S. Department of Energy, Office of Science, National Quantum Information Science Research Centers, Quantum Science Center. Support for this research was provided by the Office of the Vice Chancellor
for Research and Graduate Education at the University
of Wisconsin–Madison with funding from the Wisconsin
Alumni Research Foundation. This research was supported in part by grants NSF PHY-1748958 and PHY-
2309135 to the Kavli Institute for Theoretical Physics
(KITP). EJK acknowledges hospitality
by the KITP.

\bibliography{apssamp}

\begin{thebibliography}{71}%
\makeatletter
\providecommand \@ifxundefined [1]{%
 \@ifx{#1\undefined}
}%
\providecommand \@ifnum [1]{%
 \ifnum #1\expandafter \@firstoftwo
 \else \expandafter \@secondoftwo
 \fi
}%
\providecommand \@ifx [1]{%
 \ifx #1\expandafter \@firstoftwo
 \else \expandafter \@secondoftwo
 \fi
}%
\providecommand \natexlab [1]{#1}%
\providecommand \enquote  [1]{``#1''}%
\providecommand \bibnamefont  [1]{#1}%
\providecommand \bibfnamefont [1]{#1}%
\providecommand \citenamefont [1]{#1}%
\providecommand \href@noop [0]{\@secondoftwo}%
\providecommand \href [0]{\begingroup \@sanitize@url \@href}%
\providecommand \@href[1]{\@@startlink{#1}\@@href}%
\providecommand \@@href[1]{\endgroup#1\@@endlink}%
\providecommand \@sanitize@url [0]{\catcode `\\12\catcode `\$12\catcode
  `\&12\catcode `\#12\catcode `\^12\catcode `\_12\catcode `\%12\relax}%
\providecommand \@@startlink[1]{}%
\providecommand \@@endlink[0]{}%
\providecommand \url  [0]{\begingroup\@sanitize@url \@url }%
\providecommand \@url [1]{\endgroup\@href {#1}{\urlprefix }}%
\providecommand \urlprefix  [0]{URL }%
\providecommand \Eprint [0]{\href }%
\providecommand \doibase [0]{https://doi.org/}%
\providecommand \selectlanguage [0]{\@gobble}%
\providecommand \bibinfo  [0]{\@secondoftwo}%
\providecommand \bibfield  [0]{\@secondoftwo}%
\providecommand \translation [1]{[#1]}%
\providecommand \BibitemOpen [0]{}%
\providecommand \bibitemStop [0]{}%
\providecommand \bibitemNoStop [0]{.\EOS\space}%
\providecommand \EOS [0]{\spacefactor3000\relax}%
\providecommand \BibitemShut  [1]{\csname bibitem#1\endcsname}%
\let\auto@bib@innerbib\@empty
\bibitem [{\citenamefont {Kondo}(1964)}]{kondo1964resistance}%
  \BibitemOpen
  \bibfield  {author} {\bibinfo {author} {\bibfnamefont {J.}~\bibnamefont
  {Kondo}},\ }\bibfield  {title} {\bibinfo {title} {Resistance minimum in
  dilute magnetic alloys},\ }\href@noop {} {\bibfield  {journal} {\bibinfo
  {journal} {Progress of theoretical physics}\ }\textbf {\bibinfo {volume}
  {32}},\ \bibinfo {pages} {37} (\bibinfo {year} {1964})}\BibitemShut {NoStop}%
\bibitem [{\citenamefont {{Nozi\`eres, Ph.}}\ and\ \citenamefont {{Blandin,
  A.}}(1980)}]{Nozieres1980}%
  \BibitemOpen
  \bibfield  {author} {\bibinfo {author} {\bibnamefont {{Nozi\`eres, Ph.}}}\
  and\ \bibinfo {author} {\bibnamefont {{Blandin, A.}}},\ }\bibfield  {title}
  {\bibinfo {title} {Kondo effect in real metals},\ }\href
  {https://doi.org/10.1051/jphys:01980004103019300} {\bibfield  {journal}
  {\bibinfo  {journal} {J. Phys. France}\ }\textbf {\bibinfo {volume} {41}},\
  \bibinfo {pages} {193} (\bibinfo {year} {1980})}\BibitemShut {NoStop}%
\bibitem [{\citenamefont {Goldhaber-Gordon}\ \emph {et~al.}(1998)\citenamefont
  {Goldhaber-Gordon}, \citenamefont {Shtrikman}, \citenamefont {Mahalu},
  \citenamefont {Abusch-Magder}, \citenamefont {Meirav},\ and\ \citenamefont
  {Kastner}}]{GoldhaberGordon1998}%
  \BibitemOpen
  \bibfield  {author} {\bibinfo {author} {\bibfnamefont {D.}~\bibnamefont
  {Goldhaber-Gordon}}, \bibinfo {author} {\bibfnamefont {H.}~\bibnamefont
  {Shtrikman}}, \bibinfo {author} {\bibfnamefont {D.}~\bibnamefont {Mahalu}},
  \bibinfo {author} {\bibfnamefont {D.}~\bibnamefont {Abusch-Magder}}, \bibinfo
  {author} {\bibfnamefont {U.}~\bibnamefont {Meirav}},\ and\ \bibinfo {author}
  {\bibfnamefont {M.~A.}\ \bibnamefont {Kastner}},\ }\bibfield  {title}
  {\bibinfo {title} {Kondo effect in a single-electron transistor},\ }\href
  {https://doi.org/10.1038/34373} {\bibfield  {journal} {\bibinfo  {journal}
  {Nature}\ }\textbf {\bibinfo {volume} {391}},\ \bibinfo {pages} {156}
  (\bibinfo {year} {1998})}\BibitemShut {NoStop}%
\bibitem [{\citenamefont {Pustilnik}\ and\ \citenamefont
  {Glazman}(2004)}]{Pustilnik_2004}%
  \BibitemOpen
  \bibfield  {author} {\bibinfo {author} {\bibfnamefont {M.}~\bibnamefont
  {Pustilnik}}\ and\ \bibinfo {author} {\bibfnamefont {L.}~\bibnamefont
  {Glazman}},\ }\bibfield  {title} {\bibinfo {title} {Kondo effect in quantum
  dots},\ }\href {https://doi.org/10.1088/0953-8984/16/16/r01} {\bibfield
  {journal} {\bibinfo  {journal} {Journal of Physics: Condensed Matter}\
  }\textbf {\bibinfo {volume} {16}},\ \bibinfo {pages} {R513} (\bibinfo {year}
  {2004})}\BibitemShut {NoStop}%
\bibitem [{\citenamefont {{Potok}}\ \emph {et~al.}(2007)\citenamefont
  {{Potok}}, \citenamefont {{Rau}}, \citenamefont {{Shtrikman}}, \citenamefont
  {{Oreg}},\ and\ \citenamefont {{Goldhaber-Gordon}}}]{2007Natur.446..167P}%
  \BibitemOpen
  \bibfield  {author} {\bibinfo {author} {\bibfnamefont {R.~M.}\ \bibnamefont
  {{Potok}}}, \bibinfo {author} {\bibfnamefont {I.~G.}\ \bibnamefont {{Rau}}},
  \bibinfo {author} {\bibfnamefont {H.}~\bibnamefont {{Shtrikman}}}, \bibinfo
  {author} {\bibfnamefont {Y.}~\bibnamefont {{Oreg}}},\ and\ \bibinfo {author}
  {\bibfnamefont {D.}~\bibnamefont {{Goldhaber-Gordon}}},\ }\bibfield  {title}
  {\bibinfo {title} {{Observation of the two-channel Kondo effect}},\ }\href
  {https://doi.org/10.1038/nature05556} {\bibfield  {journal} {\bibinfo
  {journal} {\nat}\ }\textbf {\bibinfo {volume} {446}},\ \bibinfo {pages} {167}
  (\bibinfo {year} {2007})}\BibitemShut {NoStop}%
\bibitem [{\citenamefont {{Keller}}\ \emph {et~al.}(2015)\citenamefont
  {{Keller}}, \citenamefont {{Peeters}}, \citenamefont {{Moca}}, \citenamefont
  {{Weymann}}, \citenamefont {{Mahalu}}, \citenamefont {{Umansky}},
  \citenamefont {{Zar{\'a}nd}},\ and\ \citenamefont
  {{Goldhaber-Gordon}}}]{KellerGoldhaberGordon2015}%
  \BibitemOpen
  \bibfield  {author} {\bibinfo {author} {\bibfnamefont {A.~J.}\ \bibnamefont
  {{Keller}}}, \bibinfo {author} {\bibfnamefont {L.}~\bibnamefont {{Peeters}}},
  \bibinfo {author} {\bibfnamefont {C.~P.}\ \bibnamefont {{Moca}}}, \bibinfo
  {author} {\bibfnamefont {I.}~\bibnamefont {{Weymann}}}, \bibinfo {author}
  {\bibfnamefont {D.}~\bibnamefont {{Mahalu}}}, \bibinfo {author}
  {\bibfnamefont {V.}~\bibnamefont {{Umansky}}}, \bibinfo {author}
  {\bibfnamefont {G.}~\bibnamefont {{Zar{\'a}nd}}},\ and\ \bibinfo {author}
  {\bibfnamefont {D.}~\bibnamefont {{Goldhaber-Gordon}}},\ }\bibfield  {title}
  {\bibinfo {title} {{Universal Fermi liquid crossover and quantum criticality
  in a mesoscopic system}},\ }\href {https://doi.org/10.1038/nature15261}
  {\bibfield  {journal} {\bibinfo  {journal} {\nat}\ }\textbf {\bibinfo
  {volume} {526}},\ \bibinfo {pages} {237} (\bibinfo {year}
  {2015})}\BibitemShut {NoStop}%
\bibitem [{\citenamefont {{Iftikhar}}\ \emph {et~al.}(2015)\citenamefont
  {{Iftikhar}}, \citenamefont {{Jezouin}}, \citenamefont {{Anthore}},
  \citenamefont {{Gennser}}, \citenamefont {{Parmentier}}, \citenamefont
  {{Cavanna}},\ and\ \citenamefont {{Pierre}}}]{2015Natur.526..233I}%
  \BibitemOpen
  \bibfield  {author} {\bibinfo {author} {\bibfnamefont {Z.}~\bibnamefont
  {{Iftikhar}}}, \bibinfo {author} {\bibfnamefont {S.}~\bibnamefont
  {{Jezouin}}}, \bibinfo {author} {\bibfnamefont {A.}~\bibnamefont
  {{Anthore}}}, \bibinfo {author} {\bibfnamefont {U.}~\bibnamefont
  {{Gennser}}}, \bibinfo {author} {\bibfnamefont {F.~D.}\ \bibnamefont
  {{Parmentier}}}, \bibinfo {author} {\bibfnamefont {A.}~\bibnamefont
  {{Cavanna}}},\ and\ \bibinfo {author} {\bibfnamefont {F.}~\bibnamefont
  {{Pierre}}},\ }\bibfield  {title} {\bibinfo {title} {{Two-channel Kondo
  effect and renormalization flow with macroscopic quantum charge states}},\
  }\href {https://doi.org/10.1038/nature15384} {\bibfield  {journal} {\bibinfo
  {journal} {\nat}\ }\textbf {\bibinfo {volume} {526}},\ \bibinfo {pages} {233}
  (\bibinfo {year} {2015})}\BibitemShut {NoStop}%
\bibitem [{\citenamefont {{Iftikhar}}\ \emph {et~al.}(2018)\citenamefont
  {{Iftikhar}}, \citenamefont {{Anthore}}, \citenamefont {{Mitchell}},
  \citenamefont {{Parmentier}}, \citenamefont {{Gennser}}, \citenamefont
  {{Ouerghi}}, \citenamefont {{Cavanna}}, \citenamefont {{Mora}}, \citenamefont
  {{Simon}},\ and\ \citenamefont {{Pierre}}}]{2018Sci...360.1315I}%
  \BibitemOpen
  \bibfield  {author} {\bibinfo {author} {\bibfnamefont {Z.}~\bibnamefont
  {{Iftikhar}}}, \bibinfo {author} {\bibfnamefont {A.}~\bibnamefont
  {{Anthore}}}, \bibinfo {author} {\bibfnamefont {A.~K.}\ \bibnamefont
  {{Mitchell}}}, \bibinfo {author} {\bibfnamefont {F.~D.}\ \bibnamefont
  {{Parmentier}}}, \bibinfo {author} {\bibfnamefont {U.}~\bibnamefont
  {{Gennser}}}, \bibinfo {author} {\bibfnamefont {A.}~\bibnamefont
  {{Ouerghi}}}, \bibinfo {author} {\bibfnamefont {A.}~\bibnamefont
  {{Cavanna}}}, \bibinfo {author} {\bibfnamefont {C.}~\bibnamefont {{Mora}}},
  \bibinfo {author} {\bibfnamefont {P.}~\bibnamefont {{Simon}}},\ and\ \bibinfo
  {author} {\bibfnamefont {F.}~\bibnamefont {{Pierre}}},\ }\bibfield  {title}
  {\bibinfo {title} {{Tunable quantum criticality and super-ballistic transport
  in a {\textquotedblleft}charge{\textquotedblright} Kondo circuit}},\ }\href
  {https://doi.org/10.1126/science.aan5592} {\bibfield  {journal} {\bibinfo
  {journal} {Science}\ }\textbf {\bibinfo {volume} {360}},\ \bibinfo {pages}
  {1315} (\bibinfo {year} {2018})}\BibitemShut {NoStop}%
\bibitem [{\citenamefont {Pouse}\ \emph {et~al.}(2023)\citenamefont {Pouse},
  \citenamefont {Peeters}, \citenamefont {Hsueh}, \citenamefont {Gennser},
  \citenamefont {Cavanna}, \citenamefont {Kastner}, \citenamefont {Mitchell},\
  and\ \citenamefont {Goldhaber-Gordon}}]{PouseGoldhaberGordon2022}%
  \BibitemOpen
  \bibfield  {author} {\bibinfo {author} {\bibfnamefont {W.}~\bibnamefont
  {Pouse}}, \bibinfo {author} {\bibfnamefont {L.}~\bibnamefont {Peeters}},
  \bibinfo {author} {\bibfnamefont {C.~L.}\ \bibnamefont {Hsueh}}, \bibinfo
  {author} {\bibfnamefont {U.}~\bibnamefont {Gennser}}, \bibinfo {author}
  {\bibfnamefont {A.}~\bibnamefont {Cavanna}}, \bibinfo {author} {\bibfnamefont
  {M.~A.}\ \bibnamefont {Kastner}}, \bibinfo {author} {\bibfnamefont {A.~K.}\
  \bibnamefont {Mitchell}},\ and\ \bibinfo {author} {\bibfnamefont
  {D.}~\bibnamefont {Goldhaber-Gordon}},\ }\bibfield  {title} {\bibinfo {title}
  {{Quantum simulation of an exotic quantum critical point in a two-site charge
  Kondo circuit}},\ }\href {https://www.nature.com/articles/s41567-022-01905-4}
  {\bibfield  {journal} {\bibinfo  {journal} {Nature Physics}\ }\textbf
  {\bibinfo {volume} {19}},\ \bibinfo {pages} {492} (\bibinfo {year}
  {2023})}\BibitemShut {NoStop}%
\bibitem [{\citenamefont {Affleck}(1995)}]{affleck1995conformal}%
  \BibitemOpen
  \bibfield  {author} {\bibinfo {author} {\bibfnamefont {I.}~\bibnamefont
  {Affleck}},\ }\bibfield  {title} {\bibinfo {title} {Conformal field theory
  approach to the kondo effect},\ }\href@noop {} {\bibfield  {journal}
  {\bibinfo  {journal} {Acta Phys.~Polon.}\ }\textbf {\bibinfo {volume}
  {B26}},\ \bibinfo {pages} {1869} (\bibinfo {year} {1995})}\BibitemShut
  {NoStop}%
\bibitem [{\citenamefont {Lopes}\ \emph {et~al.}(2020)\citenamefont {Lopes},
  \citenamefont {Affleck},\ and\ \citenamefont {Sela}}]{LopesSela2020}%
  \BibitemOpen
  \bibfield  {author} {\bibinfo {author} {\bibfnamefont {P.~L.~S.}\
  \bibnamefont {Lopes}}, \bibinfo {author} {\bibfnamefont {I.}~\bibnamefont
  {Affleck}},\ and\ \bibinfo {author} {\bibfnamefont {E.}~\bibnamefont
  {Sela}},\ }\bibfield  {title} {\bibinfo {title} {Anyons in multichannel kondo
  systems},\ }\href {https://doi.org/10.1103/PhysRevB.101.085141} {\bibfield
  {journal} {\bibinfo  {journal} {Phys. Rev. B}\ }\textbf {\bibinfo {volume}
  {101}},\ \bibinfo {pages} {085141} (\bibinfo {year} {2020})}\BibitemShut
  {NoStop}%
\bibitem [{\citenamefont {Komijani}(2020)}]{Komijani2020}%
  \BibitemOpen
  \bibfield  {author} {\bibinfo {author} {\bibfnamefont {Y.}~\bibnamefont
  {Komijani}},\ }\bibfield  {title} {\bibinfo {title} {Isolating kondo anyons
  for topological quantum computation},\ }\href
  {https://doi.org/10.1103/PhysRevB.101.235131} {\bibfield  {journal} {\bibinfo
   {journal} {Phys. Rev. B}\ }\textbf {\bibinfo {volume} {101}},\ \bibinfo
  {pages} {235131} (\bibinfo {year} {2020})}\BibitemShut {NoStop}%
\bibitem [{\citenamefont {Gabay}\ \emph {et~al.}(2022)\citenamefont {Gabay},
  \citenamefont {Han}, \citenamefont {Lopes}, \citenamefont {Affleck},\ and\
  \citenamefont {Sela}}]{PhysRevB.105.035151}%
  \BibitemOpen
  \bibfield  {author} {\bibinfo {author} {\bibfnamefont {D.}~\bibnamefont
  {Gabay}}, \bibinfo {author} {\bibfnamefont {C.}~\bibnamefont {Han}}, \bibinfo
  {author} {\bibfnamefont {P.~L.~S.}\ \bibnamefont {Lopes}}, \bibinfo {author}
  {\bibfnamefont {I.}~\bibnamefont {Affleck}},\ and\ \bibinfo {author}
  {\bibfnamefont {E.}~\bibnamefont {Sela}},\ }\bibfield  {title} {\bibinfo
  {title} {Multi-impurity chiral kondo model: Correlation functions and anyon
  fusion rules},\ }\href {https://doi.org/10.1103/PhysRevB.105.035151}
  {\bibfield  {journal} {\bibinfo  {journal} {Phys. Rev. B}\ }\textbf {\bibinfo
  {volume} {105}},\ \bibinfo {pages} {035151} (\bibinfo {year}
  {2022})}\BibitemShut {NoStop}%
\bibitem [{\citenamefont {Lotem}\ \emph {et~al.}(2022)\citenamefont {Lotem},
  \citenamefont {Sela},\ and\ \citenamefont
  {Goldstein}}]{PhysRevLett.129.227703}%
  \BibitemOpen
  \bibfield  {author} {\bibinfo {author} {\bibfnamefont {M.}~\bibnamefont
  {Lotem}}, \bibinfo {author} {\bibfnamefont {E.}~\bibnamefont {Sela}},\ and\
  \bibinfo {author} {\bibfnamefont {M.}~\bibnamefont {Goldstein}},\ }\bibfield
  {title} {\bibinfo {title} {Manipulating non-abelian anyons in a chiral
  multichannel kondo model},\ }\href
  {https://doi.org/10.1103/PhysRevLett.129.227703} {\bibfield  {journal}
  {\bibinfo  {journal} {Phys. Rev. Lett.}\ }\textbf {\bibinfo {volume} {129}},\
  \bibinfo {pages} {227703} (\bibinfo {year} {2022})}\BibitemShut {NoStop}%
\bibitem [{\citenamefont {Ren}\ \emph {et~al.}(2024)\citenamefont {Ren},
  \citenamefont {K\"onig},\ and\ \citenamefont {Tsvelik}}]{RanTsvelik2024}%
  \BibitemOpen
  \bibfield  {author} {\bibinfo {author} {\bibfnamefont {T.}~\bibnamefont
  {Ren}}, \bibinfo {author} {\bibfnamefont {E.~J.}\ \bibnamefont {K\"onig}},\
  and\ \bibinfo {author} {\bibfnamefont {A.~M.}\ \bibnamefont {Tsvelik}},\
  }\bibfield  {title} {\bibinfo {title} {Topological quantum computation on a
  chiral kondo chain},\ }\href {https://doi.org/10.1103/PhysRevB.109.075145}
  {\bibfield  {journal} {\bibinfo  {journal} {Phys. Rev. B}\ }\textbf {\bibinfo
  {volume} {109}},\ \bibinfo {pages} {075145} (\bibinfo {year}
  {2024})}\BibitemShut {NoStop}%
\bibitem [{\citenamefont {Gan}\ \emph {et~al.}(1993)\citenamefont {Gan},
  \citenamefont {Andrei},\ and\ \citenamefont {Coleman}}]{PhysRevLett.70.686}%
  \BibitemOpen
  \bibfield  {author} {\bibinfo {author} {\bibfnamefont {J.}~\bibnamefont
  {Gan}}, \bibinfo {author} {\bibfnamefont {N.}~\bibnamefont {Andrei}},\ and\
  \bibinfo {author} {\bibfnamefont {P.}~\bibnamefont {Coleman}},\ }\bibfield
  {title} {\bibinfo {title} {{Perturbative approach to the non-Fermi-liquid
  fixed point of the overscreened Kondo problem}},\ }\href
  {https://doi.org/10.1103/PhysRevLett.70.686} {\bibfield  {journal} {\bibinfo
  {journal} {Phys. Rev. Lett.}\ }\textbf {\bibinfo {volume} {70}},\ \bibinfo
  {pages} {686} (\bibinfo {year} {1993})}\BibitemShut {NoStop}%
\bibitem [{\citenamefont {Andrei}\ and\ \citenamefont
  {Destri}(1984)}]{AndreiDestri1984}%
  \BibitemOpen
  \bibfield  {author} {\bibinfo {author} {\bibfnamefont {N.}~\bibnamefont
  {Andrei}}\ and\ \bibinfo {author} {\bibfnamefont {C.}~\bibnamefont
  {Destri}},\ }\bibfield  {title} {\bibinfo {title} {Solution of the
  multichannel kondo problem},\ }\href
  {https://doi.org/10.1103/PhysRevLett.52.364} {\bibfield  {journal} {\bibinfo
  {journal} {Phys. Rev. Lett.}\ }\textbf {\bibinfo {volume} {52}},\ \bibinfo
  {pages} {364} (\bibinfo {year} {1984})}\BibitemShut {NoStop}%
\bibitem [{\citenamefont {Tsvelick}\ and\ \citenamefont
  {Wiegmann}(1985)}]{TsvelickWiegmann1985}%
  \BibitemOpen
  \bibfield  {author} {\bibinfo {author} {\bibfnamefont {A.}~\bibnamefont
  {Tsvelick}}\ and\ \bibinfo {author} {\bibfnamefont {P.}~\bibnamefont
  {Wiegmann}},\ }\bibfield  {title} {\bibinfo {title} {Exact solution of the
  multichannel kondo problem, scaling, and integrability},\ }\href@noop {}
  {\bibfield  {journal} {\bibinfo  {journal} {Journal of Statistical Physics}\
  }\textbf {\bibinfo {volume} {38}},\ \bibinfo {pages} {125} (\bibinfo {year}
  {1985})}\BibitemShut {NoStop}%
\bibitem [{\citenamefont {Emery}\ and\ \citenamefont
  {Kivelson}(1992)}]{PhysRevB.46.10812}%
  \BibitemOpen
  \bibfield  {author} {\bibinfo {author} {\bibfnamefont {V.~J.}\ \bibnamefont
  {Emery}}\ and\ \bibinfo {author} {\bibfnamefont {S.}~\bibnamefont
  {Kivelson}},\ }\bibfield  {title} {\bibinfo {title} {Mapping of the
  two-channel kondo problem to a resonant-level model},\ }\href
  {https://doi.org/10.1103/PhysRevB.46.10812} {\bibfield  {journal} {\bibinfo
  {journal} {Phys. Rev. B}\ }\textbf {\bibinfo {volume} {46}},\ \bibinfo
  {pages} {10812} (\bibinfo {year} {1992})}\BibitemShut {NoStop}%
\bibitem [{\citenamefont {{Sengupta}}\ and\ \citenamefont
  {{Georges}}(1994)}]{1994PhRvB..4910020S}%
  \BibitemOpen
  \bibfield  {author} {\bibinfo {author} {\bibfnamefont {A.~M.}\ \bibnamefont
  {{Sengupta}}}\ and\ \bibinfo {author} {\bibfnamefont {A.}~\bibnamefont
  {{Georges}}},\ }\bibfield  {title} {\bibinfo {title} {{Emery-Kivelson
  solution of the two-channel Kondo problem}},\ }\href
  {https://doi.org/10.1103/PhysRevB.49.10020} {\bibfield  {journal} {\bibinfo
  {journal} {\prb}\ }\textbf {\bibinfo {volume} {49}},\ \bibinfo {pages}
  {10020} (\bibinfo {year} {1994})},\ \Eprint
  {https://arxiv.org/abs/cond-mat/9312016} {arXiv:cond-mat/9312016 [cond-mat]}
  \BibitemShut {NoStop}%
\bibitem [{\citenamefont {Coleman}\ \emph {et~al.}(1995)\citenamefont
  {Coleman}, \citenamefont {Ioffe},\ and\ \citenamefont
  {Tsvelik}}]{Coleman_1995}%
  \BibitemOpen
  \bibfield  {author} {\bibinfo {author} {\bibfnamefont {P.}~\bibnamefont
  {Coleman}}, \bibinfo {author} {\bibfnamefont {L.~B.}\ \bibnamefont {Ioffe}},\
  and\ \bibinfo {author} {\bibfnamefont {A.~M.}\ \bibnamefont {Tsvelik}},\
  }\bibfield  {title} {\bibinfo {title} {Simple formulation of the two-channel
  kondo model},\ }\href {https://doi.org/10.1103/physrevb.52.6611} {\bibfield
  {journal} {\bibinfo  {journal} {Physical Review B}\ }\textbf {\bibinfo
  {volume} {52}},\ \bibinfo {pages} {6611} (\bibinfo {year}
  {1995})}\BibitemShut {NoStop}%
\bibitem [{\citenamefont {Rozhkov}(1998)}]{Rozhkov_1998}%
  \BibitemOpen
  \bibfield  {author} {\bibinfo {author} {\bibfnamefont {A.~V.}\ \bibnamefont
  {Rozhkov}},\ }\bibfield  {title} {\bibinfo {title} {Impurity entropy for the
  two-channel kondo model},\ }\href {https://doi.org/10.1142/s0217979298002805}
  {\bibfield  {journal} {\bibinfo  {journal} {International Journal of Modern
  Physics B}\ }\textbf {\bibinfo {volume} {12}},\ \bibinfo {pages} {3457}
  (\bibinfo {year} {1998})}\BibitemShut {NoStop}%
\bibitem [{\citenamefont {Hartman}\ \emph {et~al.}(2018)\citenamefont
  {Hartman}, \citenamefont {Olsen}, \citenamefont {Lüscher}, \citenamefont
  {Samani}, \citenamefont {Fallahi}, \citenamefont {Gardner}, \citenamefont
  {Manfra},\ and\ \citenamefont {Folk}}]{hartman2018directentropymeasurement}%
  \BibitemOpen
  \bibfield  {author} {\bibinfo {author} {\bibfnamefont {N.}~\bibnamefont
  {Hartman}}, \bibinfo {author} {\bibfnamefont {C.}~\bibnamefont {Olsen}},
  \bibinfo {author} {\bibfnamefont {S.}~\bibnamefont {Lüscher}}, \bibinfo
  {author} {\bibfnamefont {M.}~\bibnamefont {Samani}}, \bibinfo {author}
  {\bibfnamefont {S.}~\bibnamefont {Fallahi}}, \bibinfo {author} {\bibfnamefont
  {G.~C.}\ \bibnamefont {Gardner}}, \bibinfo {author} {\bibfnamefont
  {M.}~\bibnamefont {Manfra}},\ and\ \bibinfo {author} {\bibfnamefont
  {J.}~\bibnamefont {Folk}},\ }\bibfield  {title} {\bibinfo {title} {Direct
  entropy measurement in a mesoscopic quantum system},\ }\href
  {https://doi.org/10.1038/s41567-018-0250-5} {\bibfield  {journal} {\bibinfo
  {journal} {Nature Physics}\ }\textbf {\bibinfo {volume} {14}},\ \bibinfo
  {pages} {1083–1086} (\bibinfo {year} {2018})}\BibitemShut {NoStop}%
\bibitem [{\citenamefont {Child}\ \emph {et~al.}(2022)\citenamefont {Child},
  \citenamefont {Sheekey}, \citenamefont {L\"uscher}, \citenamefont {Fallahi},
  \citenamefont {Gardner}, \citenamefont {Manfra}, \citenamefont {Mitchell},
  \citenamefont {Sela}, \citenamefont {Kleeorin}, \citenamefont {Meir},\ and\
  \citenamefont {Folk}}]{child2022qdotentropymeasurement}%
  \BibitemOpen
  \bibfield  {author} {\bibinfo {author} {\bibfnamefont {T.}~\bibnamefont
  {Child}}, \bibinfo {author} {\bibfnamefont {O.}~\bibnamefont {Sheekey}},
  \bibinfo {author} {\bibfnamefont {S.}~\bibnamefont {L\"uscher}}, \bibinfo
  {author} {\bibfnamefont {S.}~\bibnamefont {Fallahi}}, \bibinfo {author}
  {\bibfnamefont {G.~C.}\ \bibnamefont {Gardner}}, \bibinfo {author}
  {\bibfnamefont {M.}~\bibnamefont {Manfra}}, \bibinfo {author} {\bibfnamefont
  {A.}~\bibnamefont {Mitchell}}, \bibinfo {author} {\bibfnamefont
  {E.}~\bibnamefont {Sela}}, \bibinfo {author} {\bibfnamefont {Y.}~\bibnamefont
  {Kleeorin}}, \bibinfo {author} {\bibfnamefont {Y.}~\bibnamefont {Meir}},\
  and\ \bibinfo {author} {\bibfnamefont {J.}~\bibnamefont {Folk}},\ }\bibfield
  {title} {\bibinfo {title} {Entropy measurement of a strongly coupled quantum
  dot},\ }\href {https://doi.org/10.1103/PhysRevLett.129.227702} {\bibfield
  {journal} {\bibinfo  {journal} {Phys. Rev. Lett.}\ }\textbf {\bibinfo
  {volume} {129}},\ \bibinfo {pages} {227702} (\bibinfo {year}
  {2022})}\BibitemShut {NoStop}%
\bibitem [{\citenamefont {Han}\ \emph {et~al.}(2022)\citenamefont {Han},
  \citenamefont {Iftikhar}, \citenamefont {Kleeorin}, \citenamefont {Anthore},
  \citenamefont {Pierre}, \citenamefont {Meir}, \citenamefont {Mitchell},\ and\
  \citenamefont {Sela}}]{han2022fractionalentropy}%
  \BibitemOpen
  \bibfield  {author} {\bibinfo {author} {\bibfnamefont {C.}~\bibnamefont
  {Han}}, \bibinfo {author} {\bibfnamefont {Z.}~\bibnamefont {Iftikhar}},
  \bibinfo {author} {\bibfnamefont {Y.}~\bibnamefont {Kleeorin}}, \bibinfo
  {author} {\bibfnamefont {A.}~\bibnamefont {Anthore}}, \bibinfo {author}
  {\bibfnamefont {F.}~\bibnamefont {Pierre}}, \bibinfo {author} {\bibfnamefont
  {Y.}~\bibnamefont {Meir}}, \bibinfo {author} {\bibfnamefont {A.~K.}\
  \bibnamefont {Mitchell}},\ and\ \bibinfo {author} {\bibfnamefont
  {E.}~\bibnamefont {Sela}},\ }\bibfield  {title} {\bibinfo {title} {Fractional
  entropy of multichannel kondo systems from conductance-charge relations},\
  }\href {https://doi.org/10.1103/PhysRevLett.128.146803} {\bibfield  {journal}
  {\bibinfo  {journal} {Phys. Rev. Lett.}\ }\textbf {\bibinfo {volume} {128}},\
  \bibinfo {pages} {146803} (\bibinfo {year} {2022})}\BibitemShut {NoStop}%
\bibitem [{\citenamefont {Affleck}\ and\ \citenamefont
  {Ludwig}(1991{\natexlab{a}})}]{PhysRevLett.67.161}%
  \BibitemOpen
  \bibfield  {author} {\bibinfo {author} {\bibfnamefont {I.}~\bibnamefont
  {Affleck}}\ and\ \bibinfo {author} {\bibfnamefont {A.~W.~W.}\ \bibnamefont
  {Ludwig}},\ }\bibfield  {title} {\bibinfo {title} {Universal noninteger
  ``ground-state degeneracy'' in critical quantum systems},\ }\href
  {https://doi.org/10.1103/PhysRevLett.67.161} {\bibfield  {journal} {\bibinfo
  {journal} {Phys. Rev. Lett.}\ }\textbf {\bibinfo {volume} {67}},\ \bibinfo
  {pages} {161} (\bibinfo {year} {1991}{\natexlab{a}})}\BibitemShut {NoStop}%
\bibitem [{\citenamefont {Affleck}(1990)}]{affleck1990current}%
  \BibitemOpen
  \bibfield  {author} {\bibinfo {author} {\bibfnamefont {I.}~\bibnamefont
  {Affleck}},\ }\bibfield  {title} {\bibinfo {title} {A current algebra
  approach to the kondo effect},\ }\href@noop {} {\bibfield  {journal}
  {\bibinfo  {journal} {Nuclear Physics B}\ }\textbf {\bibinfo {volume}
  {336}},\ \bibinfo {pages} {517} (\bibinfo {year} {1990})}\BibitemShut
  {NoStop}%
\bibitem [{\citenamefont {Affleck}\ and\ \citenamefont
  {Ludwig}(1991{\natexlab{b}})}]{affleck1991kondo}%
  \BibitemOpen
  \bibfield  {author} {\bibinfo {author} {\bibfnamefont {I.}~\bibnamefont
  {Affleck}}\ and\ \bibinfo {author} {\bibfnamefont {A.~W.}\ \bibnamefont
  {Ludwig}},\ }\bibfield  {title} {\bibinfo {title} {The kondo effect,
  conformal field theory and fusion rules},\ }\href@noop {} {\bibfield
  {journal} {\bibinfo  {journal} {Nuclear Physics B}\ }\textbf {\bibinfo
  {volume} {352}},\ \bibinfo {pages} {849} (\bibinfo {year}
  {1991}{\natexlab{b}})}\BibitemShut {NoStop}%
\bibitem [{\citenamefont {Affleck}\ and\ \citenamefont
  {Ludwig}(1991{\natexlab{c}})}]{affleck1991critical}%
  \BibitemOpen
  \bibfield  {author} {\bibinfo {author} {\bibfnamefont {I.}~\bibnamefont
  {Affleck}}\ and\ \bibinfo {author} {\bibfnamefont {A.~W.}\ \bibnamefont
  {Ludwig}},\ }\bibfield  {title} {\bibinfo {title} {Critical theory of
  overscreened kondo fixed points},\ }\href@noop {} {\bibfield  {journal}
  {\bibinfo  {journal} {Nuclear Physics B}\ }\textbf {\bibinfo {volume}
  {360}},\ \bibinfo {pages} {641} (\bibinfo {year}
  {1991}{\natexlab{c}})}\BibitemShut {NoStop}%
\bibitem [{\citenamefont {Affleck}\ and\ \citenamefont
  {Ludwig}(1993)}]{PhysRevB.48.7297}%
  \BibitemOpen
  \bibfield  {author} {\bibinfo {author} {\bibfnamefont {I.}~\bibnamefont
  {Affleck}}\ and\ \bibinfo {author} {\bibfnamefont {A.~W.~W.}\ \bibnamefont
  {Ludwig}},\ }\bibfield  {title} {\bibinfo {title} {Exact
  conformal-field-theory results on the multichannel kondo effect:
  Single-fermion green's function, self-energy, and resistivity},\ }\href
  {https://doi.org/10.1103/PhysRevB.48.7297} {\bibfield  {journal} {\bibinfo
  {journal} {Phys. Rev. B}\ }\textbf {\bibinfo {volume} {48}},\ \bibinfo
  {pages} {7297} (\bibinfo {year} {1993})}\BibitemShut {NoStop}%
\bibitem [{\citenamefont {Ludwig}\ and\ \citenamefont
  {Affleck}(1994)}]{ludwig1994exact}%
  \BibitemOpen
  \bibfield  {author} {\bibinfo {author} {\bibfnamefont {A.~W.}\ \bibnamefont
  {Ludwig}}\ and\ \bibinfo {author} {\bibfnamefont {I.}~\bibnamefont
  {Affleck}},\ }\bibfield  {title} {\bibinfo {title} {Exact
  conformal-field-theory results on the multi-channel kondo effect: Asymptotic
  three-dimensional space-and time-dependent multi-point and many-particle
  green's functions},\ }\href@noop {} {\bibfield  {journal} {\bibinfo
  {journal} {Nuclear Physics B}\ }\textbf {\bibinfo {volume} {428}},\ \bibinfo
  {pages} {545} (\bibinfo {year} {1994})}\BibitemShut {NoStop}%
\bibitem [{\citenamefont {Ludwig}\ and\ \citenamefont
  {Affleck}(1991)}]{PhysRevLett.67.3160}%
  \BibitemOpen
  \bibfield  {author} {\bibinfo {author} {\bibfnamefont {A.~W.~W.}\
  \bibnamefont {Ludwig}}\ and\ \bibinfo {author} {\bibfnamefont
  {I.}~\bibnamefont {Affleck}},\ }\bibfield  {title} {\bibinfo {title} {Exact,
  asymptotic, three-dimensional, space- and time-dependent, green's functions
  in the multichannel kondo effect},\ }\href
  {https://doi.org/10.1103/PhysRevLett.67.3160} {\bibfield  {journal} {\bibinfo
   {journal} {Phys. Rev. Lett.}\ }\textbf {\bibinfo {volume} {67}},\ \bibinfo
  {pages} {3160} (\bibinfo {year} {1991})}\BibitemShut {NoStop}%
\bibitem [{\citenamefont {Parcollet}\ \emph {et~al.}(1998)\citenamefont
  {Parcollet}, \citenamefont {Georges}, \citenamefont {Kotliar},\ and\
  \citenamefont {Sengupta}}]{PhysRevB.58.3794}%
  \BibitemOpen
  \bibfield  {author} {\bibinfo {author} {\bibfnamefont {O.}~\bibnamefont
  {Parcollet}}, \bibinfo {author} {\bibfnamefont {A.}~\bibnamefont {Georges}},
  \bibinfo {author} {\bibfnamefont {G.}~\bibnamefont {Kotliar}},\ and\ \bibinfo
  {author} {\bibfnamefont {A.}~\bibnamefont {Sengupta}},\ }\bibfield  {title}
  {\bibinfo {title} {{Overscreened multichannel $\mathrm{SU}(N)$ Kondo model:
  Large-$N$ solution and conformal field theory}},\ }\href
  {https://doi.org/10.1103/PhysRevB.58.3794} {\bibfield  {journal} {\bibinfo
  {journal} {Phys. Rev. B}\ }\textbf {\bibinfo {volume} {58}},\ \bibinfo
  {pages} {3794} (\bibinfo {year} {1998})}\BibitemShut {NoStop}%
\bibitem [{\citenamefont
  {Kimura}(2021{\natexlab{a}})}]{doi:10.7566/JPSJ.90.024708}%
  \BibitemOpen
  \bibfield  {author} {\bibinfo {author} {\bibfnamefont {T.}~\bibnamefont
  {Kimura}},\ }\bibfield  {title} {\bibinfo {title} {Abcd of kondo effect},\
  }\href {https://doi.org/10.7566/JPSJ.90.024708} {\bibfield  {journal}
  {\bibinfo  {journal} {Journal of the Physical Society of Japan}\ }\textbf
  {\bibinfo {volume} {90}},\ \bibinfo {pages} {024708} (\bibinfo {year}
  {2021}{\natexlab{a}})}\BibitemShut {NoStop}%
\bibitem [{\citenamefont {B\'eri}\ and\ \citenamefont
  {Cooper}(2012)}]{PhysRevLett.109.156803}%
  \BibitemOpen
  \bibfield  {author} {\bibinfo {author} {\bibfnamefont {B.}~\bibnamefont
  {B\'eri}}\ and\ \bibinfo {author} {\bibfnamefont {N.~R.}\ \bibnamefont
  {Cooper}},\ }\bibfield  {title} {\bibinfo {title} {Topological kondo effect
  with majorana fermions},\ }\href
  {https://doi.org/10.1103/PhysRevLett.109.156803} {\bibfield  {journal}
  {\bibinfo  {journal} {Phys. Rev. Lett.}\ }\textbf {\bibinfo {volume} {109}},\
  \bibinfo {pages} {156803} (\bibinfo {year} {2012})}\BibitemShut {NoStop}%
\bibitem [{\citenamefont {B\'eri}(2013)}]{PhysRevLett.110.216803}%
  \BibitemOpen
  \bibfield  {author} {\bibinfo {author} {\bibfnamefont {B.}~\bibnamefont
  {B\'eri}},\ }\bibfield  {title} {\bibinfo {title} {Majorana-klein
  hybridization in topological superconductor junctions},\ }\href
  {https://doi.org/10.1103/PhysRevLett.110.216803} {\bibfield  {journal}
  {\bibinfo  {journal} {Phys. Rev. Lett.}\ }\textbf {\bibinfo {volume} {110}},\
  \bibinfo {pages} {216803} (\bibinfo {year} {2013})}\BibitemShut {NoStop}%
\bibitem [{\citenamefont {Li}\ \emph {et~al.}(2023{\natexlab{a}})\citenamefont
  {Li}, \citenamefont {Oreg},\ and\ \citenamefont
  {V\"ayrynen}}]{PhysRevLett.130.066302}%
  \BibitemOpen
  \bibfield  {author} {\bibinfo {author} {\bibfnamefont {G.}~\bibnamefont
  {Li}}, \bibinfo {author} {\bibfnamefont {Y.}~\bibnamefont {Oreg}},\ and\
  \bibinfo {author} {\bibfnamefont {J.~I.}\ \bibnamefont {V\"ayrynen}},\
  }\bibfield  {title} {\bibinfo {title} {Multichannel topological kondo
  effect},\ }\href {https://doi.org/10.1103/PhysRevLett.130.066302} {\bibfield
  {journal} {\bibinfo  {journal} {Phys. Rev. Lett.}\ }\textbf {\bibinfo
  {volume} {130}},\ \bibinfo {pages} {066302} (\bibinfo {year}
  {2023}{\natexlab{a}})}\BibitemShut {NoStop}%
\bibitem [{\citenamefont {Bollmann}\ \emph {et~al.}(2024)\citenamefont
  {Bollmann}, \citenamefont {V\"ayrynen},\ and\ \citenamefont
  {K\"onig}}]{BollmannKoenig2024}%
  \BibitemOpen
  \bibfield  {author} {\bibinfo {author} {\bibfnamefont {S.}~\bibnamefont
  {Bollmann}}, \bibinfo {author} {\bibfnamefont {J.~I.}\ \bibnamefont
  {V\"ayrynen}},\ and\ \bibinfo {author} {\bibfnamefont {E.~J.}\ \bibnamefont
  {K\"onig}},\ }\bibfield  {title} {\bibinfo {title} {Topological kondo effect
  with spinful majorana fermions},\ }\href
  {https://doi.org/10.1103/PhysRevB.110.035136} {\bibfield  {journal} {\bibinfo
   {journal} {Phys. Rev. B}\ }\textbf {\bibinfo {volume} {110}},\ \bibinfo
  {pages} {035136} (\bibinfo {year} {2024})}\BibitemShut {NoStop}%
\bibitem [{\citenamefont {Li}\ \emph {et~al.}(2023{\natexlab{b}})\citenamefont
  {Li}, \citenamefont {K\"onig},\ and\ \citenamefont
  {V\"ayrynen}}]{PhysRevB.107.L201401}%
  \BibitemOpen
  \bibfield  {author} {\bibinfo {author} {\bibfnamefont {G.}~\bibnamefont
  {Li}}, \bibinfo {author} {\bibfnamefont {E.~J.}\ \bibnamefont {K\"onig}},\
  and\ \bibinfo {author} {\bibfnamefont {J.~I.}\ \bibnamefont {V\"ayrynen}},\
  }\bibfield  {title} {\bibinfo {title} {Topological symplectic kondo effect},\
  }\href {https://doi.org/10.1103/PhysRevB.107.L201401} {\bibfield  {journal}
  {\bibinfo  {journal} {Phys. Rev. B}\ }\textbf {\bibinfo {volume} {107}},\
  \bibinfo {pages} {L201401} (\bibinfo {year}
  {2023}{\natexlab{b}})}\BibitemShut {NoStop}%
\bibitem [{\citenamefont {König}\ and\ \citenamefont
  {Tsvelik}(2023)}]{KONIG2023169231}%
  \BibitemOpen
  \bibfield  {author} {\bibinfo {author} {\bibfnamefont {E.~J.}\ \bibnamefont
  {König}}\ and\ \bibinfo {author} {\bibfnamefont {A.~M.}\ \bibnamefont
  {Tsvelik}},\ }\bibfield  {title} {\bibinfo {title} {Exact solution of the
  topological symplectic kondo problem},\ }\href
  {https://doi.org/https://doi.org/10.1016/j.aop.2023.169231} {\bibfield
  {journal} {\bibinfo  {journal} {Annals of Physics}\ }\textbf {\bibinfo
  {volume} {456}},\ \bibinfo {pages} {169231} (\bibinfo {year}
  {2023})}\BibitemShut {NoStop}%
\bibitem [{\citenamefont {Lotem}\ \emph
  {et~al.}(2024{\natexlab{a}})\citenamefont {Lotem}, \citenamefont {Sankar},
  \citenamefont {Ren}, \citenamefont {Goldstein}, \citenamefont {K\"onig},
  \citenamefont {Weichselbaum}, \citenamefont {Sela},\ and\ \citenamefont
  {Tsvelik}}]{LotemAnisotropyTSK}%
  \BibitemOpen
  \bibfield  {author} {\bibinfo {author} {\bibfnamefont {M.}~\bibnamefont
  {Lotem}}, \bibinfo {author} {\bibfnamefont {S.}~\bibnamefont {Sankar}},
  \bibinfo {author} {\bibfnamefont {T.}~\bibnamefont {Ren}}, \bibinfo {author}
  {\bibfnamefont {M.}~\bibnamefont {Goldstein}}, \bibinfo {author}
  {\bibfnamefont {E.~J.}\ \bibnamefont {K\"onig}}, \bibinfo {author}
  {\bibfnamefont {A.}~\bibnamefont {Weichselbaum}}, \bibinfo {author}
  {\bibfnamefont {E.}~\bibnamefont {Sela}},\ and\ \bibinfo {author}
  {\bibfnamefont {A.~M.}\ \bibnamefont {Tsvelik}},\ }\bibfield  {title}
  {\bibinfo {title} {Relevance of anisotropy in the kondo effect: Lessons from
  the symplectic case},\ }\href {https://doi.org/10.1103/PhysRevB.110.235122}
  {\bibfield  {journal} {\bibinfo  {journal} {Phys. Rev. B}\ }\textbf {\bibinfo
  {volume} {110}},\ \bibinfo {pages} {235122} (\bibinfo {year}
  {2024}{\natexlab{a}})}\BibitemShut {NoStop}%
\bibitem [{\citenamefont {V\"ayrynen}\ \emph {et~al.}(2020)\citenamefont
  {V\"ayrynen}, \citenamefont {Feiguin},\ and\ \citenamefont
  {Lutchyn}}]{PhysRevResearch.2.043228}%
  \BibitemOpen
  \bibfield  {author} {\bibinfo {author} {\bibfnamefont {J.~I.}\ \bibnamefont
  {V\"ayrynen}}, \bibinfo {author} {\bibfnamefont {A.~E.}\ \bibnamefont
  {Feiguin}},\ and\ \bibinfo {author} {\bibfnamefont {R.~M.}\ \bibnamefont
  {Lutchyn}},\ }\bibfield  {title} {\bibinfo {title} {Signatures of topological
  ground state degeneracy in majorana islands},\ }\href
  {https://doi.org/10.1103/PhysRevResearch.2.043228} {\bibfield  {journal}
  {\bibinfo  {journal} {Phys. Rev. Research}\ }\textbf {\bibinfo {volume}
  {2}},\ \bibinfo {pages} {043228} (\bibinfo {year} {2020})}\BibitemShut
  {NoStop}%
\bibitem [{\citenamefont {qiang Bao}\ and\ \citenamefont
  {Zhang}(2017)}]{bao2017quantumhallchargekondo}%
  \BibitemOpen
  \bibfield  {author} {\bibinfo {author} {\bibfnamefont {Z.}~\bibnamefont
  {qiang Bao}}\ and\ \bibinfo {author} {\bibfnamefont {F.}~\bibnamefont
  {Zhang}},\ }\href {https://arxiv.org/abs/1708.09139} {\bibinfo {title}
  {Quantum hall charge kondo criticality}} (\bibinfo {year} {2017}),\ \Eprint
  {https://arxiv.org/abs/1708.09139} {arXiv:1708.09139 [cond-mat.str-el]}
  \BibitemShut {NoStop}%
\bibitem [{\citenamefont {Ludwig}(1994)}]{ludwig1994field}%
  \BibitemOpen
  \bibfield  {author} {\bibinfo {author} {\bibfnamefont {A.~W.}\ \bibnamefont
  {Ludwig}},\ }\bibfield  {title} {\bibinfo {title} {Field theory approach to
  critical quantum impurity problems and applications to the multi-channel
  kondo effect},\ }\href@noop {} {\bibfield  {journal} {\bibinfo  {journal}
  {International Journal of Modern Physics B}\ }\textbf {\bibinfo {volume}
  {8}},\ \bibinfo {pages} {347} (\bibinfo {year} {1994})}\BibitemShut {NoStop}%
\bibitem [{\citenamefont {Lotem}\ \emph
  {et~al.}(2024{\natexlab{b}})\citenamefont {Lotem}, \citenamefont {Sankar},
  \citenamefont {Ren}, \citenamefont {Goldstein}, \citenamefont {K\"onig},
  \citenamefont {Weichselbaum}, \citenamefont {Sela},\ and\ \citenamefont
  {Tsvelik}}]{matanrelevanceofanisotropy2024}%
  \BibitemOpen
  \bibfield  {author} {\bibinfo {author} {\bibfnamefont {M.}~\bibnamefont
  {Lotem}}, \bibinfo {author} {\bibfnamefont {S.}~\bibnamefont {Sankar}},
  \bibinfo {author} {\bibfnamefont {T.}~\bibnamefont {Ren}}, \bibinfo {author}
  {\bibfnamefont {M.}~\bibnamefont {Goldstein}}, \bibinfo {author}
  {\bibfnamefont {E.~J.}\ \bibnamefont {K\"onig}}, \bibinfo {author}
  {\bibfnamefont {A.}~\bibnamefont {Weichselbaum}}, \bibinfo {author}
  {\bibfnamefont {E.}~\bibnamefont {Sela}},\ and\ \bibinfo {author}
  {\bibfnamefont {A.~M.}\ \bibnamefont {Tsvelik}},\ }\bibfield  {title}
  {\bibinfo {title} {Relevance of anisotropy in the kondo effect: Lessons from
  the symplectic case},\ }\href {https://doi.org/10.1103/PhysRevB.110.235122}
  {\bibfield  {journal} {\bibinfo  {journal} {Phys. Rev. B}\ }\textbf {\bibinfo
  {volume} {110}},\ \bibinfo {pages} {235122} (\bibinfo {year}
  {2024}{\natexlab{b}})}\BibitemShut {NoStop}%
\bibitem [{\citenamefont {Tsvelik}(2014)}]{tsvelik2014topological}%
  \BibitemOpen
  \bibfield  {author} {\bibinfo {author} {\bibfnamefont {A.}~\bibnamefont
  {Tsvelik}},\ }\bibfield  {title} {\bibinfo {title} {Topological kondo effect
  in star junctions of ising magnetic chains: exact solution},\ }\href@noop {}
  {\bibfield  {journal} {\bibinfo  {journal} {New Journal of Physics}\ }\textbf
  {\bibinfo {volume} {16}},\ \bibinfo {pages} {033003} (\bibinfo {year}
  {2014})}\BibitemShut {NoStop}%
\bibitem [{\citenamefont {Mitchell}\ \emph {et~al.}(2021)\citenamefont
  {Mitchell}, \citenamefont {Liberman}, \citenamefont {Sela},\ and\
  \citenamefont {Affleck}}]{MitchellAffleck2021}%
  \BibitemOpen
  \bibfield  {author} {\bibinfo {author} {\bibfnamefont {A.~K.}\ \bibnamefont
  {Mitchell}}, \bibinfo {author} {\bibfnamefont {A.}~\bibnamefont {Liberman}},
  \bibinfo {author} {\bibfnamefont {E.}~\bibnamefont {Sela}},\ and\ \bibinfo
  {author} {\bibfnamefont {I.}~\bibnamefont {Affleck}},\ }\bibfield  {title}
  {\bibinfo {title} {{SO(5) Non-Fermi Liquid in a Coulomb Box Device}},\ }\href
  {https://doi.org/10.1103/PhysRevLett.126.147702} {\bibfield  {journal}
  {\bibinfo  {journal} {Phys. Rev. Lett.}\ }\textbf {\bibinfo {volume} {126}},\
  \bibinfo {pages} {147702} (\bibinfo {year} {2021})}\BibitemShut {NoStop}%
\bibitem [{\citenamefont {Liberman}\ \emph {et~al.}(2021)\citenamefont
  {Liberman}, \citenamefont {Mitchell}, \citenamefont {Affleck},\ and\
  \citenamefont {Sela}}]{LibermanSela2021}%
  \BibitemOpen
  \bibfield  {author} {\bibinfo {author} {\bibfnamefont {A.}~\bibnamefont
  {Liberman}}, \bibinfo {author} {\bibfnamefont {A.~K.}\ \bibnamefont
  {Mitchell}}, \bibinfo {author} {\bibfnamefont {I.}~\bibnamefont {Affleck}},\
  and\ \bibinfo {author} {\bibfnamefont {E.}~\bibnamefont {Sela}},\ }\bibfield
  {title} {\bibinfo {title} {{SO(5) critical point in a spin-flavor Kondo
  device: Bosonization and refermionization solution}},\ }\href
  {https://doi.org/10.1103/PhysRevB.103.195131} {\bibfield  {journal} {\bibinfo
   {journal} {Phys. Rev. B}\ }\textbf {\bibinfo {volume} {103}},\ \bibinfo
  {pages} {195131} (\bibinfo {year} {2021})}\BibitemShut {NoStop}%
\bibitem [{\citenamefont {Georgi}(2000)}]{georgi2000lie}%
  \BibitemOpen
  \bibfield  {author} {\bibinfo {author} {\bibfnamefont {H.}~\bibnamefont
  {Georgi}},\ }\href@noop {} {\emph {\bibinfo {title} {Lie algebras in particle
  physics: from isospin to unified theories}}}\ (\bibinfo  {publisher} {Taylor
  \& Francis},\ \bibinfo {year} {2000})\BibitemShut {NoStop}%
\bibitem [{\citenamefont {Ma}(2007)}]{ma2007group}%
  \BibitemOpen
  \bibfield  {author} {\bibinfo {author} {\bibfnamefont {Z.-Q.}\ \bibnamefont
  {Ma}},\ }\href@noop {} {\emph {\bibinfo {title} {Group theory for
  physicists}}}\ (\bibinfo  {publisher} {World Scientific},\ \bibinfo {year}
  {2007})\BibitemShut {NoStop}%
\bibitem [{\citenamefont {Zee}(2016)}]{zee2016group}%
  \BibitemOpen
  \bibfield  {author} {\bibinfo {author} {\bibfnamefont {A.}~\bibnamefont
  {Zee}},\ }\href@noop {} {\emph {\bibinfo {title} {Group theory in a nutshell
  for physicists}}},\ Vol.~\bibinfo {volume} {17}\ (\bibinfo  {publisher}
  {Princeton University Press},\ \bibinfo {year} {2016})\BibitemShut {NoStop}%
\bibitem [{\citenamefont {Feger}\ \emph {et~al.}(2020)\citenamefont {Feger},
  \citenamefont {Kephart},\ and\ \citenamefont {Saskowski}}]{LieARTFeger_2020}%
  \BibitemOpen
  \bibfield  {author} {\bibinfo {author} {\bibfnamefont {R.}~\bibnamefont
  {Feger}}, \bibinfo {author} {\bibfnamefont {T.~W.}\ \bibnamefont {Kephart}},\
  and\ \bibinfo {author} {\bibfnamefont {R.~J.}\ \bibnamefont {Saskowski}},\
  }\bibfield  {title} {\bibinfo {title} {Lieart 2.0 – a mathematica
  application for lie algebras and representation theory},\ }\href
  {https://doi.org/10.1016/j.cpc.2020.107490} {\bibfield  {journal} {\bibinfo
  {journal} {Computer Physics Communications}\ }\textbf {\bibinfo {volume}
  {257}},\ \bibinfo {pages} {107490} (\bibinfo {year} {2020})}\BibitemShut
  {NoStop}%
\bibitem [{\citenamefont {Fabrizio}\ and\ \citenamefont
  {Gogolin}(1994)}]{PhysRevB.50.17732}%
  \BibitemOpen
  \bibfield  {author} {\bibinfo {author} {\bibfnamefont {M.}~\bibnamefont
  {Fabrizio}}\ and\ \bibinfo {author} {\bibfnamefont {A.~O.}\ \bibnamefont
  {Gogolin}},\ }\bibfield  {title} {\bibinfo {title} {Toulouse limit for the
  overscreened four-channel kondo problem},\ }\href
  {https://doi.org/10.1103/PhysRevB.50.17732} {\bibfield  {journal} {\bibinfo
  {journal} {Phys. Rev. B}\ }\textbf {\bibinfo {volume} {50}},\ \bibinfo
  {pages} {17732} (\bibinfo {year} {1994})}\BibitemShut {NoStop}%
\bibitem [{\citenamefont {Sengupta}\ and\ \citenamefont
  {Kim}(1996)}]{PhysRevB.54.14918}%
  \BibitemOpen
  \bibfield  {author} {\bibinfo {author} {\bibfnamefont {A.~M.}\ \bibnamefont
  {Sengupta}}\ and\ \bibinfo {author} {\bibfnamefont {Y.~B.}\ \bibnamefont
  {Kim}},\ }\bibfield  {title} {\bibinfo {title} {Overscreened single-channel
  kondo problem},\ }\href {https://doi.org/10.1103/PhysRevB.54.14918}
  {\bibfield  {journal} {\bibinfo  {journal} {Phys. Rev. B}\ }\textbf {\bibinfo
  {volume} {54}},\ \bibinfo {pages} {14918} (\bibinfo {year}
  {1996})}\BibitemShut {NoStop}%
\bibitem [{\citenamefont {Altland}\ \emph
  {et~al.}(2014{\natexlab{a}})\citenamefont {Altland}, \citenamefont {B\'eri},
  \citenamefont {Egger},\ and\ \citenamefont
  {Tsvelik}}]{PhysRevLett.113.076401}%
  \BibitemOpen
  \bibfield  {author} {\bibinfo {author} {\bibfnamefont {A.}~\bibnamefont
  {Altland}}, \bibinfo {author} {\bibfnamefont {B.}~\bibnamefont {B\'eri}},
  \bibinfo {author} {\bibfnamefont {R.}~\bibnamefont {Egger}},\ and\ \bibinfo
  {author} {\bibfnamefont {A.~M.}\ \bibnamefont {Tsvelik}},\ }\bibfield
  {title} {\bibinfo {title} {Multichannel kondo impurity dynamics in a majorana
  device},\ }\href {https://doi.org/10.1103/PhysRevLett.113.076401} {\bibfield
  {journal} {\bibinfo  {journal} {Phys. Rev. Lett.}\ }\textbf {\bibinfo
  {volume} {113}},\ \bibinfo {pages} {076401} (\bibinfo {year}
  {2014}{\natexlab{a}})}\BibitemShut {NoStop}%
\bibitem [{\citenamefont {Verlinde}(1988)}]{VERLINDE1988360}%
  \BibitemOpen
  \bibfield  {author} {\bibinfo {author} {\bibfnamefont {E.}~\bibnamefont
  {Verlinde}},\ }\bibfield  {title} {\bibinfo {title} {Fusion rules and modular
  transformations in 2d conformal field theory},\ }\href
  {https://doi.org/https://doi.org/10.1016/0550-3213(88)90603-7} {\bibfield
  {journal} {\bibinfo  {journal} {Nuclear Physics B}\ }\textbf {\bibinfo
  {volume} {300}},\ \bibinfo {pages} {360} (\bibinfo {year}
  {1988})}\BibitemShut {NoStop}%
\bibitem [{\citenamefont {Hung}\ \emph {et~al.}(2018)\citenamefont {Hung},
  \citenamefont {Wu},\ and\ \citenamefont {Zhou}}]{hung2018linking}%
  \BibitemOpen
  \bibfield  {author} {\bibinfo {author} {\bibfnamefont {L.-Y.}\ \bibnamefont
  {Hung}}, \bibinfo {author} {\bibfnamefont {Y.-S.}\ \bibnamefont {Wu}},\ and\
  \bibinfo {author} {\bibfnamefont {Y.}~\bibnamefont {Zhou}},\ }\bibfield
  {title} {\bibinfo {title} {Linking entanglement and discrete anomaly},\
  }\href {https://doi.org/10.1007/JHEP05(2018)008} {\bibfield  {journal}
  {\bibinfo  {journal} {Journal of High Energy Physics}\ }\textbf {\bibinfo
  {volume} {2018}},\ \bibinfo {pages} {8} (\bibinfo {year} {2018})}\BibitemShut
  {NoStop}%
\bibitem [{\citenamefont {Kong}\ and\ \citenamefont
  {Zhang}(2022)}]{kong2022invitationtopologicalorderscategory}%
  \BibitemOpen
  \bibfield  {author} {\bibinfo {author} {\bibfnamefont {L.}~\bibnamefont
  {Kong}}\ and\ \bibinfo {author} {\bibfnamefont {Z.-H.}\ \bibnamefont
  {Zhang}},\ }\href {https://arxiv.org/abs/2205.05565} {\bibinfo {title} {An
  invitation to topological orders and category theory}} (\bibinfo {year}
  {2022}),\ \Eprint {https://arxiv.org/abs/2205.05565} {arXiv:2205.05565
  [cond-mat.str-el]} \BibitemShut {NoStop}%
\bibitem [{\citenamefont {Kimura}(2021{\natexlab{b}})}]{Kimura2021}%
  \BibitemOpen
  \bibfield  {author} {\bibinfo {author} {\bibfnamefont {T.}~\bibnamefont
  {Kimura}},\ }\bibfield  {title} {\bibinfo {title} {Abcd of kondo effect},\
  }\href {https://doi.org/10.7566/JPSJ.90.024708} {\bibfield  {journal}
  {\bibinfo  {journal} {Journal of the Physical Society of Japan}\ }\textbf
  {\bibinfo {volume} {90}},\ \bibinfo {pages} {024708} (\bibinfo {year}
  {2021}{\natexlab{b}})}\BibitemShut {NoStop}%
\bibitem [{\citenamefont {Altland}\ \emph
  {et~al.}(2014{\natexlab{b}})\citenamefont {Altland}, \citenamefont
  {B{\'{e}}ri}, \citenamefont {Egger},\ and\ \citenamefont
  {Tsvelik}}]{Altland_2014}%
  \BibitemOpen
  \bibfield  {author} {\bibinfo {author} {\bibfnamefont {A.}~\bibnamefont
  {Altland}}, \bibinfo {author} {\bibfnamefont {B.}~\bibnamefont {B{\'{e}}ri}},
  \bibinfo {author} {\bibfnamefont {R.}~\bibnamefont {Egger}},\ and\ \bibinfo
  {author} {\bibfnamefont {A.~M.}\ \bibnamefont {Tsvelik}},\ }\bibfield
  {title} {\bibinfo {title} {Bethe ansatz solution of the topological kondo
  model},\ }\href {https://doi.org/10.1088/1751-8113/47/26/265001} {\bibfield
  {journal} {\bibinfo  {journal} {Journal of Physics A: Mathematical and
  Theoretical}\ }\textbf {\bibinfo {volume} {47}},\ \bibinfo {pages} {265001}
  (\bibinfo {year} {2014}{\natexlab{b}})}\BibitemShut {NoStop}%
\bibitem [{\citenamefont {Gogolin}\ \emph {et~al.}(2004)\citenamefont
  {Gogolin}, \citenamefont {Nersesyan},\ and\ \citenamefont
  {Tsvelik}}]{gogolin2004bosonization}%
  \BibitemOpen
  \bibfield  {author} {\bibinfo {author} {\bibfnamefont {A.~O.}\ \bibnamefont
  {Gogolin}}, \bibinfo {author} {\bibfnamefont {A.~A.}\ \bibnamefont
  {Nersesyan}},\ and\ \bibinfo {author} {\bibfnamefont {A.~M.}\ \bibnamefont
  {Tsvelik}},\ }\href@noop {} {\emph {\bibinfo {title} {Bosonization and
  strongly correlated systems}}}\ (\bibinfo  {publisher} {Cambridge university
  press},\ \bibinfo {year} {2004})\BibitemShut {NoStop}%
\bibitem [{\citenamefont {LUDWIG}(1995)}]{LudwigCFTinCMP}%
  \BibitemOpen
  \bibfield  {author} {\bibinfo {author} {\bibfnamefont {A.~W.~W.}\
  \bibnamefont {LUDWIG}},\ }\bibinfo {title} {Methods of conformal field theory
  in condensed matter physics},\ in\ \href
  {https://doi.org/10.1142/9789814447027_0007} {\emph {\bibinfo {booktitle}
  {Low-Dimensional Quantum Field Theories for Condensed Matter Physicists}}}\
  (\bibinfo {year} {1995})\ pp.\ \bibinfo {pages} {389--455}\BibitemShut
  {NoStop}%
\bibitem [{\citenamefont {Francesco}\ \emph {et~al.}(2012)\citenamefont
  {Francesco}, \citenamefont {Mathieu},\ and\ \citenamefont
  {S{\'e}n{\'e}chal}}]{francesco2012conformal}%
  \BibitemOpen
  \bibfield  {author} {\bibinfo {author} {\bibfnamefont {P.}~\bibnamefont
  {Francesco}}, \bibinfo {author} {\bibfnamefont {P.}~\bibnamefont {Mathieu}},\
  and\ \bibinfo {author} {\bibfnamefont {D.}~\bibnamefont {S{\'e}n{\'e}chal}},\
  }\href@noop {} {\emph {\bibinfo {title} {Conformal field theory}}}\ (\bibinfo
   {publisher} {Springer Science \& Business Media},\ \bibinfo {year}
  {2012})\BibitemShut {NoStop}%
\bibitem [{\citenamefont {Michaeli}\ \emph {et~al.}(2017)\citenamefont
  {Michaeli}, \citenamefont {Landau}, \citenamefont {Sela},\ and\ \citenamefont
  {Fu}}]{PhysRevB.96.205403}%
  \BibitemOpen
  \bibfield  {author} {\bibinfo {author} {\bibfnamefont {K.}~\bibnamefont
  {Michaeli}}, \bibinfo {author} {\bibfnamefont {L.~A.}\ \bibnamefont
  {Landau}}, \bibinfo {author} {\bibfnamefont {E.}~\bibnamefont {Sela}},\ and\
  \bibinfo {author} {\bibfnamefont {L.}~\bibnamefont {Fu}},\ }\bibfield
  {title} {\bibinfo {title} {Electron teleportation and statistical
  transmutation in multiterminal majorana islands},\ }\href
  {https://doi.org/10.1103/PhysRevB.96.205403} {\bibfield  {journal} {\bibinfo
  {journal} {Phys. Rev. B}\ }\textbf {\bibinfo {volume} {96}},\ \bibinfo
  {pages} {205403} (\bibinfo {year} {2017})}\BibitemShut {NoStop}%
\bibitem [{\citenamefont {Rahmani}\ \emph {et~al.}(2010)\citenamefont
  {Rahmani}, \citenamefont {Hou}, \citenamefont {Feiguin}, \citenamefont
  {Chamon},\ and\ \citenamefont {Affleck}}]{PhysRevLett.105.226803}%
  \BibitemOpen
  \bibfield  {author} {\bibinfo {author} {\bibfnamefont {A.}~\bibnamefont
  {Rahmani}}, \bibinfo {author} {\bibfnamefont {C.-Y.}\ \bibnamefont {Hou}},
  \bibinfo {author} {\bibfnamefont {A.}~\bibnamefont {Feiguin}}, \bibinfo
  {author} {\bibfnamefont {C.}~\bibnamefont {Chamon}},\ and\ \bibinfo {author}
  {\bibfnamefont {I.}~\bibnamefont {Affleck}},\ }\bibfield  {title} {\bibinfo
  {title} {How to find conductance tensors of quantum multiwire junctions
  through static calculations: Application to an interacting $y$ junction},\
  }\href {https://doi.org/10.1103/PhysRevLett.105.226803} {\bibfield  {journal}
  {\bibinfo  {journal} {Phys. Rev. Lett.}\ }\textbf {\bibinfo {volume} {105}},\
  \bibinfo {pages} {226803} (\bibinfo {year} {2010})}\BibitemShut {NoStop}%
\bibitem [{\citenamefont {Rahmani}\ \emph {et~al.}(2012)\citenamefont
  {Rahmani}, \citenamefont {Hou}, \citenamefont {Feiguin}, \citenamefont
  {Oshikawa}, \citenamefont {Chamon},\ and\ \citenamefont
  {Affleck}}]{PhysRevB.85.045120}%
  \BibitemOpen
  \bibfield  {author} {\bibinfo {author} {\bibfnamefont {A.}~\bibnamefont
  {Rahmani}}, \bibinfo {author} {\bibfnamefont {C.-Y.}\ \bibnamefont {Hou}},
  \bibinfo {author} {\bibfnamefont {A.}~\bibnamefont {Feiguin}}, \bibinfo
  {author} {\bibfnamefont {M.}~\bibnamefont {Oshikawa}}, \bibinfo {author}
  {\bibfnamefont {C.}~\bibnamefont {Chamon}},\ and\ \bibinfo {author}
  {\bibfnamefont {I.}~\bibnamefont {Affleck}},\ }\bibfield  {title} {\bibinfo
  {title} {General method for calculating the universal conductance of strongly
  correlated junctions of multiple quantum wires},\ }\href
  {https://doi.org/10.1103/PhysRevB.85.045120} {\bibfield  {journal} {\bibinfo
  {journal} {Phys. Rev. B}\ }\textbf {\bibinfo {volume} {85}},\ \bibinfo
  {pages} {045120} (\bibinfo {year} {2012})}\BibitemShut {NoStop}%
\bibitem [{Note1()}]{Note1}%
  \BibitemOpen
  \bibinfo {note} {See the supplementary material of Ref.~\cite
  {PhysRevB.107.L201401} for similar calculations in full detail.}\BibitemShut
  {Stop}%
\bibitem [{\citenamefont {Galpin}\ \emph {et~al.}(2014)\citenamefont {Galpin},
  \citenamefont {Mitchell}, \citenamefont {Temaismithi}, \citenamefont {Logan},
  \citenamefont {B\'eri},\ and\ \citenamefont {Cooper}}]{PhysRevB.89.045143}%
  \BibitemOpen
  \bibfield  {author} {\bibinfo {author} {\bibfnamefont {M.~R.}\ \bibnamefont
  {Galpin}}, \bibinfo {author} {\bibfnamefont {A.~K.}\ \bibnamefont
  {Mitchell}}, \bibinfo {author} {\bibfnamefont {J.}~\bibnamefont
  {Temaismithi}}, \bibinfo {author} {\bibfnamefont {D.~E.}\ \bibnamefont
  {Logan}}, \bibinfo {author} {\bibfnamefont {B.}~\bibnamefont {B\'eri}},\ and\
  \bibinfo {author} {\bibfnamefont {N.~R.}\ \bibnamefont {Cooper}},\ }\bibfield
   {title} {\bibinfo {title} {Conductance fingerprint of majorana fermions in
  the topological kondo effect},\ }\href
  {https://doi.org/10.1103/PhysRevB.89.045143} {\bibfield  {journal} {\bibinfo
  {journal} {Phys. Rev. B}\ }\textbf {\bibinfo {volume} {89}},\ \bibinfo
  {pages} {045143} (\bibinfo {year} {2014})}\BibitemShut {NoStop}%
\bibitem [{\citenamefont {Herviou}\ \emph {et~al.}(2016)\citenamefont
  {Herviou}, \citenamefont {Le~Hur},\ and\ \citenamefont
  {Mora}}]{PhysRevB.94.235102}%
  \BibitemOpen
  \bibfield  {author} {\bibinfo {author} {\bibfnamefont {L.}~\bibnamefont
  {Herviou}}, \bibinfo {author} {\bibfnamefont {K.}~\bibnamefont {Le~Hur}},\
  and\ \bibinfo {author} {\bibfnamefont {C.}~\bibnamefont {Mora}},\ }\bibfield
  {title} {\bibinfo {title} {Many-terminal majorana island: From topological to
  multichannel kondo model},\ }\href
  {https://doi.org/10.1103/PhysRevB.94.235102} {\bibfield  {journal} {\bibinfo
  {journal} {Phys. Rev. B}\ }\textbf {\bibinfo {volume} {94}},\ \bibinfo
  {pages} {235102} (\bibinfo {year} {2016})}\BibitemShut {NoStop}%
\bibitem [{\citenamefont {Kimura}\ and\ \citenamefont
  {Ozaki}(2017)}]{doi:10.7566/JPSJ.86.084703}%
  \BibitemOpen
  \bibfield  {author} {\bibinfo {author} {\bibfnamefont {T.}~\bibnamefont
  {Kimura}}\ and\ \bibinfo {author} {\bibfnamefont {S.}~\bibnamefont {Ozaki}},\
  }\bibfield  {title} {\bibinfo {title} {Fermi/non-fermi mixing in su(n) kondo
  effect},\ }\href {https://doi.org/10.7566/JPSJ.86.084703} {\bibfield
  {journal} {\bibinfo  {journal} {Journal of the Physical Society of Japan}\
  }\textbf {\bibinfo {volume} {86}},\ \bibinfo {pages} {084703} (\bibinfo
  {year} {2017})}\BibitemShut {NoStop}%
\bibitem [{\citenamefont {Oshikawa}\ \emph {et~al.}(2006)\citenamefont
  {Oshikawa}, \citenamefont {Chamon},\ and\ \citenamefont
  {Affleck}}]{Oshikawa_2006}%
  \BibitemOpen
  \bibfield  {author} {\bibinfo {author} {\bibfnamefont {M.}~\bibnamefont
  {Oshikawa}}, \bibinfo {author} {\bibfnamefont {C.}~\bibnamefont {Chamon}},\
  and\ \bibinfo {author} {\bibfnamefont {I.}~\bibnamefont {Affleck}},\
  }\bibfield  {title} {\bibinfo {title} {Junctions of three quantum wires},\
  }\href {https://doi.org/10.1088/1742-5468/2006/02/p02008} {\bibfield
  {journal} {\bibinfo  {journal} {Journal of Statistical Mechanics: Theory and
  Experiment}\ }\textbf {\bibinfo {volume} {2006}},\ \bibinfo {pages} {P02008}
  (\bibinfo {year} {2006})}\BibitemShut {NoStop}%
\end{thebibliography}%

\begin{appendix}


\begin{widetext}

\section{$\text{SO}(2m)$ Lie algebra and spinor representation}\label{ap:so2m}
In this section, we introduce the basics of the $\text{SO}(2m)$ Lie algebra and show that the ground states of the $\text{SO}(2m)$ topological Kondo interaction (no kinetic energy) are at half-filling (the $m$-particle sector).
\subsection{$\text{SO}(2m)$ Lie algebra}\label{ap:so2mliealgebra}
The group $\text{SO}(2m)$ is generated by $m(2m-1)$ generators. In our case, they can be Majorana bilinears $-\mi \gamma_{\alpha} \gamma_{\beta}/2$ with $\alpha<\beta$. The rank of $\mathfrak{so}(2m)$ Lie algebra is $m$, which means that we have $m$ Cartan generators. The corresponding Cartan generators and the $2m(m-1)$ roots are \cite{georgi2000lie,zee2016group}
\begin{equation}
\begin{aligned}
H_i=-\mi \gamma_{2i-1} \gamma_{2i}/2,\ \ E_{ij}^{s s'}=-\mi(\gamma_{2i-1}+\mi s  \gamma_{2i})(\gamma_{2j-1}+\mi s'  \gamma_{2j})/4 
\end{aligned}
\end{equation} 
where $i<j=1,2,\dots,m$ and $s,s'=\pm 1$. We can check the commutation relations are \begin{equation}
[H_j,E_{j k}^{s s'}]=s E_{j k}^{s s'}, \ \ [H_k,E_{j k}^{s s'}]=s' E_{j k}^{s s'}.
\end{equation} 
Other $H_i$s with $i\neq j,k$ will commute with $E_{j k}^{s s'}$. From above, we obtain the simple roots, which are 
\begin{align}
&\alpha_1=(1,-1,0,\dots,0),\\ 
&\alpha_2=(0,1,-1,0,\dots,0),  \ \dots, \\
&\alpha_{m-1}=(0,\dots,0,1,-1),\\  
&\alpha_m=(0,\dots,0,1,1).
\end{align}
Cartan matrix is defined as $C_{ij}=2(\alpha_i\cdot\alpha_j)/(\alpha_i\cdot\alpha_i)$. The fundamental weights are 
\begin{align}
&\mu_1=(1,0,\dots,0), \\ 
&\mu_2=(1,1,0,\dots,0), \ \dots, \\
&\mu_{m-2}=(1,\dots,1,0,0),\\ 
&\mu_{m-1}=(1,\dots,1,-1)/2,\label{apeq:mumm1}\\
&\mu_m=(1,\dots,1)/2. \label{apeq:mum}
\end{align}
Any representation can be written as $\mu=\sum_{i=1}^m a_i \mu_i$ where $a_i$s are Dynkin's labels. The corresponding Casimir invariant for this representation $\mu$ is given by \cite{ma2007group}
\begin{equation}
c(R)=\sum_{i,j=1}^{m}(\alpha_i\cdot \alpha_i)(a_i/2+1)(C^{-1})_{ij}a_j. \label{apeq:casimirinv}  
\end{equation}

\subsection{Representation of $S^A$, $J^A$ and decomposition of $R(S^A)\otimes R(J^A)$}\label{ap:repSJ}
The spinor representations $S^A $ using Majorana operators can be further written in terms of Pauli matrices $\sigma_{i}, i=1,2,3$:
\begin{align}
&\gamma_1=\sigma_1\otimes \sigma_3 \otimes \dots \otimes \sigma_3 ,\ \gamma_2=\sigma_2\otimes \sigma_3 \otimes \dots \otimes \sigma_3 , \nonumber \\
&\gamma_3=\mathbb{I}\otimes \sigma_1 \otimes \sigma_3 \otimes \dots \otimes \sigma_3 , \ \gamma_4=\mathbb{I} \otimes \sigma_2 \otimes \sigma_3 \otimes \dots \otimes \sigma_3 , \ \dots, \nonumber \\ 
&\gamma_{2m-1}= \mathbb{I}  \otimes \dots \otimes \mathbb{I}  \otimes \sigma_1, \ \gamma_{2m}= \mathbb{I}  \otimes \dots \otimes \mathbb{I}  \otimes \sigma_2. \label{apeq:gammapauli}
\end{align}
The natural basis for the above $m$ spin-1/2 operators is the spin-$\uparrow,\downarrow$ states. Then, the Cartan generators measure the $z$-component of each spin because $H_i\sim\sigma_{3}^{(i)}$. The root $E_{jk}^{ss'}$ is nearly $\sigma_{s}^{(j)}\sigma_{s'}^{(k)}$ where $\sigma_{s}=\sigma_1 + \mi s \sigma_2 $. Representations are defined through their highest weights. We can easily find that the highest weights of this spinor representation are $\ket{\uparrow \dots \uparrow \uparrow }$ and $\ket{\uparrow \dots \uparrow \downarrow }$. These two highest-weight states are eigenstates of Cartan generators, and the eigenvalues are $\mu_{m-1}$ [Eq.~(\ref{apeq:mumm1})] and $\mu_{m}$ [Eq.~(\ref{apeq:mum})], which gives Eq.~(\ref{eq:RSA}): $R(S^A)=(\mu_m)\oplus(\mu_{m-1})$ in the main text. The difference between the $(\mu_{m-1})$ and $(\mu_{m})$ subspaces is their parities, which can be measured by $(-1)^m\prod_{i=1}^{2m}\gamma_i=\sigma_3 \otimes \dots \otimes \sigma_3$. They are both $2^{m-1}$-dimensional. 

When it comes to the representations of $J^A$, we need to consider the particle sectors separately. 
The dimension of the $K$-particle sector is equal to the dimension of the linear space in which the states live. As we defined in Sec. \ref{sec:existence}, the number of states that live in the $K$-particle sector is $\binom{2 m}{K}$. The total dimension for $J^A$ is thus the sum of the dimensions of all $K$-particle sectors $\sum_{i=0}^{2m}\binom{2 m}{K}=2^{2m}$, which verifies our conclusion that the representation of $J^A$ is the same as the representation given by $R(S^A)\otimes R(S^A)$ [Eq.~(\ref{eq:JAtwoSA})] in terms of its dimensions. We can decompose $R(J^A)=R(S^A)\otimes R(S^A)$ as $\bigoplus_{i=0}^{2m} R(J_n^{A,i})$ with
\begin{align}
& R(J^{A,i})=R(J^{A,2m-i})=(\mu_i)\  \text{for} \ 0 \le i \le m-2;\\
& R(J^{A,m-1})=(\mu_{m-1}+\mu_{m}); \ \ R(J^{A,m})=(2\mu_{m-1})\oplus (2\mu_{m}).
\end{align}
 We used the following three decomposition rules:
\begin{align}
&(\mu_{m-1})\otimes (\mu_{m-1})=(2\mu_{m-1})\bigoplus_{i=1}^{[m/2]} (\mu_{m-2i});\ \ (\mu_{m})\otimes (\mu_{m})=(2\mu_{m})\bigoplus_{i=1}^{[m/2]} (\mu_{m-2i});\\
&(\mu_{m-1})\otimes (\mu_{m})=(\mu_{m-1}+\mu_m)\bigoplus_{j=1}^{[(m-1)/2]} (\mu_{m-2j-1}).
\end{align}
Note that $\mu_0=\bm{0}$ and the above equations are symmetric for $\mu_{m-1}$ and $\mu_m$. We can verify using LieART \cite{LieARTFeger_2020} that $R(S^A)\otimes R(J^{A,K})$ contains $(\mu_m)\oplus(\mu_{m-1})$ for every $0 \le K \le 2m$. The representation $\mu_{m-1}$ and $\mu_{m}$ have the smallest Casimir invariant according to Eq.~(\ref{apeq:casimirinv}). Thus, the lowest Kondo energy [Eq.~(\ref{eq:casimirops})] is at $K=m$ (half-filling), which maximizes $c[R(J_n^{A,K})]=K(2m-K)$ [see also Eq.~(\ref{eq:casimirJA})].

\section{Effective Hamiltonian of SO(4) topological Kondo at strong coupling} \label{ap:so4Heff}
In this section, we calculate the ground states of the $\text{SO}(4)$ topological Kondo interaction and find the strong coupling effective Hamiltonian of it, i.e., Eq.~(\ref{apeq:effectTKso4}).

\subsection{Ground states of the $\text{SO}(4)$ topological Kondo interaction}\label{ap:so4gs}
The simple roots and fundamental weights of $\text{SO}(4)$ are \cite{georgi2000lie,zee2016group}
\begin{equation}
\begin{aligned}
\alpha_1=(1,-1), \ \alpha_2=(1,1); \ \ \mu_1=(1/2,-1/2), \  \mu_2=(1/2,1/2).
\end{aligned}
\end{equation}
The Casimir invariant is thus
\begin{equation}\label{eq:casimirso4}
c(a_1\mu_1+a_2\mu_2)=\frac{1}{2}a_1\qty(a_2+2) + \frac{1}{2}a_1\qty(a_2+2).
\end{equation}
The representation $R(S^A)=(\mu_1) \oplus (\mu_2)$ gives that $c(R(S^A))=3/2$ which agrees with Eq.~(\ref{eq:casimirSA}). The dimension of the $K$-particle sector $R(J^{A,K})$ is $\binom{4}{K}$. The corresponding representations $R(J^{A})=R(S^A)\otimes R(S^A)=[(\mu_1) \oplus (\mu_2)]\otimes[(\mu_1) \oplus (\mu_2)]$ can be written as $\bigoplus_{K=0}^{4} R(J^{A,K})$ where
\begin{equation}
\begin{aligned}
&R(J_n^{A,0})=R(J_n^{A,4})=(\bm{0}), \ R(J_n^{A,1})=R(J_n^{A,3})=(\mu_1+\mu_2) \ \text{and}\  R(J_n^{A,2})=(2\mu_1)\oplus (2\mu_2).\label{apeq:JArepso4}
\end{aligned}
\end{equation}
Here, we used a fact that $(\mu_1)\otimes(\mu_1)=(\bm{0})\oplus(2\mu_1)$, $(\mu_2)\otimes(\mu_2)=(\bm{0})\oplus(2\mu_2)$ and $(\mu_1)\otimes(\mu_2)=(\mu_1+\mu_2)$. We can check that $c(R(J_n^{A,K}))=K(4-K)$ which agrees with Eq.~(\ref{eq:casimirJA}).  The decompositions of $R(S^A)\otimes R(J^{A})=\bigoplus_{K=0}^{4} R(S^A) \otimes R(J^{A,K})$ where
\begin{align}
&R(S^A)\otimes R(J^{A,0,4})=[(\mu_1) \oplus (\mu_2)]\otimes (\bm{0}) =(\mu_1) \oplus (\mu_2), \\ 
&R(S^A) \otimes R(J_n^{A,1,3})=[(\mu_1) \oplus (\mu_2)]\otimes (\mu_1+\mu_2)=(\mu2)\oplus(2\mu_1+\mu_2)\oplus (\mu_1+2\mu_2)\oplus (\mu_1), \\ 
&R(S^A) \otimes R(J_n^{A,2})=[(\mu_1) \oplus (\mu_2)]\otimes[(2\mu_1)\otimes (2\mu_2)]=(\mu_1)\oplus(3\mu_1)\oplus(\mu_1+2\mu_2)\oplus(2\mu_1+\mu_2)\oplus(3\mu_2)\oplus(\mu_2).
\end{align}
The above expansions verify our conclusion that $(\mu_1)\oplus(\mu_2)$ always exists in $R(S^A) \otimes R(J^{A,K})$ for every $K$. The Kondo energy Eq.~(\ref{eq:casimirops}) for $\text{SO}(4)$ is
\begin{align}
E_{\rm Kondo}=&\frac{1}{2}\qty{c[R(S^A) \otimes R(J^{A,K})]-c[R(S^A)]-c[R(J^{A,K})]}\\
=&\frac{1}{2}\qty{c[R(S^A) \otimes R(J^{A,K})]-\frac{M(M-1)}{8}-K(4-K)}.
\end{align}
The nontrivial minima of $c(a_1\mu_1 +a_2\mu_2)$ when $a_{1,2}$ are both nonnegative integers are given by $(a_1,a_2)=(1,0)$ or $(0,1)$, i.e.~$\mu_1$ or $\mu_2$, which is contained in any $R(S^A) \otimes R (J_n^{A,K})$. Thus, we conclude that the lowest $\text{SO}(4)$ Kondo energy is at $K=2$ where it maximizes the $c(R( J^{A,K}))=K(4-K)$. 

It is knowm that $\text{SO}(4)\sim \text{SU}(2)\times \text{SU}(2)$ which can be seen from Eq.~(\ref{eq:casimirso4}) where the Casimir invariant of $\text{SO}(4)$ can be reformulated as $c(a_1\mu_1 +a_2\mu_2)=2[s_1(s_1+1)]+2[s_2(s_2+1)]$ for spin-$s_{1,2}=a_{1,2}/2$. Based on the fact that $(\mu_1)\otimes(\mu_1)=(\bm{0})\oplus(2\mu_1)$, $(\mu_2)\otimes(\mu_2)=(\bm{0})\oplus(2\mu_2)$ and $(\mu_1)\otimes(\mu_2)=(\mu_1+\mu_2)$, one can consider the above decomposition as the decomposition for two independent spins. Furthermore, the conduction electrons are two independent spin-1 particles for the lowest energy case $J^{A,2}$ which is $(2\mu_1)\oplus (2\mu_2)$. The lowest energy in this $2$-particle sector is $(\mu_1)$ and $(\mu_2)$ given by $(\mu_1)\otimes(2\mu_1)=(\mu_1)\oplus(3\mu_1)$ and $(\mu_2)\otimes(2\mu_2)=(\mu_2)\oplus(3\mu_2)$. In the language of $\text{SU}(2)$ spin decompositions, they are $1/2 \otimes 1 = 1/2 \oplus 3/2 $, which is the same decomposition used for getting the ground state of the 2-channel $\text{SU}(2)$ Kondo model at the strong coupling and also half-filling.

\subsection{Representations and strong coupling solutions}\label{ap:repsSCS}
The Casimir invariant $c(a_1,a_2)$ for a representation of generators $R^{A}$ labeled by $a_1 \mu_{1}+a_2\mu_{2}$ is given by Eq.~(\ref{eq:casimirso4}), which gives $\sum_{A}(R^{A})^{2}=c\,\mathbb{I}_{\text{dim}\times\text{dim}}$.
The $K=0,4$-particle sector contain only one state $|0\rangle$ or
$|\text{full}\rangle=\psi_{1}^{\dagger}\psi_{2}^{\dagger}\psi_{3}^{\dagger}\psi_{4}^{\dagger}|0\rangle$
for electrons. This leads to $J^{A,0}=\langle0|J^{A}|0\rangle=0$
and $H_{K}=0$. For the $1,3$-particle sector, we get $J^{A,K=1,3}=T^{A}$ and the
representation is $\mu_{1}+\mu_{2}$, a $4$ dimensional representation with $\sum_{A}(T^{A})^{2}=3\,\mathbb{I}_{4\times4}$.
Next, we define the $2$-particle states as $|a\rangle=\sum_{ij}(B^{a})_{ij}\psi_{i}^{\dagger}\psi_{j}^{\dagger}|0\rangle$
with $a=1,\dots,6$, $(B^{a})^{T}=-B^{a}$ and $(B^{a})^{\dagger}=B^{a}$.
Normalization of state $|a\rangle$ requires $2\Tr(B^{a}B^{b})=\delta^{ab}$. Thus, the 2-particle matrix representation for the generator $J^A$ is
\begin{equation}
[J^{A,K=2}]_{ba}=\langle b|J^{A}|a\rangle=\sum_{ii'jj'\rho\sigma}(B^{b})_{i'j'}^{*}(T^{A})_{\rho\sigma}(B^{a})_{ij}\langle0|\psi_{j'}\psi_{i'}\psi_{\rho}^{\dagger}\psi_{\sigma}\psi_{i}^{\dagger}\psi_{j}^{\dagger}|0\rangle=4\Tr\big(B^{b}T^{A}B^{a}\big).
\end{equation}
Let us consider the following $B$ tensor
\begin{align}
[B^{b=(b_{1},b_{2})}]_{ij} & =-\frac{\mathrm{i}}{2!}\delta_{i}^{b_{1}}\delta_{j}^{b_{2}}+\frac{\mathrm{i}}{2!}\delta_{j}^{b_{1}}\delta_{i}^{b_{2}}\ \text{with}\ b_{1}<b_{2}.
\end{align}
Here, $b_{1},b_{2},i,j=1,2,3,4$. This six-dimensional space spanned by $|a\rangle$ is reducible and the representation is $2\mu_{1}\oplus2\mu_{2}$ because we checked
that $\sum_{A}[J^{A,K=2}]^{2}=4\,\mathbb{I}_{6\times6}$. Each of
these two representations is three-dimensional. 

After figuring out the representations of $K$-particle $J^{A}$,
we show the decompositions for each $J^{A,K}$ with $S^{A}$ ($\mu_{1}\oplus\mu_{2}$) in the following:
$H_{K}$:
\begin{align}
 & R(S^{A}) \otimes R(J^{A,K=1,3})=(\mu_{1}\oplus\mu_{2})\otimes(\mu_{1}+\mu_{2})=\mu_{2}\oplus(2\mu_{1}+\mu_{2})\oplus(\mu_{1}+2\mu_{2})\oplus\mu_{1};\\
 & R(S^{A})\otimes R(J^{A,K=2})=(\mu_{1}\oplus\mu_{2})\otimes(2\mu_{1}\oplus2\mu_{2})=\mu_{1}\oplus3\mu_{1}\oplus(\mu_{1}+2\mu_{2})\oplus(2\mu_{1}+\mu_{2})\oplus3\mu_{2}\oplus\mu_{2},
\end{align}
where we used the following decomposition rules
\begin{equation}
\mu_{1}\otimes\mu_{1}=\bm{0}\oplus2\mu_{1};\ \mu_{2}\otimes\mu_{2}=\bm{0}\oplus2\mu_{2};\ \mu_{1}\otimes\mu_{2}=\mu_{1}+\mu_{2}.
\end{equation}
We include all the results in the following table.
\begin{table}[h]
\begin{centering}
\begin{tabular}{|c|c|c|c|c|c|}
\hline 
$K$ & $(\mu_{1}\oplus\mu_{2})\otimes(\mu_{1}+\mu_{2})\ (4\times4)$ & $\sum_{A}(S^{A}+J^{A,K=1,3})^{2}$ & $-\sum_{A}(S^{A})^{2}$ & $-\sum_{A}(J^{A,K=1,3})^{2}$ & $2H_{K}/\lambda$\tabularnewline
\hline 
\hline 
$1,3$ & $2\mu_{1}+\mu_{2}\ (6)$ & $11/2$ & $-3/2$ & $-3$ & $1$\tabularnewline
\hline 
$1,3$ & $\mu_{1}+2\mu_{2}\ (6)$ & $11/2$ & $-3/2$ & $-3$ & $1$\tabularnewline
\hline 
$1,3$ & $\mu_{2}\ (2)$ & $3/2$ & $-3/2$ & $-3$ & $-3$\tabularnewline
\hline 
$1,3$ & $\mu_{1}\ (2)$ & $3/2$ & $-3/2$ & $-3$ & $-3$\tabularnewline
\hline 
\end{tabular}
\par\end{centering}
\caption{1,3-particle decomposition with dimension and energy. The numbers
near the representations are dimensions.}
\end{table}
\begin{table}[h]
\begin{centering}
\begin{tabular}{|c|c|c|c|c|c|}
\hline 
$K$ & $(\mu_{1}\oplus\mu_{2})\otimes(2\mu_{1}\oplus2\mu_{2})\ (4\times6)$ & $\sum_{A}(S^{A}+J^{A,K=2})^{2}$ & $-\sum_{A}(S^{A})^{2}$ & $-\sum_{A}(J^{A,K=2})^{2}$ & $2H_{K}/\lambda$\tabularnewline
\hline 
\hline 
$2$ & $3\mu_{1}\ (4)$ & $15/2$ & $-3/2$ & $-4$ & $2$\tabularnewline
\hline 
$2$ & $3\mu_{2}\ (4)$ & $15/2$ & $-3/2$ & $-4$ & $2$\tabularnewline
\hline 
$2$ & $\mu_{1}+2\mu_{2}\ (6)$ & $11/2$ & $-3/2$ & $-4$ & $0$\tabularnewline
\hline 
$2$ & $2\mu_{1}+\mu_{2}\ (6)$ & $11/2$ & $-3/2$ & $-4$ & $0$\tabularnewline
\hline 
$2$ & $\mu_{1}\ (2)$ & $3/2$ & $-3/2$ & $-4$ & $-4$\tabularnewline
\hline 
$2$ & $\mu_{2}\ (2)$ & $3/2$ & $-3/2$ & $-4$ & $-4$\tabularnewline
\hline 
\end{tabular}
\par\end{centering}
\caption{2-particle decomposition with dimension and energy. The numbers near
the representations are dimensions.}
\end{table}

We conclude that the ground states are represented by $\mu_{1}$ and
$\mu_{2}$ at $2$-particle sector (half-filling). Also, the above
energies have two copies because $\mu_{1}$ and $\mu_{2}$ are interchangeable. 

\subsection{Eigenstates}

Next, we define the eigenstates of $H_{K}$ in $1,2,3$-particle sector. At $1$-particle sector, the general form for the states are
\begin{equation}
|K=1,\alpha\rangle=\sum_{ij}(t^{\alpha})_{ij}|i\rangle\otimes\psi_{j}^{\dagger}|0\rangle,\ i,j=1,\dots,4;\ \alpha=1,\dots16.
\end{equation}
Based on the above decomposition, we define that
\begin{align}
 & (2H_{K}/\lambda)|K=1,\alpha=1,\dots,12\rangle=|K=1,\alpha=1,\dots,12\rangle;\\
 & (2H_{K}/\lambda)|K=1,\alpha=13,\dots,16\rangle=-3|K=1,\alpha=13,\dots,16\rangle.
\end{align}
Similarly, at $2$-particle, we define 
\begin{equation}
|K=2,\beta\rangle=\sum_{ia}(r^{\beta})_{ia}|i\rangle\otimes|a\rangle=\sum_{ia}(r^{\beta})_{ia}|i\rangle\otimes\sum_{kl}(B^{a})_{kl}\psi_{k}^{\dagger}\psi_{l}^{\dagger}|0\rangle,\ i,k,l=1,\dots,4;\,a=1,\dots,6;\,\beta=1,\dots,24,
\end{equation}
and 
\begin{align}
 & (2H_{K}/\lambda)|K=2,\beta=1,\dots,8\rangle=2|K=2,\beta=1,\dots,8\rangle;\\
 & (2H_{K}/\lambda)|K=2,\beta=9,\dots,20\rangle=0;\\
 & (2H_{K}/\lambda)|K=2,\beta=21,\dots,24\rangle=-4|K=2,\beta=21,\dots,24\rangle.
\end{align}
At $3$-particle, we define
\begin{equation}
|K=3,\alpha'\rangle=\sum_{ij}(t'^{\alpha'})_{ij}|i\rangle\otimes\psi_{j}|\text{full}\rangle,\ i,j=1,\dots,4;\ \alpha'=1,\dots16.
\end{equation}
Under the electrons states $\psi_{j}|\text{full}\rangle$, $J^{(3),A}=J^{(1),A}=T^{A}$,
which means that $t'=t$. Thus,
\begin{align}
 & (2H_{K}/\lambda)|K=3,\alpha=1,\dots,12\rangle=|K=2,\alpha=1,\dots,12\rangle;\\
 & (2H_{K}/\lambda)|K=3,\alpha=13,\dots,16\rangle=-3|K=2,\alpha=13,\dots,16\rangle.
\end{align}

\subsection{Perturbative inclusion of the leads}

The lead fermion is $\psi_{s,i}$ where $s$ labels the site and $i=1,\dots4$
labels the flavor. The nearest site to the impurity is at $s=1$ and
we ignored this $s=1$ on the above nearest lead fermion. But now
we include the hopping energy between the first site and the second
site $H_{12}=-g\sum_{p}(\psi_{1,p}^{\dagger}\psi_{2,p}+\psi_{2,p}^{\dagger}\psi_{1,p})$:
\begin{align}
(H_{\text{eff}})_{ij}= & \sum_{\alpha=1}^{12}\frac{\langle K=2,\beta=20+i|H_{12}|K=1,\alpha\rangle\langle K=1,\alpha|H_{12}|K=2,\beta=20+j\rangle}{(-2\lambda-\lambda/2)}\\
 & +\sum_{\alpha=13}^{16}\frac{\langle K=2,\beta=20+i|H_{12}|K=1,\alpha\rangle\langle K=1,\alpha|H_{12}|K=2,\beta=20+j\rangle}{(-2\lambda+3\lambda/2)}\\
 & +\sum_{\alpha=1}^{12}\frac{\langle K=2,\beta=20+i|H_{12}|K=3,\alpha\rangle\langle K=3,\alpha|H_{12}|K=2,\beta=20+j\rangle}{(-2\lambda-\lambda/2)}\\
 & +\sum_{\alpha=13}^{16}\frac{\langle K=2,\beta=20+i|H_{12}|K=3,\alpha\rangle\langle K=3,\alpha|H_{12}|K=2,\beta=20+j\rangle}{(-2\lambda+3\lambda/2)}.
\end{align}
We need to calculate the following terms:
\begin{align}
 & \langle K=1,\alpha|H_{12}|K=2,\beta\rangle=-g\sum_{p}\sum_{i'j'}(t^{\alpha})_{i'j'}^{*}\sum_{ia}(r^{\beta})_{ia}\sum_{kl}(B^{a})_{kl}\langle i'|i\rangle\langle0|\psi_{j'}\psi_{p}\psi_{k}^{\dagger}\psi_{l}^{\dagger}|0\rangle\psi_{2,p}^{\dagger}\\
= & 2g\sum_{p}\sum_{ila}(t^{\alpha})_{il}^{*}(r^{\beta})_{ia}(B^{a})_{lp}\psi_{2,p}^{\dagger};\\
 & \langle K=2,\beta|H_{12}|K=1,\alpha\rangle=2g\sum_{p}\sum_{ila}(t^{\alpha})_{il}(r^{\beta})_{ia}^{*}(B^{a})_{lp}^{*}\psi_{2,p};\\
 & \langle K=3,\alpha|H_{12}|K=2,\beta\rangle=-g\sum_{p}\sum_{i'j'}(t^{\alpha})_{i'j'}^{*}\sum_{ia}(r^{\beta})_{ia}\sum_{kl}(B^{a})_{kl}\langle i'|i\rangle\langle\text{full}|\psi_{j'}^{\dagger}\psi_{p}^{\dagger}\psi_{k}^{\dagger}\psi_{l}^{\dagger}|0\rangle\psi_{2,p}\\
= & -g\sum_{p}\sum_{ija}(t^{\alpha})_{ij}^{*}(r^{\beta})_{ia}\sum_{kl}\epsilon_{jpkl}(B^{a})_{kl}\psi_{2,p}=2g\sum_{p}\sum_{ija}(t^{\alpha})_{ij}^{*}(r^{\beta})_{ia}(C^{a})_{jp}\psi_{2,p};\\
 & \langle K=2,\beta|H_{12}|K=3,\alpha\rangle=2g\sum_{p}\sum_{ija}(t^{\alpha})_{ij}(r^{\beta})_{ia}^{*}(C^{a})_{jp}^{*}\psi_{2,p}^{\dagger}
\end{align}
where we define $(C^{a})_{jp}=-\frac{1}{2}\sum_{kl}\epsilon_{jpkl}(B^{a})_{kl}$.
Thus,
\begin{align}
(H_{\text{eff}})_{11}= & -\frac{16g^{2}}{5\lambda}-\frac{22g^{2}}{15\lambda}\sum_{pq}(T^{12}+T^{34})_{pq}\psi_{2,p}^{\dagger}\psi_{2,q};\ (H_{\text{eff}})_{14}=\frac{22g^{2}}{15\lambda}\sum_{pq}(T^{13}-T^{24}+\mathrm{i}T^{14}+\mathrm{i}T^{23})_{pq}\psi_{2,p}^{\dagger}\psi_{2,q};\\
(H_{\text{eff}})_{41}= & \frac{22g^{2}}{15\lambda}\sum_{pq}(T^{13}-T^{24}-\mathrm{i}T^{14}-\mathrm{i}T^{23})_{pq}\psi_{2,p}^{\dagger}\psi_{2,q};\ (H_{\text{eff}})_{22}=-\frac{16g^{2}}{5\lambda}-\frac{22g^{2}}{15\lambda}\sum_{pq}(T^{12}-T^{34})_{pq}\psi_{2,p}^{\dagger}\psi_{2,q};\\
(H_{\text{eff}})_{33}= & -\frac{16g^{2}}{5\lambda}+\frac{22g^{2}}{15\lambda}\sum_{pq}(T^{12}-T^{34})_{pq}\psi_{2,p}^{\dagger}\psi_{2,q};\ (H_{\text{eff}})_{23}=\frac{22g^{2}}{15\lambda}\sum_{pq}(T^{13}+T^{24}-\mathrm{i}T^{14}+\mathrm{i}T^{23})_{pq}\psi_{2,p}^{\dagger}\psi_{2,q};\\
(H_{\text{eff}})_{32}= & \frac{22g^{2}}{15\lambda}\sum_{pq}(T^{13}+T^{24}+\mathrm{i}T^{14}-\mathrm{i}T^{23})_{pq}\psi_{2,p}^{\dagger}\psi_{2,q};\ (H_{\text{eff}})_{44}=-\frac{16g^{2}}{5\lambda}+\frac{22g^{2}}{15\lambda}\sum_{pq}(T^{12}+T^{34})_{pq}\psi_{2,p}^{\dagger}\psi_{2,q}.
\end{align}
Before calculating the whole $H_{\text{eff}}$, we write down all
matrix $S^{A=(\alpha,\beta)}=-\frac{\mathrm{i}}{2}\gamma_{\alpha}\gamma_{\beta}$
according to the definitions Eq.~(\ref{apeq:gammapauli}):
\begin{align}
 & S^{12}=\left(\begin{array}{cccc}
\frac{1}{2}\\
 & \frac{1}{2}\\
 &  & -\frac{1}{2}\\
 &  &  & -\frac{1}{2}
\end{array}\right);\ S^{13}=\left(\begin{array}{cccc}
 &  &  & -\frac{\mathrm{i}}{2}\\
 &  & \frac{\mathrm{i}}{2}\\
 & -\frac{\mathrm{i}}{2}\\
\frac{\mathrm{i}}{2}
\end{array}\right);\ S^{14}=\left(\begin{array}{cccc}
 &  &  & -\frac{1}{2}\\
 &  & -\frac{1}{2}\\
 & -\frac{1}{2}\\
-\frac{1}{2}
\end{array}\right);\\
 & S^{23}=\left(\begin{array}{cccc}
 &  &  & -\frac{1}{2}\\
 &  & \frac{1}{2}\\
 & \frac{1}{2}\\
-\frac{1}{2}
\end{array}\right);\ S^{24}=\left(\begin{array}{cccc}
 &  &  & \frac{\mathrm{i}}{2}\\
 &  & \frac{\mathrm{i}}{2}\\
 & -\frac{\mathrm{i}}{2}\\
-\frac{\mathrm{i}}{2}
\end{array}\right);\ S^{34}=\left(\begin{array}{cccc}
\frac{1}{2}\\
 & -\frac{1}{2}\\
 &  & \frac{1}{2}\\
 &  &  & -\frac{1}{2}
\end{array}\right).
\end{align}
Thus,
\begin{align}
 & S^{12}+S^{34}=\left(\begin{array}{cccc}
1\\
 & 0\\
 &  & 0\\
 &  &  & -1
\end{array}\right);\ S^{12}-S^{34}=\left(\begin{array}{cccc}
0\\
 & 1\\
 &  & -1\\
 &  &  & 0
\end{array}\right);\ \frac{1}{2}(\mathrm{i}S^{13}-\mathrm{i}S^{24}-S^{14}-S^{23})=\left(\begin{array}{cccc}
 &  &  & 1\\
 &  & 0\\
 & 0\\
0
\end{array}\right);\\
 & \frac{1}{2}(-\mathrm{i}S^{13}-\mathrm{i}S^{24}-S^{14}+S^{23})=\left(\begin{array}{cccc}
 &  &  & 0\\
 &  & 1\\
 & 0\\
0
\end{array}\right);\ \frac{1}{2}(\mathrm{i}S^{13}+\mathrm{i}S^{24}-S^{14}+S^{23})=\left(\begin{array}{cccc}
 &  &  & 0\\
 &  & 0\\
 & 1\\
0
\end{array}\right);\\
 & \ \frac{1}{2}(-\mathrm{i}S^{13}+\mathrm{i}S^{24}-S^{14}-S^{23})=\left(\begin{array}{cccc}
 &  &  & 0\\
 &  & 0\\
 & 0\\
1
\end{array}\right).
\end{align}
Finally,
\begin{align}
H_{\text{eff}}= & \left(\begin{array}{cccc}
1\\
 & 0\\
 &  & 0\\
 &  &  & 0
\end{array}\right)(H_{\text{eff}})_{11}+\left(\begin{array}{cccc}
0\\
 & 1\\
 &  & 0\\
 &  &  & 0
\end{array}\right)(H_{\text{eff}})_{22}+\left(\begin{array}{cccc}
0\\
 & 0\\
 &  & 1\\
 &  &  & 0
\end{array}\right)(H_{\text{eff}})_{33}+\left(\begin{array}{cccc}
0\\
 & 0\\
 &  & 0\\
 &  &  & 1
\end{array}\right)(H_{\text{eff}})_{44}\\
 & +(H_{\text{eff}})_{14}\left(\begin{array}{cccc}
 &  &  & 1\\
 &  & 0\\
 & 0\\
0
\end{array}\right)+(H_{\text{eff}})_{23}\left(\begin{array}{cccc}
 &  &  & 0\\
 &  & 1\\
 & 0\\
0
\end{array}\right)+(H_{\text{eff}})_{32}\left(\begin{array}{cccc}
 &  &  & 0\\
 &  & 0\\
 & 1\\
0
\end{array}\right)+(H_{\text{eff}})_{41}\left(\begin{array}{cccc}
 &  &  & 0\\
 &  & 0\\
 & 0\\
1
\end{array}\right)\\
= & -\frac{16g^{2}}{5\lambda}\mathbb{I}_{4\times4}-\frac{22g^{2}}{15\lambda}(S^{12}+S^{34})\sum_{pq}(T^{12}+T^{34})_{pq}\psi_{2,p}^{\dagger}\psi_{2,q}-\frac{22g^{2}}{15\lambda}(S^{12}-S^{34})\sum_{pq}(T^{12}-T^{34})_{pq}\psi_{2,p}^{\dagger}\psi_{2,q}\\
 & +\frac{11g^{2}}{15\lambda}(\mathrm{i}S^{13}-\mathrm{i}S^{24}-S^{14}-S^{23})\sum_{pq}(T^{13}-T^{24}+\mathrm{i}T^{14}+\mathrm{i}T^{23})_{pq}\psi_{2,p}^{\dagger}\psi_{2,q}\\
 & +\frac{11g^{2}}{15\lambda}(-\mathrm{i}S^{13}+\mathrm{i}S^{24}-S^{14}-S^{23})\sum_{pq}(T^{13}-T^{24}-\mathrm{i}T^{14}-\mathrm{i}T^{23})_{pq}\psi_{2,p}^{\dagger}\psi_{2,q}\\
 & +\frac{11g^{2}}{15\lambda}(-\mathrm{i}S^{13}-\mathrm{i}S^{24}-S^{14}+S^{23})\sum_{pq}(T^{13}+T^{24}-\mathrm{i}T^{14}+\mathrm{i}T^{23})_{pq}\psi_{2,p}^{\dagger}\psi_{2,q}\\
 & +\frac{11g^{2}}{15\lambda}(\mathrm{i}S^{13}+\mathrm{i}S^{24}-S^{14}+S^{23})\sum_{pq}(T^{13}+T^{24}+\mathrm{i}T^{14}-\mathrm{i}T^{23})_{pq}\psi_{2,p}^{\dagger}\psi_{2,q}\\
= & -\frac{16g^{2}}{5\lambda}\mathbb{I}_{4\times4}-\frac{44g^{2}}{15\lambda}\Big[S^{12}\psi^{\dagger}T^{12}\psi+S^{34}\psi^{\dagger}T^{34}\psi+S^{14}\psi^{\dagger}T^{13}\psi+S^{13}\psi^{\dagger}T^{14}\psi-S^{24}\psi^{\dagger}T^{23}\psi-S^{23}\psi^{\dagger}T^{24}\psi\Big].
\end{align}
We can perform the following rotations
\begin{align}
& \exp\left(-\frac{\mathrm{i}\pi}{2}T^{12}\right)T^{13}\exp\left(\frac{\mathrm{i}\pi}{2}T^{12}\right)=T^{23},\ \exp\left(-\frac{\mathrm{i}\pi}{2}T^{12}\right)T^{14}\exp\left(\frac{\mathrm{i}\pi}{2}T^{12}\right)=T^{24},\\
& \exp\left(-\frac{\mathrm{i}\pi}{2}T^{12}\right)T^{23}\exp\left(\frac{\mathrm{i}\pi}{2}T^{12}\right)=-T^{13},\ \exp\left(-\frac{\mathrm{i}\pi}{2}T^{12}\right)T^{24}\exp\left(\frac{\mathrm{i}\pi}{2}T^{12}\right)=-T^{14}
\end{align}
such that 
\begin{equation}
H_{\text{eff}}=-\frac{16g^{2}}{5\lambda}\mathbb{I}_{4\times4}-\frac{44g^{2}}{15\lambda}\Big[S^{12}\psi^{\dagger}T^{12}\psi+S^{34}\psi^{\dagger}T^{34}\psi+S^{14}\psi^{\dagger}T^{23}\psi+S^{13}\psi^{\dagger}T^{24}\psi+S^{24}\psi^{\dagger}T^{13}\psi+S^{23}\psi^{\dagger}T^{14}\psi\Big]. \label{apeq:effectTKso4}
\end{equation}
One can interchange the definitions of $\gamma_{1},\gamma_{2}$ and $\gamma_{3},\gamma_{4}$ in $S^A$ such that the second term in Eq.~(\ref{apeq:effectTKso4}) becomes the $N=1,M=4$ topological Kondo model.

\section{Current algebra of the $N$-channel topological Kondo model}\label{ap:SOM2N}
In this section, we derive the Kac-Moody algebra for the $N$-channel topological $\text{SO}(M)$ Kondo model.
The current operator for $N$-channel topological $\text{SO}(M)$ Kondo model is defined as
\begin{equation}
\widetilde{J}^{A}(x)=\sum_{n=1}^{N}J_{n}^{A}=\sum_{n=1}^{N}\sum_{\alpha,\beta}^{M}\psi_{n,\alpha}^{\dagger}(x)(T^{A})_{\alpha\beta}\psi_{n,\beta}(x)
\end{equation}
where $T^{A}$s are the traceless generators of $\text{SO}(M)$ group. They
are defined as 
\begin{equation}
(T^{A})_{\alpha\beta}\equiv(T^{rs})_{\alpha\beta}=\mathrm{i}(\delta_{\alpha}^{r}\delta_{\beta}^{s}-\delta_{\beta}^{r}\delta_{\alpha}^{s}).
\end{equation}
By definition, $1\le r<s\le M$. Thus, $A=1,\dots,M(M-1)/2$. They
satisfy $\mathrm{Tr}(T^{A}T^{B})=2\delta^{AB}$ and $[T^{A},T^{B}]=\mathrm{i}\sum_{C}f^{ABC}T^{C}$
where $f^{ABC}=(-\mathrm{i}/2)\mathrm{Tr}([T^{A},T^{B}]T^{C})$ is
the structure constant. We define the normal order with respect to the (zero-temperature) Fermi sea, which means that we move creation (annihilation) operators to the right when this particle
exists (does not exist) in the Fermi sea. The normal order of the current
operator is 
\begin{equation}
:\widetilde{J}^{A}(x):=\sum_{n=1}^{N}\sum_{\alpha,\beta}^{M}(T^{A})_{\alpha\beta}:\psi_{n,\alpha}^{\dagger}(x)\psi_{n,\beta}(x):=\sum_{n=1}^{N}\sum_{\alpha,\beta}^{M}(T^{A})_{\alpha\beta}\psi_{n,\alpha}^{\dagger}(x)\psi_{n,\beta}(x)=\widetilde{J}^{A}(x)
\end{equation}
where the normal order of fermions gives an expectation value with
$\delta_{\alpha\beta}$ which was canceled by traceless $(T^{A})_{\alpha\beta}$.
The product of two current operators (with/without normal order) is
\begin{equation}
:\widetilde{J}^{A}(x)::\widetilde{J}^{B}(y):=\widetilde{J}^{A}(x)\widetilde{J}^{B}(y)=\sum_{i,j}\sum_{\alpha,\beta,\rho,\sigma}(T^{A})_{\alpha\beta}(T^{B})_{\rho\sigma}\psi_{i,\alpha}^{\dagger}(x)\psi_{i,\beta}(x)\psi_{j,\rho}^{\dagger}(y)\psi_{j,\sigma}(y).
\end{equation}
By Wick's theorem, the product for fermions is 
\begin{align}
 & \psi_{i,\alpha}^{\dagger}(x)\psi_{i,\beta}(x)\psi_{j,\rho}^{\dagger}(y)\psi_{j,\sigma}(y)\\
= & \psi_{i,\alpha}^{\dagger}(x):\psi_{i,\beta}(x)\psi_{j,\rho}^{\dagger}(y):\psi_{j,\sigma}(y)+\langle\psi_{i,\beta}(x)\psi_{j,\rho}^{\dagger}(y)\rangle\psi_{i,\alpha}^{\dagger}(x)\psi_{j,\sigma}(y)\\
= & :\psi_{i,\alpha}^{\dagger}(x)\psi_{i,\beta}(x)\psi_{j,\rho}^{\dagger}(y)\psi_{j,\sigma}(y):+\langle\psi_{i,\alpha}^{\dagger}(x)\psi_{j,\sigma}(y)\rangle:\psi_{i,\beta}(x)\psi_{j,\rho}^{\dagger}(y):\\
 & +\langle\psi_{i,\beta}(x)\psi_{j,\rho}^{\dagger}(y)\rangle:\psi_{i,\alpha}^{\dagger}(x)\psi_{j,\sigma}(y):+\langle\psi_{i,\beta}(x)\psi_{j,\rho}^{\dagger}(y)\rangle\langle\psi_{i,\alpha}^{\dagger}(x)\psi_{j,\sigma}(y)\rangle\\
= & :\psi_{i,\alpha}^{\dagger}(x)\psi_{i,\beta}(x)\psi_{j,\rho}^{\dagger}(y)\psi_{j,\sigma}(y):+G(x-y)\delta_{ij}\delta_{\alpha\sigma}:\psi_{i,\beta}(x)\psi_{j,\rho}^{\dagger}(y):\\
 & +G(x-y)\delta_{ij}\delta_{\beta\rho}:\psi_{i,\alpha}^{\dagger}(x)\psi_{j,\sigma}(y):+G^{2}(x-y)\delta_{ij}^{2}\delta_{\alpha\sigma}\delta_{\beta\rho}
\end{align}
where the contraction $\langle\psi_{i,\alpha}^{\dagger}(x)\psi_{i,\sigma}(y)\rangle=\langle\psi_{i,\alpha}(x)\psi_{j,\sigma}^{\dagger}(y)\rangle=G(x-y)\delta_{ij}\delta_{\alpha\sigma}$ is evaluated in Eq.~\eqref{eq:G(x)} below. Thus, we can find the commutator 
\begin{align}
\big[\widetilde{J}^{A}(x),\widetilde{J}^{B}(y)\big]= & \sum_{i}\sum_{\alpha,\sigma}G(x-y)(T^{B}T^{A})_{\alpha\sigma}:\psi_{i,\sigma}(x)\psi_{i,\alpha}^{\dagger}(y):-G(y-x)(T^{B}T^{A})_{\alpha\sigma}:\psi_{i,\alpha}^{\dagger}(y)\psi_{i,\sigma}(x):\label{eq:commuJ1}\\
 & +G(x-y)(T^{A}T^{B})_{\alpha\sigma}:\psi_{i,\alpha}^{\dagger}(x)\psi_{i,\sigma}(y):-G(y-x)(T^{A}T^{B})_{\alpha\sigma}:\psi_{i,\sigma}(y)\psi_{i,\alpha}^{\dagger}(x):\label{eq:commuJ2}\\
 & +[G^{2}(x-y)-G^{2}(y-x)]N\mathrm{Tr}(T^{A}T^{B}).
\end{align}
Here, as we can check $:\psi_{i,\sigma}(x)\psi_{i,\alpha}^{\dagger}(y):=-:\psi_{i,\alpha}^{\dagger}(y)\psi_{i,\sigma}(x):$
for $\alpha\neq\sigma$. When $\alpha = \sigma$, we simply have 
\begin{equation}
:\psi_{i,\alpha}(x)\psi_{i,\alpha}^{\dagger}(y):=\psi_{i,\alpha}(x)\psi_{i,\alpha}^{\dagger}(y)-G(x-y)=-:\psi_{i,\alpha}^{\dagger}(y)\psi_{i,\alpha}(x):+\delta(x-y)-G(y-x)-G(x-y).
\end{equation}
The zero-temperature fermion Green's function is evaluated as follows. 
We start with 
\begin{equation}
\langle\psi_{i,\alpha}(x)\psi_{j,\sigma}^{\dagger}(y)\rangle=\frac{1}{l}\sum_{p,p'}\mathrm{e}^{\mathrm{i}px-\mathrm{i}p'y}\langle c_{p,i\alpha}c_{p',j\sigma}^{\dagger}\rangle=\frac{1}{l}\sum_{p,p'}\mathrm{e}^{\mathrm{i}px-\mathrm{i}p'y}\delta_{ij}\delta_{\alpha\sigma}\delta_{pp'}\theta(p')=\frac{\delta_{ij}\delta_{\alpha\sigma}}{l}\sum_{p}\mathrm{e}^{\mathrm{i}p(x-y)}\theta(p).
\end{equation}
By definition, we have 
\begin{equation}
G(x-y)=\frac{1}{l}\sum_{p}\mathrm{e}^{\mathrm{i}p(x-y)}\theta(p)=\frac{1}{2\pi}\lim_{\delta\to0^{+}}\int_{0}^{\infty}\mathrm{d}p\mathrm{\,\mathrm{e}}^{\mathrm{i}p(x-y+\mathrm{i}\delta)}=\frac{1}{2\pi}\lim_{\delta\to0^{+}}\frac{\mathrm{i}}{x-y+\mathrm{i}\delta}. \label{eq:G(x)}
\end{equation}
Thus, \textbf{ 
\begin{align}
G(x-y)+G(y-x) & =\frac{1}{2\pi}\lim_{\delta\to0^{+}}\frac{\mathrm{\mathrm{i}}}{x-y+\mathrm{\mathrm{i}}\delta}+\frac{\mathrm{\mathrm{i}}}{y-x+\mathrm{i}\delta}=\delta(x-y),\\
G^{2}(x-y)-G^{2}(y-x) & =\frac{1}{4\pi^{2}}\lim_{\delta\to0^{+}}-\frac{\mathrm{1}}{(x-y+\mathrm{\mathrm{i}}\delta)^{2}}+\frac{1}{(y-x+\mathrm{i}\delta)^{2}}=\frac{\mathrm{i}}{2\pi}\partial_{x}\big[G(x-y)+G(y-x)\big] \nonumber \\
&=\frac{\mathrm{i}}{2\pi}\partial_{x}\delta(x-y).
\end{align}
}From the equation above, we have $:\psi_{i,\sigma}(x)\psi_{i,\alpha}^{\dagger}(y):=-:\psi_{i,\alpha}^{\dagger}(y)\psi_{i,\sigma}(x):$
for all $\alpha,\sigma$. This means the first two lines of Eq. ($\text{\ref{eq:commuJ1}},\text{\ref{eq:commuJ2}}$)
can be combined as a commutator of $T^{A}$ and $T^{B}$, which can
be replaced by $\mathrm{\mathrm{i}}\sum_{C}f^{ABC}T^{C}$. The last
term is simply 
\begin{equation}
[G(x-y)^{2}-G(y-x)^{2}]N\mathrm{Tr}(T^{A}T^{B})=\frac{\mathrm{i}N}{\pi}\,\partial_{x}\delta(x-y)\delta^{AB}.
\end{equation}
Finally, we have 
\begin{equation}
\big[\widetilde{J}^{A}(x),\widetilde{J}^{B}(y)\big]=\mathrm{i}\sum_{C}f^{ABC}\widetilde{J}^{C}(x)\delta(x-y)+\frac{\mathrm{i}N}{\pi}\,\partial_{x}\delta(x-y)\delta^{AB}.
\end{equation}
By a Fourier transformation,\textbf{ 
\begin{equation}
\widetilde{J}_{p}^{A}=\int_{-l}^{l}\mathrm{d}x\,\mathrm{e}^{\mathrm{i}p\pi x/l}\,\widetilde{J}^{A}(x),
\end{equation}
}the above commutator will satisfy the Kac-Moody algebra, 
\begin{align}
[\widetilde{J}_{p}^{A},\widetilde{J}_{p'}^{B}] & =\int_{-l}^{l}\int_{-l}^{l}\mathrm{d}x\,\mathrm{d}y\,\mathrm{e}^{\mathrm{i}p\pi x/l+\mathrm{i}p'\pi y/l}\,\big[\widetilde{J}^{A}(x),\widetilde{J}^{B}(y)\big]\\
 & =\int_{-l}^{l}\int_{-l}^{l}\mathrm{d}x\,\mathrm{d}y\,\mathrm{e}^{\mathrm{i}p\pi x/l+\mathrm{i}p'\pi y/l}\,\big[\mathrm{i}\sum_{C}f^{ABC}\widetilde{J}^{C}(x)\delta(x-y)+\frac{\mathrm{i}N}{\pi}\,\partial_{x}\delta(x-y)\delta^{AB}\big]\\
 & =\mathrm{i}\sum_{C}f^{ABC}\widetilde{J}_{p+p'}^{C}+2N \,p\delta^{AB}\delta_{p,-p'}.
\end{align}
This is Eq.~(\ref{eq:SOM2NKacMoody}) of the main text. 
The integral for the second term is 
\begin{align}
 & \frac{\mathrm{i}N}{\pi}\delta^{AB}\int_{-l}^{l}\int_{-l}^{l}\mathrm{d}x\,\mathrm{d}y\,\mathrm{e}^{\mathrm{i}p\pi x/l+\mathrm{i}p'\pi y/l}\,\partial_{x}\delta(x-y)\\
= & \frac{\mathrm{i}N}{\pi}\delta^{AB}\int_{-l}^{l}\int_{-l}^{l}\mathrm{d}x\,\mathrm{d}y\,\Big\{\partial_{x}\big[\mathrm{e}^{\mathrm{i}p\pi x/l+\mathrm{i}p'\pi y/l}\,\delta(x-y)\big]-\frac{\mathrm{i}p\pi}{l}\mathrm{e}^{\mathrm{i}p\pi x/l+\mathrm{i}p'\pi y/l}\,\delta(x-y)\Big\} \\
= & \frac{Nk}{l}\delta^{AB}\int_{-l}^{l}\int_{-l}^{l}\mathrm{d}x\,\mathrm{d}y\,\mathrm{e}^{\mathrm{i}p\pi x/l+\mathrm{i}p'\pi y/l}\,\delta(x-y)\\
= & \frac{Np}{l}\delta^{AB}\big(2l\delta_{p,-p'}\big)=2N\,p\delta^{AB}\delta_{p,-p'}.
\end{align}
The $\text{SO}(M)_{1}$ Kac-Moody algebra is defined to be 
\begin{equation}
[\widetilde{J}_{p}^{A},\widetilde{J}_{p'}^{B}]=\mathrm{i}\sum_{C}f^{ABC}\widetilde{J}_{p+p'}^{C}+  \, p\delta^{AB}\delta_{p,-p'}.\label{ap:KMSOM2N}
\end{equation}

\section{Primary states and fusion rules}\label{ap:PrimstatesFusion}
The primary states, Eq.~\eqref{eq:primarystates}, are the most fundamental states in the Kac-Moody algebra. Excited states can be constructed from primary states by using raising operators \cite{LudwigCFTinCMP,ludwig1994field}. Here, we use a general method to derive the allowed primary states (representations) in SO($M$). The fusion rules (including the single and double fusion with the impurity) and the energies of the allowed representations are also given for the $\text{SO}(4)_2$ case.

\subsection{The representation cut off of $\text{SO}(4)_{2}$ \label{subsec:repcutoffSO4}}
The Kac-Moody algebra for $\text{SO}(4)_{2}$ [see Eq.~(\ref{ap:KMSOM2N})] is
\begin{equation}
[\widetilde{J}_{p}^{A},\widetilde{J}_{p'}^{B}]=\mathrm{i}\sum_{C}f^{ABC}\widetilde{J}_{p+p'}^{C}+2\,p\delta^{AB}\delta_{p,-p'}
\end{equation}
where $A,B\in\big\{12,13,14,23,24,34\big\}$ and we can define $A,B=1,2,3,4,5,6$
for simplicity. This gives
\begin{align}
 & f^{1BC}=\left(\begin{array}{cccccc}
0 & 0 & 0 & 0 & 0 & 0\\
0 & 0 & 0 & 1 & 0 & 0\\
0 & 0 & 0 & 0 & 1 & 0\\
0 & -1 & 0 & 0 & 0 & 0\\
0 & 0 & -1 & 0 & 0 & 0\\
0 & 0 & 0 & 0 & 0 & 0
\end{array}\right)_{BC};f^{2BC}=\left(\begin{array}{cccccc}
0 & 0 & 0 & -1 & 0 & 0\\
0 & 0 & 0 & 0 & 0 & 0\\
0 & 0 & 0 & 0 & 0 & 1\\
1 & 0 & 0 & 0 & 0 & 0\\
0 & 0 & 0 & 0 & 0 & 0\\
0 & 0 & -1 & 0 & 0 & 0
\end{array}\right)_{BC};f^{3BC}=\left(\begin{array}{cccccc}
0 & 0 & 0 & 0 & -1 & 0\\
0 & 0 & 0 & 0 & 0 & -1\\
0 & 0 & 0 & 0 & 0 & 0\\
0 & 0 & 0 & 0 & 0 & 0\\
1 & 0 & 0 & 0 & 0 & 0\\
0 & 1 & 0 & 0 & 0 & 0
\end{array}\right)_{BC}\\
 & f^{4BC}=\left(\begin{array}{cccccc}
0 & 1 & 0 & 0 & 0 & 0\\
-1 & 0 & 0 & 0 & 0 & 0\\
0 & 0 & 0 & 0 & 0 & 0\\
0 & 0 & 0 & 0 & 0 & 0\\
0 & 0 & 0 & 0 & 0 & 1\\
0 & 0 & 0 & 0 & -1 & 0
\end{array}\right)_{BC};f^{5BC}=\left(\begin{array}{cccccc}
0 & 0 & 1 & 0 & 0 & 0\\
0 & 0 & 0 & 0 & 0 & 0\\
-1 & 0 & 0 & 0 & 0 & 0\\
0 & 0 & 0 & 0 & 0 & -1\\
0 & 0 & 0 & 0 & 0 & 0\\
0 & 0 & 0 & 1 & 0 & 0
\end{array}\right)_{BC};f^{6BC}=\left(\begin{array}{cccccc}
0 & 0 & 0 & 0 & 0 & 0\\
0 & 0 & 1 & 0 & 0 & 0\\
0 & -1 & 0 & 0 & 0 & 0\\
0 & 0 & 0 & 0 & 1 & 0\\
0 & 0 & 0 & -1 & 0 & 0\\
0 & 0 & 0 & 0 & 0 & 0
\end{array}\right)_{BC}.
\end{align}
In the following, we write $\widetilde{J}$ as $J$ for simplicity. The lowering operators are
\begin{equation}
F^{-1,1}_p  =\frac{1}{2}[J^{13}_p+iJ^{14}_p-i(J^{23}_p+iJ^{24}_p)];\
F^{-1,-1}_p  =\frac{1}{2}[J^{13}_p-iJ^{14}_p-i(J^{23}_p-iJ^{24}_p)].
\end{equation}
The raising operators are 
\begin{equation}
F^{1,1}_p  =\frac{1}{2}[J^{13}_p+iJ^{14}_p+i(J^{23}_p+iJ^{24}_p)];\
F^{1,-1}_p  =\frac{1}{2}[J^{13}_p-iJ^{14}_p+i(J^{23}_p-iJ^{24}_p)].
\end{equation}
The highest-weight primary states $|\ell_{1},\ell_{2}\rangle$ are defined
to have the following properties:
\begin{align}
 & \text{primary states:}\ J_{p}^{A}|\ell_{1},\ell_{2}\rangle=0,\ \text{for}\ p>0 \label{eq:primarystates};\\
 & \text{highest wight}:\ F_{0}^{1,1}|\ell_{1},\ell_{2}\rangle=0\ \text{and}\ F_{0}^{1,-1}|\ell_{1},\ell_{2}\rangle=0;\\
 & \text{eigenstates of Cartan generators}:H_{0}^{i}|\ell_{1},\ell_{2}\rangle=\ell_{i}|\ell_{1},\ell_{2}\rangle.
\end{align}
In order to derive the bound for the values of $\ell_{1,2}$, we consider two excited states $F_{-1}^{1,1}|\ell_{1},\ell_{2}\rangle$ and $F_{-1}^{1,-1}|\ell_{1},\ell_{2}\rangle$ by acting on all the two raising operators (positive roots) above.
Their norms should be non-negative, which are
\begin{align}
 & \langle\ell_{1},\ell_{2}|F_{1}^{-1,-1}F_{-1}^{1,1}|\ell_{1},\ell_{2}\rangle=\langle\ell_{1},\ell_{2}|[F_{1}^{-1,-1},F_{-1}^{1,1}]|\ell_{1},\ell_{2}\rangle\nonumber \\
= & \langle\ell_{1},\ell_{2}|(2-H_{0}^{1}-H_{0}^{2})|\ell_{1},\ell_{2}\rangle=2-\ell_{1}-\ell_{2}\geq0\label{eq:repcut1}
\end{align}
and
\begin{align}
 & \langle\ell_{1},\ell_{2}|F_{1}^{-1,1}F_{-1}^{1,-1}|\ell_{1},\ell_{2}\rangle=\langle\ell_{1},\ell_{2}|[F_{1}^{-1,1}F_{-1}^{1,-1}]|\ell_{1},\ell_{2}\rangle\nonumber \\
= & \langle\ell_{1},\ell_{2}|(2-H_{0}^{1}+H_{0}^{2})|\ell_{1},\ell_{2}\rangle=2-\ell_{1}+\ell_{2}\geq0\label{eq:repcut2}
\end{align}
where we used the following Kac-Moody algebra
\begin{align}
[F_{1}^{-1,-1},F_{-1}^{1,1}]= & \frac{1}{4}\big[J_{1}^{13}-\mathrm{i}J_{1}^{14}-\mathrm{i}(J_{1}^{23}-\mathrm{i}J_{1}^{24}),\,J_{-1}^{13}+\mathrm{i}J_{-1}^{14}+\mathrm{i}(J_{-1}^{23}+\mathrm{i}J_{-1}^{24})\big]\\
= & \frac{1}{4}\Big\{4\times2+\mathrm{i}[J_{1}^{13},J_{-1}^{14}]+\mathrm{i}[J_{1}^{13},J_{-1}^{23}]-[J_{1}^{13},J_{-1}^{24}]\\
 & -\mathrm{i}[J_{1}^{14},J_{-1}^{13}]+[J_{1}^{14},J_{-1}^{23}]+\mathrm{i}[J_{1}^{14},J_{-1}^{24}]\\
 & -\mathrm{i}[J_{1}^{23},J_{-1}^{13}]+[J_{1}^{23},J_{-1}^{14}]+\mathrm{i}[J_{1}^{23},J_{-1}^{24}]\\
 & -[J_{1}^{24},J_{-1}^{13}]-\mathrm{i}[J_{1}^{24},J_{-1}^{14}]-\mathrm{i}[J_{1}^{24},J_{-1}^{23}]\Big\}\\
= & 2+\frac{1}{4}\Big\{2\mathrm{i}[J_{1}^{13},J_{-1}^{14}]+2\mathrm{i}[J_{1}^{13},J_{-1}^{23}]+2\mathrm{i}[J_{1}^{14},J_{-1}^{24}]+2\mathrm{i}[J_{1}^{23},J_{-1}^{24}]\Big\}\\
= & 2-\frac{1}{2}J_{0}^{34}-\frac{1}{2}J_{0}^{12}-\frac{1}{2}J_{0}^{12}-\frac{1}{2}J_{0}^{34}=2-H_{0}^{1}-H_{0}^{2}
\end{align}
and
\begin{align}
[F_{1}^{-1,1},F_{-1}^{1,-1}]= & \frac{1}{4}\big[J_{1}^{13}+\mathrm{i}J_{1}^{14}-\mathrm{i}(J_{1}^{23}+\mathrm{i}J_{1}^{24}),\,J_{-1}^{13}-\mathrm{i}J_{-1}^{14}+\mathrm{i}(J_{-1}^{23}-\mathrm{i}J_{-1}^{24})\big]\\
= & \frac{1}{4}\Big\{4\times2-\mathrm{i}[J_{1}^{13},J_{-1}^{14}]+\mathrm{i}[J_{1}^{13},J_{-1}^{23}]-[J_{1}^{13},J_{-1}^{24}]\\
 & +\mathrm{i}[J_{1}^{14},J_{-1}^{13}]-[J_{1}^{14},J_{-1}^{23}]+\mathrm{i}[J_{1}^{14},J_{-1}^{24}]\\
 & -\mathrm{i}[J_{1}^{23},J_{-1}^{13}]-[J_{1}^{23},J_{-1}^{14}]-\mathrm{i}[J_{1}^{23},J_{-1}^{24}]\\
 & +[J_{1}^{24},J_{-1}^{13}]-\mathrm{i}[J_{1}^{24},J_{-1}^{14}]+\mathrm{i}[J_{1}^{24},J_{-1}^{23}]\Big\}\\
= & 2+\frac{1}{4}\Big\{-2\mathrm{i}[J_{1}^{13},J_{-1}^{14}]+2\mathrm{i}[J_{1}^{13},J_{-1}^{23}]+2\mathrm{i}[J_{1}^{14},J_{-1}^{24}]-2\mathrm{i}[J_{1}^{23},J_{-1}^{24}]\Big\}\\
= & 2+\frac{1}{2}J_{0}^{34}-\frac{1}{2}J_{0}^{12}-\frac{1}{2}J_{0}^{12}+\frac{1}{2}J_{0}^{34}=2-H_{0}^{1}+H_{0}^{2}.
\end{align}
The above results from the Lie algebra satisfy $[F^{\vec{\alpha}},F^{-\vec{\alpha}}]=\vec{\alpha}\cdot\vec{H}$
(see Howard Georgi's Lie algebra textbook \cite{georgi2000lie} and the Chapter 14 of Ref.~\cite{francesco2012conformal}), which can be used for further generalization with
$\text{SO}(M)_k$. In terms of the highest weights of the fundamental
representation $\mu_{1}=(1/2,1/2)$ and $\mu_{2}=(1/2,-1/2)$, we
get
\begin{equation}
(\ell_{1},\ell_{2})=(\ell_{1}+\ell_{2})\mu_{1}+(\ell_{1}-\ell_{2})\mu_{2}.
\end{equation}
As we know, the highest weight of any representation can be written
as $a_{1}\mu_{1}+a_{2}\mu_{2}$, i.e., the sum of the fundamental
ones with $a_{i}\in\mathbb{Z}_{\geq0}$. Thus, the two requirements Eq.~(\ref{eq:repcut1}) and Eq.~(\ref{eq:repcut2}) are equivalent to $a_{1,2}\leq2$, which can be considered as the
requirements of two $\text{SU}(2)_{2}$.

\subsection{The representation cut off for $\text{SO}(2m)_{k}$}\label{ap:repcutSO2mk}
For $\text{SO}(2m)_{k}$, the two fundamental spinor representations
are $\mu_{m-1}$ and $\mu_{m}$. The $m(m-1)$ positive roots are $e_{i}\pm e_{j}\,i<j=1,...,m$
where $e_{i}=(0,..,1,..,0)$ which has $1$ at the $i$th position
and $0$ elsewhere. The simple roots are $e_{i}-e_{i+1},\,i=1,m-1$
and $e_{m-1}+e_{m}$. By considering the  $m(m-1)$ lowering operators acting on the state $|\vec{\ell}\rangle$ for all the positive roots and requiring the new states to have non-negative norm [same as we did in Eq.~(\ref{eq:repcut1}) and Eq.~(\ref{eq:repcut2})], we get constrains for $\ell$s:
\begin{equation}
\ell_{i}\pm\ell_{j}\leq k,\ i<j=1,...,m.
\end{equation}
Thus, in terms of $\vec{\ell}=\sum_{i=1}^{m}a_{i}\mu_{i}$ with the fundamental weights
\begin{align}
 & \mu_{1}=(1,0,0,...,0,0,0),\\
 & \mu_{2}=(1,1,0,...,0,0,0),\\
 & \mu_{3}=(1,1,1,...,0,0,0),\\
 & \vdots\\
 & \mu_{m-2}=(1,1,1,...,1,0,0),\\
 & \mu_{m-1}=\frac{1}{2}(1,1,1,...,1,1,-1),\\
 & \mu_{m}=\frac{1}{2}(1,1,1,...,1,1,1),
\end{align}
we get
\begin{align}
 & \ell_{i}=\sum_{n=i}^{m-2}a_{i}+\frac{1}{2}a_{m-1}+\frac{1}{2}a_{m},\,i=1,m-2;\\
 & \ell_{m-1}=\frac{1}{2}a_{m-1}+\frac{1}{2}a_{m};\,\ell_{m}=-\frac{1}{2}a_{m-1}+\frac{1}{2}a_{m}
\end{align}
and inversely
\begin{equation}
a_{i}=\ell_{i}-\ell_{i+1},\,i=1,...,m-1;\ a_{m}=\ell_{m-1}+\ell_{m}.
\end{equation}
Thus, the simple roots give $a_{i=1,...,m}\leq k$.

For the case of $\text{SO}(6)_{k}$, we have 15 generators, which
are 3 Cartan generators, 6 positive roots and 6 negative roots. The
above representation cut off is generally given by considering $F_{-1}^{\alpha_{1},\alpha_{2},\alpha_{3}}|\ell_{1},\ell_{2},\ell_{3}\rangle$
when $\vec{\alpha}=(\alpha_{1},\alpha_{2},\alpha_{3})$ are the positive
roots. Here, we give all the positive roots:
\begin{align}
 & \vec{\alpha}=(1,-1,0),\,\vec{\beta}=(0,1,-1),\,\vec{\rho}=(0,1,1),\\
 & \vec{\alpha}+\vec{\beta}=(1,0,-1),\,\vec{\alpha}+\vec{\rho}=(1,0,1),\,\vec{\alpha}+\vec{\beta}+\vec{\rho}=(1,1,0)
\end{align}
where the upper line gives 3 simple roots. By calculating the norm
of $F_{-1}^{\alpha_{1},\alpha_{2},\alpha_{3}}|\ell_{1},\ell_{2},\ell_{3}\rangle$
and the commutator $[F^{\vec{\alpha}},F^{-\vec{\alpha}}]=\vec{\alpha}\cdot\vec{H}$,
we get the requirements correspondingly for each postive root:
\begin{align}
 & \ell_{1}-\ell_{2}\leq k,\,\ell_{2}-\ell_{3}\leq k,\,\ell_{2}+\ell_{3}\leq k\\
 & \ell_{1}-\ell_{3}\leq k,\,\ell_{1}+\ell_{3}\leq k,\,\ell_{1}+\ell_{2}\leq k.
\end{align}
In terms of Dynkin labels $a_{i=1,2,3}$, we have $(\ell_{1},\ell_{2},\ell_{3})=a_{1}\xi_{1}+a_{2}\xi_{2}+a_{3}\xi_{3}$
with
\begin{align}
 & \xi_{1}=(1,0,0),\,\xi_{2}=(\frac{1}{2},\frac{1}{2},\frac{1}{2}),\,\xi_{3}=(\frac{1}{2},\frac{1}{2},-\frac{1}{2});\\
 & a_{1}=\ell_{1}-\ell_{2},\,a_{2}=\ell_{2}-\ell_{3},\,a_{3}=\ell_{2}+\ell_{3}.
\end{align}
Thus, we get requirements for the Dynkin labels $a_{i=1,2,3}$:
\begin{equation}
a_{1,2,3}\leq k;\,a_{1}+a_{2}\leq k;\,a_{1}+a_{3}\leq k;\,a_{1}+a_{2}+a_{3}\leq k.
\end{equation}

\subsection{The representation cut off for $\text{SO}(2m+1)_{k}$}\label{ap:repcutSO2mp1k}
For $\text{SO}(2m+1)_{k}$, the only fundamental spinor representation
is $\xi_{m}$. The $m^{2}$ positive roots are $e_{i}\pm e_{j}$ and
$e_{j}$ with $i<j=1,...,m$. The simple roots are $e_{i}-e_{i+1},\,i=1,m-1$
and $e_{m}$. This gives the requirements for $\ell$s:
\begin{equation}
\ell_{i}\pm\ell_{j}\leq k\,\text{and}\,\mathrm{\ell}_{j}\leq k,\ i<j=1,...,m.
\end{equation}
Thus, in terms of $\vec{\ell}=\sum_{i=1}^{m}a_{i}\xi_{i}$ with the fundamental weights
\begin{align}
 & \mu_{1}=(1,0,0,...,0,0,0),\\
 & \mu_{2}=(1,1,0,...,0,0,0),\\
 & \mu_{3}=(1,1,1,...,0,0,0),\\
 & \vdots\\
 & \mu_{m-2}=(1,1,1,...,1,0,0),\\
 & \mu_{m-1}=(1,1,1,...,1,1,0),\\
 & \mu_{m}=\frac{1}{2}(1,1,1,...,1,1,1)
\end{align}
we get
\begin{align}
 & \ell_{i}=\sum_{n=i}^{m-1}a_{n}+\frac{1}{2}a_{m},\,i=1,m-1;\ \ell_{m}=\frac{1}{2}a_{m}
\end{align}
and inversely
\begin{equation}
a_{i}=\ell_{i}-\ell_{i+1},\,i=1,...,m-1;\ a_{m}=2\ell_{m}.
\end{equation}
Thus, the simple roots give $a_{i=1,...,m-1}\leq k$ and $a_{m}\leq2k$.

\subsection{The fusion rule of $\text{SO}(4)_{2}$}

According to the above cut off (see Sec.~\ref{subsec:repcutoffSO4}), the allowed representations for $\text{SO}(4)_{2}$ are
\begin{align}
\text{spinor}: & \ \mu_{1},\mu_{2},2\mu_{1}+\mu_{2},\mu_{1}+2\mu_{2};\\
\text{vector}: &\ 0,\mu_{1}+\mu_{2},2\mu_{1},2\mu_{2},2\mu_{1}+2\mu_{2}.
\end{align}
Their fusion rules are
\begin{align}
 & \mu_{1}\otimes\mu_{1}=2\mu_{1}\oplus0;\,\mu_{1}\otimes\mu_{2}=\mu_{1}+\mu_{2};\\
 & \mu_{1}\otimes(\mu_{1}+\mu_{2})=(2\mu_{1}+\mu_{2})\oplus\mu_{2};\,\mu_{1}\otimes2\mu_{1}=\cancel{3\mu_{1}}\oplus\mu_{1};\,\mu_{1}\otimes2\mu_{2}=(\mu_{1}+2\mu_{2});\\
 & \mu_{1}\otimes(2\mu_{1}+\mu_{2})=\cancel{(3\mu_{1}+\mu_{2})}\oplus(\mu_{1}+\mu_{2});\,\mu_{1}\otimes(\mu_{1}+2\mu_{2})=(2\mu_{1}+2\mu_{2})\oplus(2\mu_{2});\\
 & \mu_{1}\otimes(2\mu_{1}+2\mu_{2})=\cancel{(3\mu_{1}+2\mu_{2})}\oplus(\mu_{1}+2\mu_{2});
\end{align}
\begin{align}
 & \mu_{2}\otimes\mu_{2}=2\mu_{2}\oplus0;\\
 & \mu_{2}\otimes(\mu_{1}+\mu_{2})=(\mu_{1}+2\mu_{2})\oplus\mu_{1};\,\mu_{2}\otimes2\mu_{1}=2\mu_{1}+\mu_{2};\,\mu_{2}\otimes2\mu_{2}=\cancel{3\mu_{2}}\oplus\mu_{2};\\
 & \mu_{2}\otimes(2\mu_{1}+\mu_{2})=(2\mu_{1}+2\mu_{2})\oplus2\mu_{1};\,\mu_{2}\otimes(\mu_{1}+2\mu_{2})=\cancel{(\mu_{1}+3\mu_{2})}\oplus(\mu_{1}+\mu_{2});\\
 & \mu_{2}\otimes(2\mu_{1}+2\mu_{2})=\cancel{(2\mu_{1}+3\mu_{2})}\oplus(2\mu_{1}+\mu_{2});
\end{align}
\begin{align}
 & (\mu_{1}+\mu_{2})\otimes(\mu_{1}+\mu_{2})=(2\mu_{1}+2\mu_{2})\oplus2\mu_{1}\oplus2\mu_{2}\oplus0;\\
 & (\mu_{1}+\mu_{2})\otimes2\mu_{1}=\cancel{3\mu_{1}}\oplus\mu_{1}\oplus(2\mu_{1}+\mu_{2});\\
 & (\mu_{1}+\mu_{2})\otimes2\mu_{2}=(\mu_{1}+2\mu_{2})\oplus\cancel{3\mu_{2}}\oplus\mu_{2};\\
 & (\mu_{1}+\mu_{2})\otimes(2\mu_{1}+\mu_{2})=\cancel{(3\mu_{1}+2\mu_{2})}\oplus\cancel{3\mu_{1}}\oplus(\mu_{1}+2\mu_{2})\oplus\mu_{1};\\
 & (\mu_{1}+\mu_{2})\otimes(\mu_{1}+2\mu_{2})=\cancel{(2\mu_{1}+3\mu_{2})}\oplus\cancel{3\mu_{2}}\oplus(2\mu_{1}+\mu_{2})\oplus\mu_{2};\\
 & (\mu_{1}+\mu_{2})\otimes(2\mu_{1}+2\mu_{2})=\cancel{(3\mu_{1}+3\mu_{2})}\oplus\cancel{(3\mu_{1}+\mu_{2})}\oplus\cancel{(\mu_{1}+3\mu_{2})}\oplus(\mu_{1}+\mu_{2});
\end{align}
\begin{align}
 & 2\mu_{1}\otimes2\mu_{1}=\cancel{4\mu_{1}}\oplus0;\,2\mu_{1}\otimes2\mu_{2}=(2\mu_{1}+2\mu_{2});\\
 & 2\mu_{1}\otimes(2\mu_{1}+\mu_{2})=\cancel{(4\mu_{1}+\mu_{2})}\oplus\mu_{2};\\
 & 2\mu_{1}\otimes(\mu_{1}+2\mu_{2})=\cancel{(3\mu_{1}+2\mu_{2})}\oplus(\mu_{1}+2\mu_{2});\\
 & 2\mu_{1}\otimes(2\mu_{1}+2\mu_{2})=\cancel{(4\mu_{1}+2\mu_{2})}\oplus2\mu_{2};
\end{align}
\begin{align}
 & 2\mu_{2}\otimes2\mu_{2}=\cancel{4\mu_{2}}\oplus0;\\
 & 2\mu_{2}\otimes(2\mu_{1}+\mu_{2})=\cancel{(2\mu_{1}+3\mu_{2})}\oplus(2\mu_{1}+\mu_{2});\\
 & 2\mu_{2}\otimes(\mu_{1}+2\mu_{2})=\cancel{(\mu_{1}+4\mu_{2})}\oplus\mu_{1};\\
 & 2\mu_{2}\otimes(2\mu_{1}+2\mu_{2})=\cancel{(2\mu_{1}+4\mu_{2})}\oplus2\mu_{1};
\end{align}
\begin{align}
 & (2\mu_{1}+\mu_{2})\otimes(2\mu_{1}+\mu_{2})=\cancel{(4\mu_{1}+2\mu_{2})}\oplus\cancel{4\mu_{1}}\oplus2\mu_{2}\oplus0;\\
 & (2\mu_{1}+\mu_{2})\otimes(\mu_{1}+2\mu_{2})=\cancel{(3\mu_{1}+3\mu_{2})}\oplus\cancel{(3\mu_{1}+\mu_{2})}\oplus\cancel{(\mu_{1}+3\mu_{2})}\oplus(\mu_{1}+\mu_{2});\\
 & (2\mu_{1}+\mu_{2})\otimes(2\mu_{1}+2\mu_{2})=\cancel{(4\mu_{1}+3\mu_{2})}\oplus\cancel{(4\mu_{1}+\mu_{2})}\oplus\cancel{3\mu_{2}}\oplus\mu_{2};
\end{align}
\begin{align}
 & (\mu_{1}+2\mu_{2})\otimes(\mu_{1}+2\mu_{2})=\cancel{(2\mu_{1}+4\mu_{2})}\oplus\cancel{4\mu_{2}}\oplus2\mu_{1}\oplus0;\\
 & (\mu_{1}+2\mu_{2})\otimes(2\mu_{1}+2\mu_{2})=\cancel{(3\mu_{1}+4\mu_{2})}\oplus\cancel{(\mu_{1}+4\mu_{2})}\oplus\cancel{3\mu_{1}}\oplus\mu_{1};
\end{align}
\begin{equation}
(2\mu_{1}+2\mu_{2})\otimes(2\mu_{1}+2\mu_{2})=\cancel{(4\mu_{1}+4\mu_{2})}\oplus\cancel{4\mu_{1}}\oplus\cancel{4\mu_{2}}\oplus0.
\end{equation}
The above crossed terms will not be allowed in the fusion rules, but will appear in the representation decompositions. The (matrix) representation dimensions are always integers when the above crossed terms are kept. However, the effective ``dimensions", i.e.~the quantum dimensions are non-integers in the above fusion rules where the crossed terms are absent. In order to obtain the quantum dimensions, we first assume that the quantum dimensions of the above representations are $d_{Q,i=1,\dots,9}$ and solve them from algebraic equations. The equations are given by the above fusion rules, i.e., the product of quantum dimensions for the left-side representations equals the sum of quantum dimensions for the right-side representations. The solutions for the quantum dimensions $d_{Q,i=1,\dots,9}$ are listed in Table \ref{table:so42qdim}.
\begin{table}
\begin{centering}
\begin{tabular}{|c|c|c|c|c|c|c|c|c|c|}
\hline 
Reps & $0$ & $\mu_{1}$ & $\mu_{2}$ & $\mu_{1}+\mu_{2}$ & $2\mu_{1}$ & $2\mu_{2}$ & $2\mu_{1}+\mu_{2}$ & $\mu_{1}+2\mu_{2}$ & $2\mu_{1}+2\mu_{2}$\tabularnewline
\hline 
\hline 
Q-dim & $1$ & $\sqrt{2}$ & $\sqrt{2}$ & $2$ & $1$ & $1$ & $\sqrt{2}$ & $\sqrt{2}$ & $1$\tabularnewline
\hline 
\end{tabular}
\par\end{centering}
\caption{The quantum dimensions for all the allowed representations in $\text{SO}(4)_{2}$.}
\label{table:so42qdim}
\end{table}
The general result for the above representations $a_{1}\mu_{1}+a_{2}\mu_{2}$
is
\begin{equation}
d_{Q}(a_{1},a_{2})=\begin{cases}
1 & (\widetilde{a_{1}},\widetilde{a_{2}})=(0,0)\\
\sqrt{2} & (\widetilde{a_{1}},\widetilde{a_{2}})=(0,1)\,\text{or}\,(1,0)\\
2 & (\widetilde{a_{1}},\widetilde{a_{2}})=(1,1)
\end{cases},\,\widetilde{a_{i}}=\text{min}(a_{i},2-a_{i}).
\end{equation}
More general result can be found for $\text{SO}(4)_{k}$
\begin{align}
 & d_{Q}^{\text{SO}(4)_{k}}(a_{1},a_{2})=d^{\text{SU}(2)_{k}}(a_{1})\times d^{\text{SU}(2)_{k}}(a_{2});\,d^{\text{SU}(2)_{k}}(a_{i})=\frac{\sin[\pi(a_{i}+1)/(k+2)]}{\sin[\pi/(k+2)]}
\end{align}
which explains how the $\text{SO}(4)_{k}$ anyons are related to two
copies of $\text{SU}(2)_{k}$ anyons due to the fact $\text{SO}(4)\sim\text{SU}(2)\times\text{SU}(2)$.
This may be possible to generalize for any classical Lie group at
any level because they all have $\text{SU}(2)$ subalgebras.

\subsection{Energies of $\text{SO}(4)_{2}$}

Following Ludwig's lecture notes ~\cite{LudwigCFTinCMP,ludwig1994field} and Ref.~\cite{doi:10.7566/JPSJ.90.024708}, we get the Hamiltonian
\begin{align}
 & H_{0}=H_{\text{SO}(4)_{2}}+H_{\text{SO}(2)_{4}}\\
= & \frac{\pi v_{F}}{l}\sum_{p=-\infty}^{+\infty}\Big\{\frac{1}{2+h^{\vee}[\text{SO}(4)]}:J_{-p}^{a}J_{p}^{a}:+\frac{1}{4+h^{\vee}[\text{SO}(2)]}:J_{-p}^{A}J_{p}^{A}:\Big\}\\
= & \frac{\pi v_{F}}{l}\sum_{p=-\infty}^{+\infty}\Big\{\frac{1}{2+2}:J_{-p}^{a}J_{p}^{a}:+\frac{1}{4+0}:J_{-p}^{A}J_{p}^{A}:\Big\}.
\end{align}
The $\text{SO}(2)$ has only one generator $J_{0}^{A=1}$ that is a Cartan generator, which is like $\text{U}(1)$. Thus, we define the highest wight primary state as $|Q\rangle$ and get $:J_{-p}^{A}J_{p}^{A}:|Q\rangle=Q^{2}|Q\rangle$
where $Q$ is the eigenvalue of $J_{0}^{A=1}$. If the representation
of $\text{SO}(4)$ are given by $a_{1}\mu_{1}+a_{2}\mu_{2}$, we get
\begin{equation}
E-E^{(0)}=\frac{\pi v_{F}}{l}\Big\{\frac{1}{4}\frac{1}{2}\big[a_{1}+\frac{a_{1}^{2}}{2}+a_{2}+\frac{a_{2}^{2}}{2}\big]+\frac{1}{4}\frac{1}{2}Q^{2}\Big\}+\dots \ .
\end{equation}
If we take $a_{1}=2j$ and $a_{2}=2j_{f}$, we get
\begin{equation}
E-E^{(0)}=\frac{\pi v_{F}}{l}\Big\{\frac{1}{4}\big(j+j^{2}+j_{f}+j_{f}^{2}\big)+\frac{Q^{2}}{8}\Big\}+\dots
\end{equation}
which is identical to the $\text{SU}(2)_{2}$ result.
\begin{table}
\begin{centering}
\begin{tabular}{|c|c|c|c|c|}
\hline 
$a_{1}$ & $a_{2}$ & $Q\,\text{mod}4$ & $l\big[E-E^{(0)}\big]/\pi v_{F}$ & $n$\tabularnewline
\hline 
\hline 
$0$ & $0$ & $0$ & $0$ & $1$\tabularnewline
\hline 
$1$ & $1$ & $1$ & $1/2$ & $1$\tabularnewline
\hline 
$2$ & $0$ & $2$ & $1$ & $1$\tabularnewline
\hline 
$0$ & $2$ & $2$ & $1$ & $1$\tabularnewline
\hline 
$2$ & $2$ & $0$ & $1$ & $1$\tabularnewline
\hline 
$1$ & $1$ & $3$ & $3/2$ & $1$\tabularnewline
\hline 
\end{tabular}
\par\end{centering}
\caption{Table for the primary states and their energies/scaling dimensions at the free case for $\text{SO(4)}_2$. }\label{aptab:SO42free}
\end{table}

\subsection{Single fusion with the impurity for $\text{SO(4)}_2$}\label{ap:singlefusion}
The representation for the impurity is $\mu_{1}\oplus\mu_{2}$. We
consider the single fusion of the impurity $\mu_{1}$ ($\mu_{2}$
is similar) with the above primary states. The results are listed in Table~\ref{aptab:SFmu1}.
\begin{table}[h]
\begin{centering}
\begin{tabular}{|c|c|c|c|c|}
\hline 
$a_{1}$ & $a_{2}$ & $Q\,\text{mod}4$ & $l\big[E-E^{(0)}\big]/\pi v_{F}$ & $n$\tabularnewline
\hline 
\hline 
$1$ & $0$ & $0$ & $3/16$ & $1$\tabularnewline
\hline 
$0$ & $1$ & $1$ & $5/16$ & $1$\tabularnewline
\hline 
$2$ & $1$ & $1$ & $13/16$ & $1$\tabularnewline
\hline 
$1$ & $0$ & $2$ & $11/16$ & $1$\tabularnewline
\hline 
$1$ & $2$ & $2$ & $19/16$ & $1$\tabularnewline
\hline 
$1$ & $2$ & $0$ & $11/16$ & $1$\tabularnewline
\hline 
$0$ & $1$ & $3$ & $21/16$ & $1$\tabularnewline
\hline 
$2$ & $1$ & $3$ & $29/16$ & $1$\tabularnewline
\hline 
\end{tabular}
\par\end{centering}
\caption{Table for the primary boundary operators and their energies/scaling dimensions at the Kondo fixed point for $\text{SO(4)}_2$ obtained by single fusion with the impurity $\mu_1$. The single fusion results with impurity $\mu_2$ can be found by interchanging $a_1 \leftrightarrow a_2$.}\label{aptab:SFmu1}
\end{table}

\subsection{Double fusion with the impurity for $\text{SO(4)}_2$}\label{ap:doublefusion}
The representation for the impurity is $\mu_{1}\oplus\mu_{2}$. We
consider the double fusion of the impurity with the above primary states shown in Table~\ref{aptab:SO42free} and allow the cross terms like $\mu_{1,2}$ from the first fusion with the impurity $\mu_{1}\oplus\mu_{2}$ and $\mu_{2,1}$ from the second fusion with the impurity $\mu_{1}\oplus\mu_{2}$. This is equivalent to not fixing the charge parity of the impurity. The double fusion results are in the following:
\begin{equation}
[\mu_{1}\oplus\mu_{2}]\otimes[\mu_{1}\oplus\mu_{2}]\otimes0=2\mu_{1}\oplus0\oplus2\mu_{2}\oplus0\oplus\red{(\mu_{1}+\mu_{2})}\oplus\red{(\mu_{1}+\mu_{2})};
\end{equation}
\begin{align}
 & [\mu_{1}\oplus\mu_{2}]\otimes[\mu_{1}\oplus\mu_{2}]\otimes(\mu_{1}+\mu_{2})=[2\mu_{1}\oplus0\oplus2\mu_{2}\oplus0\oplus\red{(\mu_{1}+\mu_{2})}\oplus\red{(\mu_{1}+\mu_{2})}]\otimes(\mu_{1}+\mu_{2})\\
= & \mu_{1}\oplus(2\mu_{1}+\mu_{2})\
  \oplus(\mu_{1}+\mu_{2})\
  \oplus(\mu_{1}+2\mu_{2})\oplus\mu_{2}\
  \oplus(\mu_{1}+\mu_{2})\
  \oplus\red{(2\mu_{1}+2\mu_{2})}\oplus\red{2\mu_{1}}\oplus\red{2\mu_{2}}\oplus\red{0}\\
 & \oplus\red{(2\mu_{1}+2\mu_{2})}\oplus\red{2\mu_{1}}\oplus\red{2\mu_{2}}\oplus\red{0};
\end{align}
\begin{align}
 & [\mu_{1}\oplus\mu_{2}]\otimes[\mu_{1}\oplus\mu_{2}]\otimes2\mu_{1}=[2\mu_{1}\oplus0\oplus2\mu_{2}\oplus0\oplus\red{(\mu_{1}+\mu_{2})}\oplus\red{(\mu_{1}+\mu_{2})}]\otimes2\mu_{1}\\
= & 0
  \oplus2\mu_{1}
  \oplus(2\mu_{1}+2\mu_{2})
  \oplus2\mu_{1}
  \oplus\red{\mu_{1}}\oplus\red{(2\mu_{1}+\mu_{2})}
  \oplus\red{\mu_{1}}\oplus\red{(2\mu_{1}+\mu_{2})};
\end{align}
\begin{align}
 & [\mu_{1}\oplus\mu_{2}]\otimes[\mu_{1}\oplus\mu_{2}]\otimes2\mu_{2}=[2\mu_{1}\oplus0\oplus2\mu_{2}\oplus0\oplus\red{(\mu_{1}+\mu_{2})}\oplus\red{(\mu_{1}+\mu_{2})}]\otimes2\mu_{2}\\
= & (2\mu_{1}+2\mu_{2})
  \oplus2\mu_{2}
  \oplus0
  \oplus2\mu_{2}
  \oplus\red{(\mu_{1}+2\mu_{2})}\oplus\red{\mu_{2}}
  \oplus\red{(\mu_{1}+2\mu_{2})}\oplus\red{\mu_{2}};
\end{align}
\begin{align}
 & [\mu_{1}\oplus\mu_{2}]\otimes[\mu_{1}\oplus\mu_{2}]\otimes(2\mu_{1}+2\mu_{2})=[2\mu_{1}\oplus0\oplus2\mu_{2}\oplus0\oplus\red{(\mu_{1}+\mu_{2})}\oplus\red{(\mu_{1}+\mu_{2})}]\otimes(2\mu_{1}+2\mu_{2})\\
= & 2\mu_{2}
  \oplus(2\mu_{1}+2\mu_{2})
  \oplus2\mu_{1}
  \oplus(2\mu_{1}+2\mu_{2})
  \oplus\red{(\mu_{1}+\mu_{2})}
  \oplus\red{(\mu_{1}+\mu_{2})}.
\end{align}
Thus, we get Table~\ref{aptab:DFnoparityfixing}. The red representations will not appear in the double fusion results if we fix the parity.
\begin{table}
\begin{centering}
\begin{tabular}{|c|c|c|c|c|}
\hline 
$a_{1}$ & $a_{2}$ & $Q\,\text{mod}4$ & $l\big[E-E^{(0)}\big]/\pi v_{F}$ & $n$\tabularnewline
\hline 
\hline 
$0$ & $0$ & $0$ & $0$ & $2$\tabularnewline
\hline 
\red{$1$} & \red{$1$} & $0$ & $3/8$ & $2$\tabularnewline
\hline 
$2$ & $0$ & $0$ & $1/2$ & $1$\tabularnewline
\hline 
$0$ & $2$ & $0$ & $1/2$ & $1$\tabularnewline
\hline 
 &  &  &  & \tabularnewline
\hline 
\red{$0$} & \red{$0$} & $1$ & $1/8$ & $2$\tabularnewline
\hline 
$1$ & $0$ & $1$ & $5/16$ & $1$\tabularnewline
\hline 
$0$ & $1$ & $1$ & $5/16$ & $1$\tabularnewline
\hline 
$1$ & $1$ & $1$ & $1/2$ & $2$\tabularnewline
\hline 
\red{$2$} & \red{$0$} & $1$ & $5/8$ & $2$\tabularnewline
\hline 
\red{$0$} & \red{$2$} & $1$ & $5/8$ & $2$\tabularnewline
\hline 
$2$ & $1$ & $1$ & $13/16$ & $1$\tabularnewline
\hline 
$1$ & $2$ & $1$ & $13/16$ & $1$\tabularnewline
\hline 
\red{$2$} & \red{$2$} & $1$ & $9/8$ & $2$\tabularnewline
\hline 
 &  &  &  & \tabularnewline
\hline 
$0$ & $0$ & $2$ & $1/2$ & $1$\tabularnewline
\hline 
\red{$1$} & \red{$0$} & $2$ & $11/16$ & $2$\tabularnewline
\hline 
$2$ & $0$ & $2$ & $1$ & $2$\tabularnewline
\hline 
\red{$2$} & \red{$1$} & $2$ & $19/16$ & $2$\tabularnewline
\hline 
$2$ & $2$ & $2$ & $3/2$ & $1$\tabularnewline
\hline 
 &  &  &  & \tabularnewline
\hline 
$0$ & $0$ & $2$ & $1/2$ & $1$\tabularnewline
\hline 
\red{$0$} & \red{$1$} & $2$ & $11/16$ & $2$\tabularnewline
\hline 
$0$ & $2$ & $2$ & $1$ & $2$\tabularnewline
\hline 
\red{$1$} & \red{$2$} & $2$ & $19/16$ & $2$\tabularnewline
\hline 
$2$ & $2$ & $2$ & $3/2$ & $1$\tabularnewline
\hline 
 &  &  &  & \tabularnewline
\hline 
\red{$1$} & \red{$1$} & $0$ & $3/8$ & $2$\tabularnewline
\hline 
$2$ & $0$ & $0$ & $1/2$ & $1$\tabularnewline
\hline 
$0$ & $2$ & $0$ & $1/2$ & $1$\tabularnewline
\hline 
$2$ & $2$ & $0$ & $1$ & $2$\tabularnewline
\hline 
 &  &  &  & \tabularnewline
\hline 
\red{$0$} & \red{$0$} & $3$ & $1+1/8$ & $2$\tabularnewline
\hline 
$1$ & $0$ & $3$ & $1+5/16$ & $1$\tabularnewline
\hline 
$0$ & $1$ & $3$ & $1+5/16$ & $1$\tabularnewline
\hline 
$1$ & $1$ & $3$ & $1+1/2$ & $2$\tabularnewline
\hline 
\red{$2$} & \red{$0$} & $3$ & $1+5/8$ & $2$\tabularnewline
\hline 
\red{$0$} & \red{$2$} & $3$ & $1+5/8$ & $2$\tabularnewline
\hline 
$2$ & $1$ & $3$ & $1+13/16$ & $1$\tabularnewline
\hline 
$1$ & $2$ & $3$ & $1+13/16$ & $1$\tabularnewline
\hline 
\red{$2$} & \red{$2$} & $3$ & $1+9/8$ & $2$\tabularnewline
\hline 
\end{tabular}
\par\end{centering}
\caption{Table for the primary boundary operators and their energies/scaling dimensions at the Kondo fixed point for $\text{SO(4)}_2$ obtained by double fusion with the impurity. The red representations will not appear in the double fusion results if we fix the parity.} \label{aptab:DFnoparityfixing}
\end{table}
If we do not fix the parity, the primary operators $\phi_{\alpha}^{(a_1=1,a_2=1,Q=0)}$ have the smallest scaling dimension $3/8$ in the table. However, we can not construct a set of operators that from this representation $a_1=a_2=1$ using the generators $\vec{J}^{(0,2)}$. Thus, the leading irrelevant operator will be $\vec{J}^{(0,2)}_{-1}\cdot\vec{\phi}^{(0,2)}+\vec{J}^{(2,0)}_{-1}\cdot\vec{\phi}^{(2,0)}$ where $\vec{\phi}^{(2,0)}$ are the three $2\mu_{1}$-primary operators and $\vec{J}^{(2,0)}$ are the $2\mu_{1}$-operators [three
generators of all the six generators for $\text{SO}(4)$ and the other three operators are $\vec{J}^{(0,2)}$]. This leading irrelevant operator has scaling dimension $1+1/2=3/2$ becayse the primary operators $\vec{\phi}^{(0,2)}$ and $\vec{\phi}^{(2,0)}$ both have scaling dimension $1/2$. 
If we fix the parity, we still use these two primary operators $\vec{\phi}^{(0,2)}$ and $\vec{\phi}^{(2,0)}$ to form the same leading irrelevant operator.

\section{The representation of the density $J^{(d)}$} \label{ap:rep2mu1}
In Sec.~\ref{sec:T0conduc}, we defined $J^{(d)}_{\chi}(x)=\sum_{j}\vec{\psi}^\dagger_{\chi;j}(x)D_1\vec{\psi}_{\chi;j}(x)$ where $D_1=\text{diag}(1,-1,0,\dots,0)$. The sum over channel index $j$ means that $J^{(d)}$ is a singlet in the channel sector. In the flavor [$\text{SO}(M)$] sector, a general form for a density operator is $\sum_{ab} c^\dagger_a B_{ab}^{\alpha} c_b$ where $a,b=1,\dots,M$. One finds that there are three sets of them: the first one is given by $B^{1}=\mathbb{I}_{M\times M}$, a symmetric traceful matrix; the second set of densities are the $[M(M-1)/2]$ antisymmetric traceless matrices; the last set of densities are the $[(M-1)+M(M-1)/2]$ symmetric traceless matrices. The matrix $D_1$ is one of these $M-1+M(M-1)/2$ symmetric traceless $M\times M$ matrices. The representation of these three sets of densities can be determined by decomposition of two 1-particle representations (the 1-particle representation has dimension $M$), which are two $2\mu_1$s (spin-1s) for SO($3$), two $\mu_1+\mu_2$s for SO($4$) [see Eq.~(\ref{apeq:JArepso4})] and two $\mu_1$s for SO($M\geq 5$). These decompositions are listed in the following equation.
\begin{equation}
\begin{cases}
2\mu_1 \otimes 2\mu_1 = 4\mu_1 \oplus 2\mu_1\oplus \bm{0}, & \text{for}\ \text{SO}(3)\\
(\mu_1 + \mu_2) \otimes (\mu_1 + \mu_2) = (2\mu_1 + 2\mu_2) \oplus 2\mu_1 \oplus 2\mu_2 \oplus \bm{0}, & \text{for}\ \text{SO}(4)\\
\mu_1 \otimes \mu_1 = 2\mu_1 \oplus 2\mu_2 \oplus \bm{0}, & \text{for}\ \text{SO}(5)\\
\mu_1 \otimes \mu_1 = 2\mu_1 \oplus (\mu_2+\mu_3) \oplus \bm{0}, & \text{for}\ \text{SO}(6)\\
\mu_1 \otimes \mu_1 = 2\mu_1 \oplus \mu_2 \oplus \bm{0}, & \text{for}\ \text{SO}(M\geq 7).
\end{cases}
\end{equation}
The $\bm{0}$ is the identity $B^1$. The first representations on the right side (also the representation of $J^{(d)}$) are for the $M-1+M(M-1)/2$ real traceless $M\times M$ matrices; The rest of them are for the $[M(M-1)/2]$ antisymmetric traceless (the adjoint) $M\times M$ matrices.

\end{widetext}

\end{appendix}

\end{document}